\documentclass{aastex631}
\usepackage{csquotes}

\shorttitle{Scylla III}
\shortauthors{Cohen et al.}

\begin{document}
\title{Scylla III. The Outside-In Radial Age Gradient in the Small Magellanic Cloud and the Star Formation Histories of the Main Body, Wing and Outer Regions} 

\correspondingauthor{Roger E. Cohen}
\email{rc1273@physics.rutgers.edu}

\author[0000-0002-2970-7435]{Roger E. Cohen}
\affiliation{Department of Physics and Astronomy, Rutgers the State University of New Jersey, 136 Frelinghuysen Rd., Piscataway, NJ, 08854, USA}

\author[0000-0001-5538-2614]{Kristen B. W. McQuinn}
\affiliation{Department of Physics and Astronomy, Rutgers the State University of New Jersey, 136 Frelinghuysen Rd., Piscataway, NJ, 08854, USA}
\affiliation{Space Telescope Science Institute, 3700 San Martin Drive, Baltimore, MD 21218, USA}

\author[0000-0002-7743-8129]{Claire E. Murray}
\affiliation{Space Telescope Science Institute, 3700 San Martin Drive, Baltimore, MD 21218, USA}
\affiliation{The William H. Miller III Department of Physics \& Astronomy, Bloomberg Center for Physics and Astronomy, Johns Hopkins University, 3400 N. Charles Street, Baltimore, MD 21218, USA}

\author{Benjamin F. Williams}
\affiliation{Department of Astronomy, University of Washington, Box 351580, U. W., Seattle, WA 98195-1580, USA}

\author[0000-0003-1680-1884]{Yumi Choi}
\affiliation{NSF National Optical-Infrared Astronomy Research Laboratory, 950 N. Cherry Avenue, Tucson, AZ 85719 USA}

\author[0000-0003-0588-7360]{Christina W. Lindberg}
\affiliation{The William H. Miller III Department of Physics \& Astronomy, Bloomberg Center for Physics and Astronomy, Johns Hopkins University, 3400 N. Charles Street, Baltimore, MD 21218, USA}
\affiliation{Space Telescope Science Institute, 3700 San Martin Drive, Baltimore, MD 21218, USA}

\author[0009-0005-0339-015X]{Clare Burhenne}
\affiliation{Department of Physics and Astronomy, Rutgers the State University of New Jersey, 136 Frelinghuysen Rd., Piscataway, NJ, 08854, USA}

\author[0000-0001-5340-6774]{Karl~D.~Gordon}
\affiliation{Space Telescope Science Institute, 3700 San Martin Drive, Baltimore, MD 21218, USA}

\author{Petia Yanchulova Merica-Jones}
\affiliation{Space Telescope Science Institute, 3700 San Martin Drive, Baltimore, MD 21218, USA}

\author[0000-0001-6118-2985]{Caroline Bot}
\affiliation{Observatoire Astronomique de Strasbourg, Universit\'e de Strasbourg, UMR 7550, 11 rue de l'Universit\'e, F-67000 Strasbourg, France}

\author{Andrew E. Dolphin}
\affiliation{Raytheon, 1151 E. Hermans Road, Tucson, AZ 85756, USA} 
\affiliation{University of Arizona, Steward Observatory, 933 North Cherry Avenue, Tucson, AZ 85721, USA}

\author[0000-0003-0394-8377]{Karoline M. Gilbert}
\affiliation{Space Telescope Science Institute, 3700 San Martin Drive, Baltimore, MD 21218, USA}
\affiliation{The William H. Miller III Department of Physics \& Astronomy, Bloomberg Center for Physics and Astronomy, Johns Hopkins University, 3400 N. Charles Street, Baltimore, MD 21218, USA}

\author[0000-0002-8937-3844]{Steven Goldman}
\affiliation{Space Telescope Science Institute, 3700 San Martin Drive, Baltimore, MD 21218, USA}

\author{Alec S. Hirschauer}
\affiliation{Space Telescope Science Institute, 3700 San Martin Drive, Baltimore, MD 21218, USA}

\author[0000-0002-4378-8534]{Karin M. Sandstrom}
\affil{Department of Astronomy \& Astrophysics, University of California San Diego, 9500 Gilman Drive, La Jolla, CA 92093, USA}

\author{O. Grace Telford}
\altaffiliation{Carnegie-Princeton Fellow}
\affiliation{Department of Astrophysical Sciences, Princeton University, 4 Ivy Lane, Princeton, NJ 08544, USA}
\affiliation{The Observatories of the Carnegie Institution for Science, 813 Santa Barbara Street, Pasadena, CA 91101, USA}
\affiliation{Department of Physics and Astronomy, Rutgers the State University of New Jersey, 136 Frelinghuysen Rd., Piscataway, NJ, 08854, USA}

\begin{abstract}
     
The proximity of the Large and Small Magellanic Clouds (LMC and SMC) provides the opportunity to study the impact of dwarf-dwarf interactions on their mass assembly with a unique level of detail.  To this end, we analyze two-filter broadband imaging of 83 Hubble Space Telescope (HST) pointings covering 0.203 deg$^2$ towards the SMC, extending out to $\sim$3.5 kpc in projection from its optical center.  
Lifetime star formation histories (SFHs) fit to each pointing independently reveal an outside-in age gradient such that fields in the SMC outskirts are older on average.  We measure radial gradients of the lookback time to form 90\%, 75\% and 50\% of the cumulative stellar mass for the first time, finding $\delta$($\tau_{90}$, $\tau_{75}$, $\tau_{50}$)/$\delta$R  
= (0.61$^{+0.08}_{-0.07}$, 0.65$^{+0.09}_{-0.08}$, 0.82$^{+0.12}_{-0.16}$) Gyr/kpc 
assuming PARSEC evolutionary models and a commonly used elliptical geometry of the SMC, although our results are robust to these assumptions. 
The wing of the SMC deviates from this trend, forming 25\% of its cumulative mass over the most recent 3 Gyr due to a best-fit star formation rate that remains approximately constant.  
Our results are consistent with chemodynamical evidence of a tidally stripped SMC component in the foreground, and imply contributions to the observed SFH from multiple previous LMC-SMC interactions.  We also compare our SMC SFH with results from a companion study of the LMC, finding that while the two galaxies present different internal, spatially resolved SFH trends, both the LMC and SMC have similar near-constant lifetime SFHs when viewed globally.  

\end{abstract}

\section{Introduction}

The evolution of dwarf galaxies is inextricably tied to their environment.  The timeline of their mass assembly is fundamentally affected by their interactions, both with each other and with more massive hosts in a complex, multifaceted way \citep[e.g.,][]{weisz15,geha17,engler23}.  A particularly prominent role, over cosmic time, is played by dwarf-dwarf interactions.  Dwarfs contributed to galaxy growth via mergers in a hierarchical accretion scenario \citep[e.g.,][]{wetzel15a}, including \enquote{satellites of satellites} \citep[e.g.,][]{martinezdelgado12,deason15,annibali16} and the coalescence of isolated groups of dwarfs that were more common in the past \citep{stierwalt17}.  Through the present day, both observations and simulations suggest that interactions between gas-rich dwarfs trigger star formation.  Observationally, paired dwarfs (7$\leq$Log$_{\rm 10}  M_{\star}/M_{\odot}$$\leq$9.7) show enhanced star formation rates compared to their single counterparts regardless of environment (\citealt{stierwalt15} and references therein, also see \citealt{kadofong24}).  From a theoretical standpoint, cosmological simulations suggest that increased star formation in dwarfs (7$\leq$Log$_{\rm 10}  M_{\star}/M_{\odot}$$\leq$9.5) can be caused by flybys without requiring a merger \citep{martin21}.

The Large and Small Magellanic Clouds (LMC and SMC), because of their proximity ($\approx$50 and $\approx$62 kpc respectively; \citealt{degrijsbono14,degrijsbono15}), provide us with a uniquely detailed view of an ongoing dwarf-dwarf interaction.  The LMC-SMC interaction has likely been a prolonged one, lasting as long as $\sim$6$-$7 Gyr based on simulations \citep{diazbekki12,besla12}.  Importantly, this interaction substantially predates their infall into the Milky Way, entering its virial radius only $\sim$1$-$2 Gyr ago \citep[][but see \citealt{vasiliev24}]{kallivayalil06a,besla07,besla12,patel17}.  Accordingly, many of the features we observe in the vicinity of the LMC and SMC are likely due to their interaction with each other rather than with the Milky Way, including a recent ($\sim$140$-$160 Myr ago), near-direct (impact parameter $\sim$5 kpc) collision \citep{zivick18,choi22}.  

Among the plethora of interaction signatures imprinted in the morphology and chemodynamics of the LMC and SMC, some have been traced to particular past interactions.  
The Magellanic Bridge \citep{hindman63} is thought to result from the most recent LMC-SMC collision $\sim$150 Myr ago based on its morphology, chemistry and kinematics \citep{besla12,diazbekki12,zivick18,zivick19}, although the presence of metal-poor massive O stars has been used to argue for some role of earlier interactions \citep{ramachandran21}.  
In contrast to the Bridge, the Magellanic Stream is thought to have been formed $\sim$1.5$-$2.5 Gyr ago, through some combination of tidal stripping from the LMC and SMC \citep{nidever08,besla12,diazbekki12} and/or ram pressure stripping from the LMC and Milky Way \citep[e.g.,][]{meurer85,lucchini20}.
In addition to the Bridge and the Stream, initially characterized based on the large-scale structure of their gaseous components, 
multiple LMC-SMC interactions have left many additional signatures in the morphology, chemistry and kinematics of the Clouds \citep[e.g.,][]{olsen02,olsen11,dobbie14a,mackey18,deleo20,grady21,massana24}.  The LMC, as the more massive of the two, has survived repeated perturbations by the SMC with its disk-like 
morphology and kinematics relatively intact \citep{yumiring,choi22}.  However, the structure of the less massive SMC is much more complex, with tidal stripping by the LMC likely playing an important role \citep[e.g.,][]{noel13,zivick18,james21,niederhofer21,omkumar21,zivick21,murray23}.  

The highly disturbed nature of the SMC is underscored by its gas and dust properties: Neutral hydrogen has a complex, position-dependent distance and velocity distribution \citep[e.g.,][]{deleo20,gaskap,murray23}, while the dust distribution is  separate from the stellar distribution along the line of sight \citep{petia21}.  The stellar distribution of the SMC shows plentiful evidence of past disturbances as well.  Various stellar substructures have been discovered in the periphery of the SMC \citep{mackey18,belokuroverkal19,elyoussoufi21,grady21,gaia21}, some of which show bimodalities in their abundances and/or kinematics \citep{parisi15,parisi16,pieres17,parisi22,cullinane23,massana24}, in addition to the shell-like structure $\sim$2$^{\circ}$ to the northeast \citep{martinezdelgado19,sakowska23}.   

Perhaps the most prominent stellar substructure of the SMC is its wing to the east in the direction of the Bridge \citep{shapley40}.  The wing hosts composite stellar populations \citep{noel13} and is spatially and kinematically distinct from both the Bridge \citep[e.g.,][]{muraveva18,omkumar21} and the inner SMC, showing residual motions towards the LMC beyond the tidal expansion seen also on the opposite (western) side of the SMC \citep{oey18,deleo20,zivick21,niederhofer21,gaia21,elyoussoufi23}.  The structure of the wing is complex along the line of sight as well: although a substantial ($\gtrsim$5 kpc) and position-dependent line-of-sight depth is seen throughout the SMC \citep{sub09,subsub12,jd16,jd17,tatton21,petia21,elyoussoufi21}, the distance distribution towards the wing is bimodal.  In particular, the magnitude distribution of red clump stars reveals a feature $\sim$10$-$12 kpc in front of the main body of the SMC along the line of sight \citep{hh89,nidever13,sub17,tatton21,elyoussoufi21}, and a bimodality is also seen in kinematics of red clump and red giant stars towards the wing  \citep{omkumar21,james21,elyoussoufi23}.  The most recent LMC-SMC collision, thought to have created the Bridge, is also likely responsible for the bimodalities in stellar distances and kinematics towards the wing.  In particular, \citet{almeida23a} claim that the foreground component originated in the inner SMC, and was tidally stripped during the most recent LMC-SMC collision.  Their claim is supported by trends in radial velocities and distances for the foreground component.  The foreground component examined by \citet{almeida23a} may extend from the innermost regions of the LMC, suggesting an approach towards the LMC at increasingly larger distances from the SMC center.  

None of the bimodalities seen in stellar distances and chemodynamics towards the wing are present in the distribution of ancient RR Lyrae variables, although their distances are skewed towards the LMC in the vicinity of the wing \citep{jd17,muraveva18}.  Viewed together with the cylindrical rotation seen in red giants in the inner SMC ($\lesssim$2 kpc; \citealt{zivick18}), the oldest stellar populations of the SMC appear to have been the least affected by LMC-SMC interactions.  Accordingly, the stellar morphology of the SMC, viewed in projection, is highly age-dependent.  While old stars (red giants and RR Lyrae) have a shallow, negative radial metallicity gradient \citep[e.g.,][]{choudhury20,massana24} and a relatively smooth, extended, symmetric projected spatial distribution, at younger ages the stellar distribution appears increasingly asymmetric, tracing the irregular HI distribution \citep{hz04,cignoni13,rubele18,elyoussoufi19,massana20}.  

It is clear from the above evidence that the distribution of stellar populations towards the SMC is complex, since it is both age-dependent in a gross sense while also impacted, selectively, by LMC-SMC interactions.
In this context, a powerful additional avenue for disentangling the history of the SMC and its interaction with the LMC is presented by star formation histories (SFHs) ascertained from photometry of resolved stars. With this in mind, the SMC has been the target of various imaging campaigns that have already yielded crucial insights on its past assembly and interaction history, including the Magellanic Clouds Photometric Survey \citep[MCPS;][]{hz04}, the VISTA Survey of the Magellanic Clouds system \citep[VMC;][]{cioni11} and the Survey of the MAgellanic Stellar History \citep[SMASH;][]{smashdr1}.  Because such imaging campaigns can become crowding-limited in the innermost region of the SMC \citep[e.g.,][]{massana22}, additional insight has come from targeted Hubble Space Telescope (HST) imaging \citep{mccumber05,chiosi07,sabbi09,cignoni12,cignoni13,weisz13} as well as deep ground-based observations of selected fields in the outer SMC \citep{dolphin01,noelsfh}.  By trading spatial coverage for photometric depth, such observations, reaching faintward of the ancient ($\sim$13Gyr) main sequence turnoff\footnote{The main sequence turnoff occurs at absolute magnitudes of (M$_{F475W}$, M$_{F814W}$) = (4.15, 3.28) in the WFC3/UVIS filters used in our Scylla program and M$_{V}$ = 3.90 assuming 13 Gyr PARSEC evolutionary models \citep{parsec} with [M/H]=$-$1.2.}, help break age-metallicity degeneracies in fitting resolved SFHs to simultaneously measure both stellar mass assembly history and chemical evolution over cosmic timescales \citep[e.g.,][]{annibalitosi}.  
 
To date, both contiguous ground-based surveys and deep, targeted 
imaging towards the SMC have produced valuable constraints on the history of the LMC-SMC interaction.  
An interaction between the LMC and SMC $\sim$3$-$4 Gyr ago is evidenced by a global enhancement in star formation in the SMC \citep{weisz13,massana22}, with recent star formation concentrated near the center and the east towards the wing \citep{cignoni13,rubele18,massana22}, while at $\gtrsim$2 kpc from the SMC center, star formation has dramatically decreased over the last few Gyr \citep{dolphin01,noelsfh}.  Based on SFH fits to six fields with deep HST imaging, \citet{cignoni13} suggest an outside-in gradient in mean age in the SMC, such that the outskirts (except for the wing) are older on average (also see \citealt{carrera08,dobbie14}).  In contrast to these results, \citet{piatti12,piatti15} claim a null radial age gradient among the SMC field population, although they do not attempt to quantify this result.  In fact, despite various investigations of the spatially resolved SFH in the SMC field \citep{hz04,carrera08,noelsfh,piatti12,cignoni13,weisz13,piatti15,rubele15,rubele18,massana22}, a radial age gradient in the SMC has yet to be quantified.  

Given the extended and turbulent interaction history of the SMC, spatially resolved lifetime SFH measurements can provide important insights into the role of dwarf-dwarf interactions on the mass assembly of dwarf galaxies.  Although current models of the LMC-SMC-Milky Way interaction lack realistic star formation prescriptions \citep{diazbekki12,besla12}, measurements of spatial trends in stellar age in the SMC (and LMC) provide quantitative constraints that must be met by future models.  In addition, the latest zoom-in cosmological hydrodynamical simulations make predictions for radial age gradients in dwarf galaxies \citep{graus}, but these predictions remain untested.  While these simulations currently lack direct LMC-SMC-Milky Way analogs, empirical measurements of radial stellar age gradients in the SMC and LMC again provide important tests of such models for the limiting case of extended dwarf-dwarf interactions moving forward.  Lastly, a comparison between SFH trends in the wing of the SMC versus other fields in the outer SMC as well as the main body can help discern which previous LMC-SMC interactions contributed to which present-day structures observed in the SMC (i.e., wing, main body).  

This paper is the third in a series presenting results from Scylla, a multi-cycle pure parallel HST campaign imaging 96 fields towards the LMC and SMC.  In addition to an initial paper presenting an overview of the Scylla program (Murray et al.~2024), in a companion paper we present the spatially resolved SFH of the LMC (Cohen et al.~2024), combining our Scylla imaging with a plethora of archival HST imaging to detect and characterize spatial trends in the SFH of the LMC over its entire lifetime.  Here, we take a similar approach with the SMC, fitting SFHs individually to an ensemble of HST pointings and using the results to characterize spatial trends in the SFH of the SMC.  This paper is organized as follows: We describe our observations and data reduction in Sect.~\ref{datasect}, and our methodology for SFH fitting in Sect.~\ref{sfhsect}.  In Sect.~\ref{resultsect}, we present our results, broken down both into subsets based on location in the SMC as well as for individual fields.  In Sect.~\ref{discusssect} we discuss constraints on the LMC-SMC interaction and place our SMC results in the context of other Local Group dwarfs, and our main conclusions are summarized in the final section.  Throughout our analysis, we assume an SMC distance of 62 kpc \citep[e.g.,][]{degrijsbono15}, and an elliptical geometry with an axis ratio of 2:1 and a position angle of 45$^{\circ}$ east of north \citep[e.g.,][]{piatti05,piatti12,dias14,parisi22} to assign a projected semi-major axis equivalent distance R$_{ell}^{SMC}$ to each of our observed fields.       

\section{Data \label{datasect}}
\subsection{Observations \label{obssect}}
\begin{figure}
\gridline{\fig{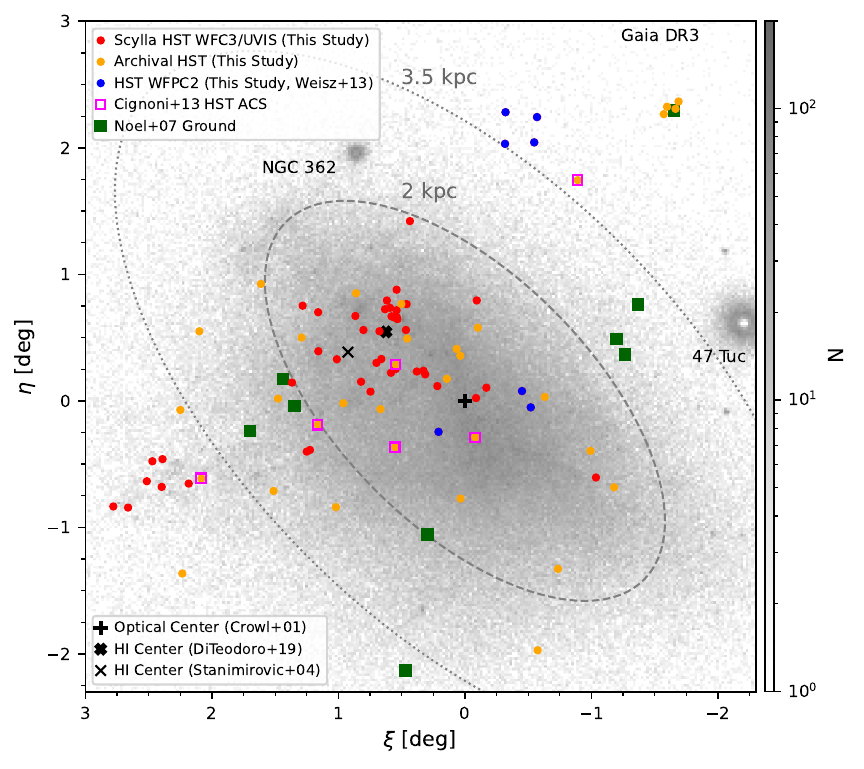}{0.6\textwidth}{}}
\vspace{-1.11cm}
\gridline{\fig{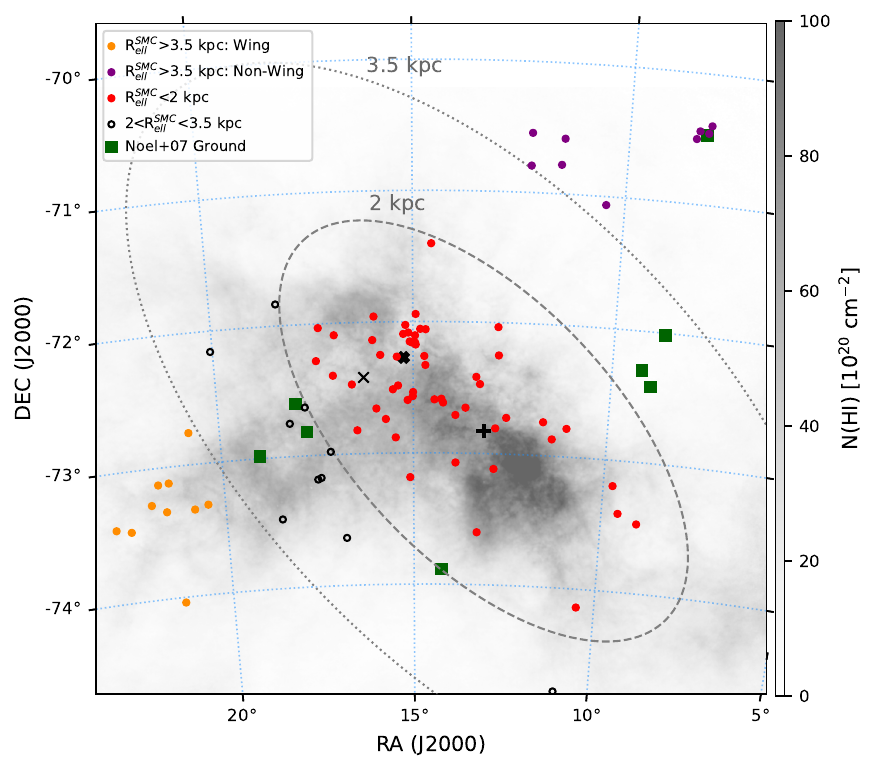}{0.6\textwidth}{}}
\vspace{-1cm}
\caption{\textbf{Top:} Map of the fields analyzed here, overplotted on a stellar density map from \textit{Gaia} DR3 (see text for details).  Our fields are shown as filled circles, color-coded by the source of the imaging, including new WFC3/UVIS imaging from Scylla (red), archival ACS/WFC and WFC3/UVIS imaging (orange), and archival WFPC2 LGSPA photometry from \citet{lgspa} (blue, also analyzed by \citealt{weisz13}).  The subset of archival ACS/WFC fields also analyzed by \citet{cignoni13} are marked by magenta squares, and ground-based fields from \citet{noelphot} are shown as green squares.  Black symbols indicate different locations of the SMC center, including the optical center we adopt \citep{crowl01,subsub12} and the HI centers from \citet{diteodoro19} and \citet{stan04}.  \textbf{Bottom:} Our fields are overplotted on the GASKAP HI column density map \citep{gaskap}, now color-coded by location within the SMC, including the main body (red), wing (orange), and outer non-wing fields (purple; see text for details).  Open circles indicate fields not falling into any of these three subsets, with 2$<$R$_{ell}^{SMC}$$<$3.5 kpc.  
\label{map_fig}}
\end{figure}

To analyze spatial trends in the lifetime star formation history of the SMC, we leverage new and archival imaging from 83 individual HST pointings consisting of deep imaging in at least two broadband filters.  The spatial distribution of our fields is shown in Fig.~\ref{map_fig}, 
centered on the SMC optical center of (RA, Dec)$_{\rm J2000, optical}$ = (00:52:45, $-$72:49:43) \citep{crowl01,subsub12}.  In the upper panel of Fig.~\ref{map_fig}, our field locations are color-coded by observational source (i.e., new or archival imaging, described below), overplotted on a density map built from \textit{Gaia} DR3 \citep{gaiadr3}\footnote{For the purposes of constructing an illustrative background density map, we used only very loose parallax and color-magnitude criteria to select \textit{Gaia} sources: \texttt{parallax$<$0.1 AND parallax\_over\_error$<$ 5 AND ruwe$<$1.4 AND (phot\_bp\_mean\_mag$-$phot\_rp\_mean\_mag)$>$$-$1.0 AND phot\_g\_mean\_mag$>$10.6 AND phot\_g\_mean\_mag$<$19.0.}}.  We also indicate the locations of fields analyzed in previous targeted studies of spatial variations in the SFH of the SMC, including HST pointings from \citet{weisz13} and \citet{cignoni13} (which we reanalyze), and the ground-based fields analyzed by \citet{noelphot,noelsfh}.  An ellipse corresponding to R$_{ell}^{SMC}$=2 kpc is shown as a dotted grey line to indicate the main body of the SMC.

The morphology of the SMC varies strongly as a function of stellar age.  While old (red giants, red clump and RR Lyrae) stellar populations show a smooth elliptical spatial distribution \citep{zaritsky00,jd17}, the spatial distribution of younger stars is much more clumpy and asymmetric, with young stars concentrated in both the central region and the wing to the southeast \citep[e.g.,][]{jd16,ripepi17,elyoussoufi19}, tracing the irregular structure of neutral hydrogen \citep{gaskap} shown in the lower panel of Fig.~\ref{map_fig}.  Given this age-dependent spatial distribution, we group our fields into subsets based on location in the SMC, overplotted on the HI column density map in the lower panel of Fig.~\ref{map_fig}, including those located in the main body (red), wing (orange) and, for comparison to fields in the wing, the non-wing fields located at similarly large distances from the SMC center (purple), with projected distances $R_{ell}^{SMC}$$>$3.5 kpc.  

\subsubsection{Scylla Imaging}

The new imaging we analyze was obtained as part of the Scylla multi-cycle pure parallel program (GO-15891, GO-16235, GO-16786; PI: C.~E.~Murray).  This program accompanied the HST Ultraviolet Legacy library of Young Stars as Essential Standards (ULYSSES; \citealt{ulysses}) spectroscopic campaign, a Directors' Discretionary Time program designed to perform  ultraviolet spectroscopy of massive O and B stars towards the Magellanic Clouds.  During the ULYSSES spectroscopic primary observations, Scylla obtained parallel imaging with Wide Field Camera 3 \citep[WFC3;][]{wfc3} in as many as seven broadband filters ranging from the ultraviolet (F225W) to the near-infrared (F160W).  However, in the majority of cases, the dithering and exposure time requirements of the ULYSSES primary observations only allowed parallel exposures in a subset of filters.  Therefore, the WFC3/UVIS F475W and F814W filters were assigned top priority as a compromise between high throughput and a broad color baseline. As a result, Scylla has obtained two-filter broadband imaging in the WFC3/UVIS F475W and F814W filters across all of the observed fields.  An overview of the Scylla pure parallel imaging program, including science goals and detailed exposure-level information, is provided in Murray et al.~(2024).  Basic observational information for the fields we use is given in Table \ref{obstab}.  

\subsubsection{Archival Imaging}

To improve our spatial coverage of the SMC, we searched the \textit{HST} archive for additional archival imaging of sufficient depth ($\gtrsim$1 mag faintward of the ancient main sequence turnoff) in at least two broadband optical filters, listed in Table \ref{obstab}.  We rejected fields contaminated by known clusters (see Sect.~\ref{sfhsect} and Appendix \ref{clustersect}), including the Galactic globular clusters 47 Tuc and NGC 362 (and their tidal tails; \citealt{carballobello19}) as well as fields with severe internal differential extinction hampering SFH fits.  Specifically, we rejected fields with $\sigma_{1}$ + $\sigma_{2}$ $>$0.45 from \citealt{skowron}, where $\sigma_{1}$ and $\sigma_{2}$ represent Gaussian half-widths of the red clump distribution along the reddening vector (see the discussion in Cohen et al.~2024).  After applying these cuts (which were also applied to the aforementioned Scylla fields, causing the rejection of only one field, SMC\_50), we retained 33 archival fields, listed in Table \ref{obstab} with their field numbers preceded by the letter A.  In addition, we also reanalyze photometry from seven deep Wide Field Planetary Camera 2 (WFPC2) fields provided in the Local Group Stellar Photometry Archive \citep[LGSPA;][]{lgspa}, all of which were imaged in the F555W and F814W filters, listed in Table \ref{obstab} with field numbers beginning with the letter W.  

\startlongtable
\begin{deluxetable}{lcllcccll}
\tabletypesize{\scriptsize}
\tablecaption{Observed Fields \label{obstab}}
\tablehead{
\colhead{Field} & \colhead{Archive Name/PID} & \colhead{RA (J2000)} & \colhead{Dec (J2000)} & \colhead{Blue Filter} & \colhead{C80 (blue)} & \colhead{C80 (F814W)} & \colhead{N(obs)} & \colhead{N(used)} \\ \colhead{} & \colhead{} & \colhead{$^{\circ}$} & \colhead{$^{\circ}$} & \colhead{} & \colhead{mag} & \colhead{mag} & \colhead{} & \colhead{}}
\startdata
SMC\_1 & 15891\_SMC-3192ne-8290 & 14.946985 & -72.104922 & WFC3/UVIS F475W & 27.12 & 24.92 & 23661 & 17559 \\ 
SMC\_2 & 15891\_SMC-2750ne-8567 & 15.514903 & -72.514790 & WFC3/UVIS F475W & 27.66 & 25.37 & 21629 & 15800 \\ 
SMC\_7 & 15891\_SMC-5278ne-9802 & 14.551469 & -71.403310 & WFC3/UVIS F475W & 27.77 & 25.40 & 7229 & 5761 \\ 
SMC\_8 & 15891\_SMC-3955ne-9818 & 16.011691 & -72.137054 & WFC3/UVIS F475W & 27.01 & 24.84 & 15527 & 11867 \\ 
SMC\_9 & 15891\_SMC-3587ne-10112 & 15.186138 & -72.025194 & WFC3/UVIS F475W & 27.27 & 25.06 & 19954 & 15152 \\ 
SMC\_10 & 15891\_SMC-3149ne-12269 & 15.080012 & -72.151312 & WFC3/UVIS F475W & 26.48 & 24.34 & 8355 & 5960 \\ 
SMC\_11 & 15891\_SMC-8743se-11371 & 21.442971 & -73.124046 & WFC3/UVIS F475W & 27.51 & 25.24 & 2322 & 1759 \\ 
SMC\_13 & 15891\_SMC-2983ne-12972 & 14.931875 & -72.174660 & WFC3/UVIS F475W & 26.43 & 24.32 & 8563 & 5949 \\ 
SMC\_14 & 15891\_SMC-3669ne-13972 & 14.930241 & -71.942178 & WFC3/UVIS F475W & 27.41 & 25.15 & 17598 & 13940 \\ 
SMC\_15 & 15891\_SMC-4292sw-13841 & 9.5604038 & -73.403032 & WFC3/UVIS F475W & 26.80 & 24.66 & 18525 & 13894 \\ 
SMC\_16 & 15891\_SMC-3435ne-13258 & 15.241572 & -72.093556 & WFC3/UVIS F475W & 26.82 & 24.81 & 16438 & 13197 \\ 
SMC\_17 & 15891\_SMC-1588ne-12105 & 14.458894 & -72.592704 & WFC3/UVIS F475W & 26.74 & 24.63 & 31824 & 22077 \\ 
SMC\_18 & 15891\_SMC-3029ne-13288 & 15.942742 & -72.658791 & WFC3/UVIS F475W & 26.60 & 24.50 & 16688 & 12500 \\ 
SMC\_19 & 15891\_SMC-3032ne-13306 & 14.944591 & -72.160401 & WFC3/UVIS F475W & 26.98 & 24.80 & 24343 & 17780 \\ 
SMC\_20 & 15891\_SMC-3370ne-13459 & 15.113533 & -72.083431 & WFC3/UVIS F475W & 26.81 & 25.00 & 15413 & 12678 \\ 
SMC\_21 & 15891\_SMC-3104ne-13781 & 15.027090 & -72.159001 & WFC3/UVIS F475W & 26.48 & 24.40 & 12900 & 9334 \\ 
SMC\_22 & 15891\_SMC-9034se-13316 & 21.724082 & -73.129718 & WFC3/UVIS F475W & 27.75 & 25.39 & 2197 & 1717 \\ 
SMC\_23 & 15891\_SMC-2584ne-14274 & 14.717750 & -72.262425 & WFC3/UVIS F475W & 26.33 & 24.23 & 21414 & 15213 \\ 
SMC\_25 & 16235\_SMC-879ne-11082 & 13.921283 & -72.710449 & WFC3/UVIS F475W & 26.50 & 24.44 & 30370 & 19583 \\ 
SMC\_27 & 16235\_SMC-2668ne-11415 & 15.381839 & -72.485760 & WFC3/UVIS F475W & 26.55 & 24.65 & 21913 & 15319 \\ 
SMC\_28 & 16235\_SMC-10336se-14099 & 22.971270 & -73.435217 & WFC3/UVIS F475W & 27.00 & 24.88 & 1072 & 713 \\ 
SMC\_29 & 16235\_SMC-3870ne-14647 & 16.551180 & -72.471367 & WFC3/UVIS F475W & 27.32 & 25.09 & 16339 & 11870 \\ 
SMC\_30 & 16235\_SMC-2259ne-15609 & 15.141961 & -72.597028 & WFC3/UVIS F475W & 26.63 & 24.50 & 23868 & 16060 \\ 
SMC\_33 & 16235\_SMC-2167ne-18821 & 15.010723 & -72.568012 & WFC3/UVIS F475W & 26.20 & 24.06 & 21132 & 12376 \\ 
SMC\_34 & 16235\_SMC-2728ne-28918 & 15.705088 & -72.739956 & WFC3/UVIS F475W & 26.57 & 24.53 & 20946 & 13972 \\ 
SMC\_35 & 16235\_SMC-2773nw-32334 & 12.884764 & -72.034680 & WFC3/UVIS F475W & 27.66 & 25.32 & 14696 & 11099 \\ 
SMC\_36 & 16235\_SMC-3127ne-32138 & 15.405336 & -72.265979 & WFC3/UVIS F475W & 26.58 & 24.54 & 19671 & 13507 \\ 
SMC\_37 & 16235\_SMC-3154ne-32442 & 14.684047 & -72.058285 & WFC3/UVIS F475W & 26.85 & 24.69 & 19023 & 13440 \\ 
SMC\_38 & 16235\_SMC-8151se-32530 & 20.821645 & -73.342871 & WFC3/UVIS F475W & 26.79 & 24.64 & 2270 & 1512 \\ 
SMC\_40 & 16235\_SMC-1339ne-33009 & 14.239291 & -72.616411 & WFC3/UVIS F475W & 26.53 & 24.35 & 21651 & 13676 \\ 
SMC\_41 & 16235\_SMC-286sw-34349 & 12.890781 & -72.806140 & WFC3/UVIS F475W & 26.21 & 24.27 & 39745 & 23347 \\ 
SMC\_42 & 16235\_SMC-1443ne-34945 & 14.283263 & -72.587845 & WFC3/UVIS F475W & 26.61 & 24.47 & 27018 & 16972 \\ 
SMC\_43 & 16235\_SMC-4996ne-34726 & 17.769602 & -72.631797 & WFC3/UVIS F475W & 26.49 & 24.38 & 10184 & 6571 \\ 
SMC\_44 & 16786\_SMC-4646se-5833 & 17.422733 & -73.173429 & WFC3/UVIS F475W & 26.84 & 24.71 & 10071 & 6938 \\ 
SMC\_45 & 15891\_SMC-641nw-12753 & 12.617762 & -72.723821 & WFC3/UVIS F475W & 26.51 & 24.58 & 40480 & 24937 \\ 
SMC\_46 & 16235\_SMC-4450ne-32733 & 17.022081 & -72.399457 & WFC3/UVIS F475W & 27.15 & 24.88 & 13578 & 9611 \\ 
SMC\_47 & 16786\_SMC-9277se-14900 & 21.952307 & -73.279210 & WFC3/UVIS F475W & 27.72 & 25.36 & 2000 & 1490 \\ 
SMC\_48 & 16786\_SMC-8904se-15007 & 21.570770 & -73.340365 & WFC3/UVIS F475W & 27.13 & 24.95 & 2173 & 1614 \\ 
SMC\_49 & 16786\_SMC-4926ne-15573 & 16.962114 & -72.091641 & WFC3/UVIS F475W & 27.44 & 25.19 & 12510 & 9426 \\ 
SMC\_52 & 16786\_SMC-9946se-16175 & 22.573365 & -73.462673 & WFC3/UVIS F475W & 27.89 & 25.52 & 1956 & 1407 \\ 
SMC\_53 & 16786\_SMC-4745se-7610 & 17.512373 & -73.184838 & WFC3/UVIS F475W & 26.85 & 24.75 & 9589 & 6739 \\ 
SMC\_54 & 16786\_SMC-3529ne-15172 & 15.819982 & -72.251044 & WFC3/UVIS F475W & 27.38 & 25.09 & 19942 & 14463 \\ 
SMC\_55 & 16786\_SMC-5409ne-15524 & 17.349532 & -72.032015 & WFC3/UVIS F475W & 27.55 & 25.27 & 9651 & 7248 \\ 
SMC\_A1 & 10248 & 12.852292 & -72.249989 & ACS/WFC F555W & 25.35 & 23.97 & 27790 & 14246 \\ 
SMC\_A2 & 10248 & 14.690785 & -72.330909 & ACS/WFC F555W & 24.45 & 23.29 & 33320 & 13881 \\ 
SMC\_A3 & 10248 & 15.456643 & -72.881284 & ACS/WFC F555W & 24.74 & 23.53 & 33871 & 14614 \\ 
SMC\_A4 & 10396 & 12.905319 & -73.115748 & ACS/WFC F555W & 23.62 & 22.49 & 60252 & 19033 \\ 
SMC\_A5 & 10396 & 17.152412 & -72.977772 & ACS/WFC F555W & 25.84 & 24.31 & 25255 & 14914 \\ 
SMC\_A6 & 10396 & 15.002221 & -72.536831 & ACS/WFC F555W & 25.21 & 23.88 & 28627 & 14654 \\ 
SMC\_A7 & 10396 & 15.087749 & -73.184928 & ACS/WFC F555W & 24.93 & 23.79 & 42149 & 20397 \\ 
SMC\_A8 & 10396 & 10.440540 & -71.064289 & ACS/WFC F555W & 26.67 & 25.00 & 3791 & 2624 \\ 
SMC\_A9 & 10396 & 20.458067 & -73.315727 & ACS/WFC F555W & 26.69 & 25.02 & 5477 & 3963 \\ 
SMC\_A10 & 10766 & 13.405039 & -72.417642 & ACS/WFC F606W & 24.24 & 23.29 & 47559 & 17299 \\ 
SMC\_A11 & 10766 & 13.308628 & -72.472014 & ACS/WFC F606W & 24.34 & 23.32 & 50649 & 18751 \\ 
SMC\_A12 & 12581 & 18.175243 & -72.749376 & ACS/WFC F475W & 26.73 & 24.84 & 17648 & 12367 \\ 
SMC\_A13 & 13476 & 11.056933 & -72.785819 & ACS/WFC F606W & 24.88 & 23.73 & 39584 & 17927 \\ 
SMC\_A14 & 13476 & 13.308959 & -73.601034 & ACS/WFC F606W & 24.66 & 23.54 & 41789 & 17704 \\ 
SMC\_A15 & 13476 & 14.818366 & -72.055697 & ACS/WFC F606W & 25.12 & 24.06 & 28813 & 15101 \\ 
SMC\_A16 & 13476 & 16.444570 & -72.821259 & ACS/WFC F606W & 25.09 & 23.96 & 31695 & 16068 \\ 
SMC\_A17 & 13476 & 9.7530223 & -73.195754 & ACS/WFC F606W & 25.14 & 24.00 & 33400 & 16773 \\ 
SMC\_A18 & 13476 & 15.968124 & -71.957872 & ACS/WFC F606W & 25.48 & 24.40 & 24543 & 14410 \\ 
SMC\_A19 & 13476 & 16.808531 & -73.637940 & ACS/WFC F606W & 25.86 & 24.70 & 17670 & 11413 \\ 
SMC\_A20 & 13476 & 9.0387893 & -73.470949 & ACS/WFC F606W & 25.52 & 24.44 & 25650 & 15243 \\ 
SMC\_A21 & 13476 & 17.432008 & -72.282757 & ACS/WFC F606W & 25.79 & 24.61 & 19931 & 12667 \\ 
SMC\_A22 & 13476 & 10.495970 & -74.139501 & ACS/WFC F606W & 25.84 & 24.66 & 20911 & 13169 \\ 
SMC\_A23 & 13476 & 18.510265 & -73.473449 & ACS/WFC F606W & 26.12 & 24.94 & 10831 & 7500 \\ 
SMC\_A24 & 13476 & 18.364156 & -71.835039 & ACS/WFC F606W & 26.14 & 24.92 & 9402 & 6478 \\ 
SMC\_A25 & 13476 & 10.998826 & -74.789124 & ACS/WFC F606W & 26.27 & 25.05 & 8107 & 5769 \\ 
SMC\_A26 & 13476 & 13.665924 & -72.653847 & ACS/WFC F606W & 23.98 & 22.97 & 46906 & 15980 \\ 
SMC\_A27 & 13476 & 20.051261 & -72.158678 & ACS/WFC F606W & 26.37 & 25.12 & 5881 & 4130 \\ 
SMC\_A28 & 13476 & 20.793737 & -72.758007 & ACS/WFC F606W & 26.30 & 25.08 & 5251 & 3729 \\ 
SMC\_A29 & 13476 & 21.331453 & -74.040008 & ACS/WFC F606W & 26.39 & 25.11 & 3298 & 2332 \\ 
SMC\_A30 & 13673 & 8.4159276 & -70.441980 & ACS/WFC F606W & 27.34 & 25.93 & 2542 & 1710 \\ 
SMC\_A31 & 13673 & 8.4780615 & -70.502721 & ACS/WFC F606W & 27.34 & 25.95 & 2744 & 1820 \\ 
SMC\_A32 & 13673 & 8.2115501 & -70.453423 & WFC3/UVIS F606W & 27.30 & 25.91 & 1256 & 897 \\ 
SMC\_A33 & 13673 & 8.1553090 & -70.393972 & WFC3/UVIS F606W & 27.33 & 25.92 & 1174 & 842 \\ 
SMC\_W1 & smc\_u2o903 & 13.904166 & -73.072219 & WFPC2 F555W & 24.63 & 23.73 & 18262 & 9585 \\ 
SMC\_W2 & smc\_u37704 & 11.475 & -70.578613 & WFPC2 F555W & 26.14 & 24.21 & 1075 & 540 \\ 
SMC\_W3 & smc\_u377a4 & 11.524999 & -70.778892 & WFPC2 F555W & 26.08 & 24.18 & 1204 & 618 \\ 
SMC\_W4 & smc\_u377a6 & 12.225 & -70.545555 & WFPC2 F555W & 26.15 & 24.24 & 1078 & 521 \\ 
SMC\_W5 & smc\_u37706 & 12.225 & -70.795280 & WFPC2 F555W & 26.12 & 24.21 & 1327 & 679 \\ 
SMC\_W6 & smc\_u46c01 & 11.666666 & -72.745277 & WFPC2 F555W & 24.96 & 24.01 & 7380 & 4210 \\ 
SMC\_W7 & smc\_u65c06 & 11.420833 & -72.872222 & WFPC2 F555W & 24.46 & 23.19 & 12244 & 7061 \\ 
\enddata
\tablecomments{Columns: (1) Field name adopted here (2) Field name corresponding to data products available in the MAST archive.  For the LGSPA WFPC2 fields, we give the corresponding field name on the LGSPA website, and for the remainder of the archival imaging we simply give the program ID.  (3)-(4) J2000 coordinates of the center of each field. (5) The bluer of the two filters used in our photometry.  In all cases, the redder of the two filters was the F814W filter of the same instrument.  (6)-(7) The 80\% completeness limit in each filter, used as the faint limit of the stellar sample included in SFH fitting. (8) The total number of stars in each field passing our photometric quality cuts. (9) The number of stars used for SFH fitting, brightward of the 80\% completeness limit in both filters.}
\end{deluxetable}

In total, our sample of 83 HST pointings consists of 45 fields observed with WFC3/UVIS (43 Scylla fields and two archival fields observed in WFC3/UVIS F606W and F814W filters), with an undithered field of view of 162$\arcsec$ per side \citep{wfc3}; 31 archival ACS/WFC fields, with an undithered field of view of 200$\arcsec$ per side \citep{acs}; and the seven WFPC2 LGSPA fields each covering an area slightly (21\%) smaller than WFC3/UVIS \citep{wfpc2}.  Therefore, our resulting spatial resolution is, in a literal sense, better than any of the ground-based studies of the SMC using contiguous imaging, but comes at the cost of non-uniform spatial coverage.  For example, each of our pointings covers an area substantially smaller than either the Voronoi bins used to analyze SMASH data towards the SMC \citep{massana22} or the VMC tile subregions analyzed by \citet{rubele15,rubele18}.  Our field sizes are, however, fairly well-matched to the extinction maps of \citet{skowron}, built from contiguous, wide-field imaging of both the LMC and SMC using a self-consistent analysis technique, providing spatial resolution from 1.7$-$6.9$\arcmin$ over the area sampled by our observations.

\subsection{Photometry \label{photsect}}

The details of performing point-spread function fitting (PSF) photometry on individual science images in each field to generate clean photometric catalogs are discussed in Murray et al.~(2024).  We use the \texttt{Dolphot} software package to perform additional image preprocessing and point-spread function fitting (PSF) photometry identically to the techniques described in \citet{williams_phat,phatter}.  Therefore, we briefly summarize our photometry procedure here for convenience, noting that some previous HST studies of the SMC used earlier versions of \texttt{Dolphot} to perform PSF photometry \citep{dolphin01,weisz13} while some used other PSF photometry programs \citep[e.g.,][]{mccumber05,sabbi09,cignoni12,cignoni13}.

For each target field, individual \texttt{flc} images (corrected for charge transfer inefficiency) were combined into a deep, distortion-corrected master reference image in each filter using the \texttt{astrodrizzle} task within \texttt{drizzlepac} \citep{drizzlepac}.  Next, 
for each image, \texttt{Dolphot} masks bad pixels, applies a pixel area map, splits the image into individual chips, and calculates the sky background.  \texttt{Dolphot} includes modules specific to each imager onboard \textit{HST}, including customized spatially varying PSFs for each filter.  There are many parameters used by \texttt{Dolphot} to govern the details of the PSF photometry procedure, and we adopt the per-instrument parameters thoroughly tested and recommended by \citet{williams_phat,phatter}.  The resulting photometric catalogs are calibrated to the Vegamag photometric system, and contain several diagnostic parameters which can be used to reject poorly measured, spurious and non-stellar sources.  These parameters are described generally in the \texttt{Dolphot} manual\footnote{\url{http://americano.dolphinsim.com/dolphot/}}, and in the context of the present application in Cohen et al.~(2024), and we make cuts on the per-filter values of crowding parameter \texttt{crowd}, the absolute value of the sharpness \texttt{sharp} and the photometric quality flag to retain only well-measured stellar sources.  Given the difference in spatial resolution across the different cameras used, the values used for our photometric quality cuts are chosen on a per-camera basis, listed in Table \ref{qualcuttab}.  In the case of the WFPC2 fields, we note that while high-level merged photometry catalogs are available, we opted to perform cuts on the per-filter quality diagnostics provided in the raw catalogs\footnote{\url{http://astronomy.nmsu.edu/holtz/archival/html/lg.html}} identically as for the remainder of our imaging to keep our analysis as self-consistent as possible.  

\begin{deluxetable}{lllc}
\tablecaption{Photometric Quality Cuts \label{qualcuttab}}
\tablehead{
\colhead{Camera} & \colhead{\texttt{crowd}} & \colhead{$|$\texttt{sharp}$|$} & \colhead{Flag}}
\startdata
WFC3/UVIS & $\leq$0.25 & $\leq$0.25 & 0,2 \\
ACS/WFC & $\leq$0.1 & $\leq$0.1 & 0,2 \\
WFPC2 & $\leq$0.25 & $\leq$0.25 & N/A \\
\enddata
\tablecomments{The \texttt{crowd} and \texttt{sharp} cuts are applied to per-filter values, with stars required to pass these cuts in both filters.  Values of 0 or 2 for the photometric quality flag given in the last column reject sources with too many bad and/or saturated pixels for reliable flux measurements.}
\end{deluxetable}

\begin{figure}
\gridline{\fig{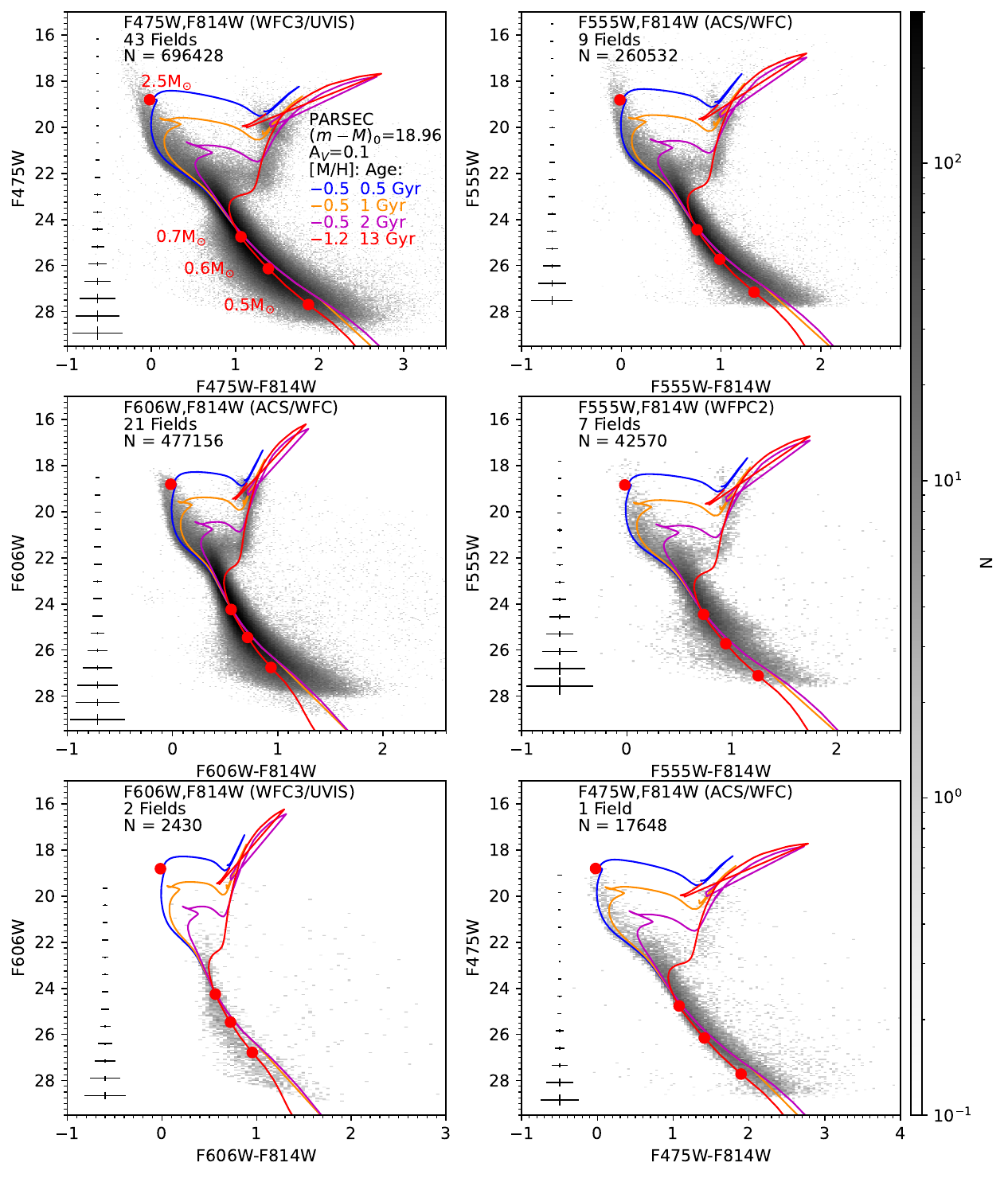}{0.95\textwidth}{}}
\vspace{-0.6cm}
\caption{Stacked Hess diagrams for the various filter combinations present in our imaging.  The total number of fields we analyze in a particular filter combination is given in the top left of each panel, along with the total number of stars passing our photometric quality cuts.  Median photometric errors are shown along the left side of each CMD.  PARSEC isochrones are overplotted to illustrate the range of stellar ages accessible in our photometry in different evolutionary phases, with representative stellar masses indicated along the main sequence.  Due to observational restrictions imposed by the ULYSSES primary imaging, the brightest SMC sources (corresponding to main sequence stellar masses $\gtrsim$2.5M$_{\odot}$) are saturated in many cases.
\label{cmd_fig}}
\end{figure}

\begin{figure}
\gridline{\fig{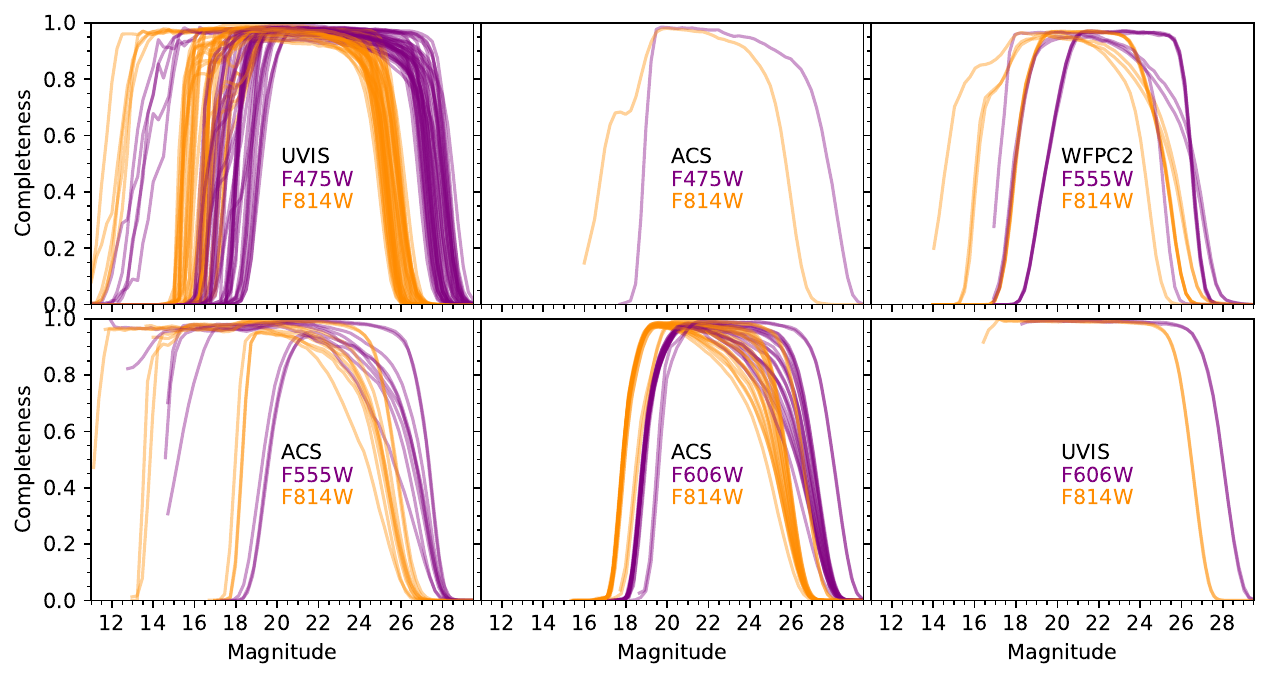}{0.73\textwidth}{}
          \fig{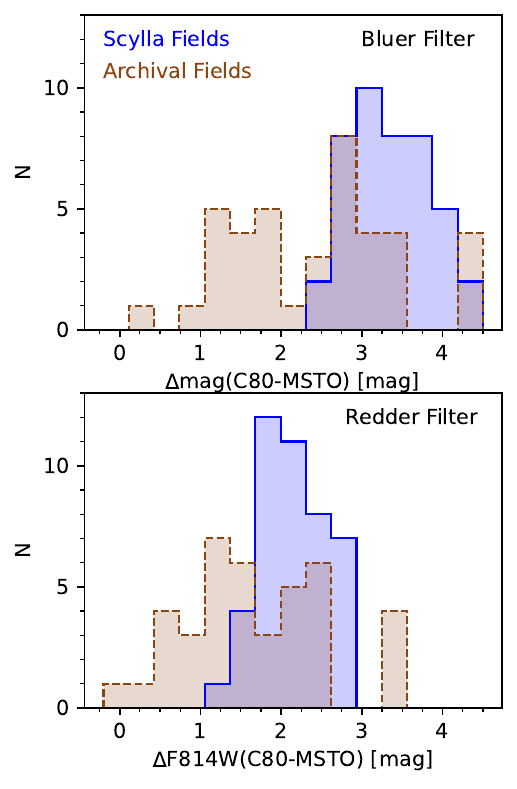}{0.25\textwidth}{}}
\vspace{-1.0cm}
\caption{\textbf{Left Columns: }Completeness versus magnitude for each individual field, with separate panels for each filter combination present in our imaging sample.  In each panel, completeness as a function of magnitude (in the Vegamag photometric system) is shown in purple for the bluer of the two filters and orange for the redder of the two filters.  \textbf{Rightmost Column: } Magnitude difference between the 80\% completeness limit, used as the faint limit for SFH fitting, and the main-sequence turnoff magnitude in each field, assessed using a 13 Gyr PARSEC isochrone with [M/H]=$-$1.2, $(m-M)_{0,SMC}$=18.96 and extinction from \citet{skowron}.  The top right and bottom right panels correspond to the bluer and redder filter available in each field, illustrating that the vast majority ($>$90\%) of our fields (and all of the Scylla fields) are 80\% complete to at least 1 mag below the ancient main sequence turnoff in both filters.
\label{completeness_fig}}
\end{figure}

Color-magnitude diagrams (CMDs) are shown in Fig.~\ref{cmd_fig} for the various filter combinations present across our imaging.  We have overlaid solar-scaled PARSEC isochrones \citep{parsec} for a representative range of masses and ages, demonstrating that our imaging typically reaches down to a mass of $\sim$0.5M$_{\odot}$ on the unevolved main sequence, as deep or deeper than previous studies of the SFH in the SMC \citep[e.g.,][]{hz04,noelsfh,cignoni12,weisz13,rubele18,massana22}.  Our SFH fitting prodecure (Sect.~\ref{sfhsect}) requires a model of observational noise and incompleteness in each field, so we perform artificial star tests by inserting $>$10$^{5}$ artificial stars per field and attempt to recover them identically as for our observations.  The artificial stars are assigned a flat input spatial distribution, and the Bayesian Extinction and Stellar Tool (BEAST; \citealt{beast}) was used to assign input magnitudes drawn from a library built with PARSEC models over a broad range of input ages (6.0$\leq$Log Age/yr$\leq$10.13), metallicities (-2.1$\leq$[M/H]$\leq$-0.3), distances (47$\leq$(d$_{\odot}$/kpc)$\leq$77) and extinctions (0$\leq$A$_{V}$$\leq$1).  The per-field photometric completeness as a function of magnitude is shown in Fig.~\ref{completeness_fig} for all of the filter combinations across our sample, with the bluer filter shown in purple and the redder filter shown in orange in each panel, demonstrating that completeness generally drops off fairly quickly as a function of magnitude close to our detection limit.  Therefore, we adopt the 80\% completeness limit, denoted C80, as the faint magnitude limit of the stellar sample used for SFH fitting, as in our companion study of the spatially resolved SFH in the LMC.  The rightmost column of Fig.~\ref{completeness_fig} illustrates the magnitude difference between C80 and the ancient main sequence turnoff (MSTO) magnitude, calculated assuming a 13 Gyr metal-poor ([M/H]$=-$1.2) PARSEC isochrone (based on our best-fit AMRs; see Sect.~\ref{litsect}) and per-field extinction values from the \citet{skowron} map, demonstrating that the vast majority of our fields ($>$90\%), including \textit{all} of our Scylla imaging (shown in blue in the right-hand column of Fig.~\ref{completeness_fig}), are 80\% complete to at least 1 mag faintward of the ancient MSTO in both filters that we use for SFH fitting.   

\section{Star Formation History Fitting \label{sfhsect}}

Before performing SFH fits to each field, we make two modifications to our photometric catalogs.  First, in cases of spatial overlap between fields, we retained the overlapping region in whichever of the two fields had deeper photometry (i.e., fainter values of C80), excising it from the shallower of the two fields.  This treatment ensures that each field can be treated as a statistically independent sample, without any sources contributing to SFH fits for multiple fields.  Second, we remove known star clusters \citep{bica20} that could bias our characterization of the SMC field population.  For each cluster we conservatively remove from our catalogs all sources lying within the cluster semi-major axis length from the cluster center listed by \citet{bica20}, which were generally estimated visually by the authors of the original source catalog, and the excised clusters and the affected fields are listed in Appendix \ref{clustersect}.  The impact on our overall sample is quite small: Across all of the affected fields, the excised fraction of the field of view has a median of only 10\%, and our total sampled area in the SMC is reduced by only 5\%. 

We use the software code \texttt{MATCH} \citep{match} to calculate the best-fitting SFH and age-metallicity relation (AMR) from the CMD of each field using the well-known CMD synthesis technique (see \citealt{tolstoy09} for a review).  \texttt{MATCH} has been used to measure the SFHs of scores of galaxies throughout the Local Volume over their entire lifetimes \citep[e.g.,][]{williams09,weisz11,williams11,radburnsmith12,weisz14,geha15,williams17,mcquinn18,savino23,mcquinn23} and produces results comparable with other CMD synthesis codes \citep{monelli10a,monelli10b,hidalgo11,skillman14,skillman17}.

To calculate the best-fitting SFH and AMR and uncertainties, \texttt{MATCH} compares an observed photometric catalog against synthetic photometry of a linear combination of simple stellar populations (SSPs) from a chosen stellar evolutionary library.  To test the sensitivity of our results to the assumed evolutionary library, we perform three fits for each field, using the PARSEC \citep{parsec}, MIST \citep{mist1,mist2} and updated BaSTI \citep[][which we denote \enquote{BaSTI18}]{basti} stellar evolutionary libraries.  All of these stellar evolutionary libraries provide synthetic photometry in the native filter systems of each of the HST cameras included in our sample (ACS/WFC, WFC3/UVIS, WFPC2), rendering field-to-field photometric calibration unneccessary. 

To compare observed and synthetic CMDs, the artificial star tests are used to apply observational noise to the synthetic SSPs, and \texttt{MATCH} varies the star formation rate and mean metallicity in each user-selected time bin, seeking the best fit SFH and AMR using a Poisson likelihood statistic.  For SFH fitting, we use the same basic input assumptions as in Cohen et al.~(2024), providing a consistent methodology across our analyses of spatially resolved SFHs in both the LMC and SMC.  Specifically, we use time bins of width $\Delta$Log Age/yr = 0.05 dex from Log Age/yr = 7.2 (the youngest age available across all three sets of evolutionary models we employ) 
to Log Age/yr = 10.15, a metallicity spread in each time bin of $\Delta$[M/H] = 0.15 dex over an allowed metallicity range of $-$2.0$\leq$[M/H]$\leq$+0.2, a \citet{kroupa} initial mass function, and an assumed binary fraction of 0.35 with a flat mass ratio distribution as a function of primary mass.  While various SFH studies of dwarf galaxies assume binary fractions ranging from 0.3$-$0.5 \citep[e.g.,][]{monelli10a,monelli10b,sacchi18,massana22}, a binary fraction of 0.35 was originally implemented for SFH fitting with \texttt{MATCH} based on multiplicity statistics for solar neighborhood dwarfs \citep{dm91}.  We retain this value for consistency with the aforementioned vast database of lifetime SFHs derived for Local Volume dwarfs using \texttt{MATCH} \citep[e.g.,][]{weisz11,williams11,weisz14} including our companion study of the LMC (Cohen et al.~2024).  In any case, the shape of the SFH is quite insensitive to variations in the assumed binary fraction (or initial mass function slope) given deep imaging extending faintward of the ancient main sequence turnoff \citep{monelli10b,hidalgo11,cignoni12,cole14}, as we have in hand for our SMC fields.  The uncertainties on the resultant SFHs include both random (statistical) and systematic contributions, detailed in \citet{dolphin_randerr} and \citet{dolphin_syserr} respectively.

When searching for the best-fit SFH, \texttt{MATCH} also has the capability to float distance modulus $(m-M)_{0}$, foreground Galactic extinction A$_{V,fg}$ and additional internal differential extinction $\delta$A$_{V}$ as free parameters.  We allow \texttt{MATCH} to search over a grid with a grid spacing of 0.05 mag in all three of these parameters.  A similar grid search technique has been implemented using \texttt{MATCH} in the past, but searching over two parameters  (A$_{V,fg}$ and $\delta$A$_{V}$) rather than three, using an identical grid spacing \citep{lewis15,lazzarini22}.  Our strategy is a direct extension of this approach, simply adding $(m-M)_{0}$ as a third parameter in the grid search, and has been applied identically to the LMC (Cohen et al.~2024).  In our companion study of the LMC, we exploited the wealth of independent line-of-sight distance and extinction measurements \citep[e.g.,][]{yumimap,skowron} to assess the ability of \texttt{MATCH} to recover the three parameters using our grid search strategy, finding that it performed quite well, recovering per-field distances with a median precision of $\lesssim$3\% (see Fig.~5 of Cohen et al.~2024).  The LMC is an optimal testbed for such a comparison, since the three-dimensional distribution of its red clump stars is well-fit by a disk with a thickness of only $\sim$1 kpc \citep{yumiring,yumimap}.  For the SMC, the situation is more complicated, since \texttt{MATCH} assumes a single fixed distance for SFH fitting while the SMC is known to have a line-of-sight depth that is both significant and position-dependent \citep{nidever13,sub17,jd17,muraveva18,tatton21,elyoussoufi21,petia21}.  Therefore, we performed a set of simulations in which we used SMC-like SFHs to generate input synthetic stellar populations, and then assigned them various line-of-sight distance distributions based on extant analyses of the SMC before attempting to recover their SFHs with \texttt{MATCH} identically as for our observations.  Further details are given in Appendix \ref{simsect}, where we demonstrate that our results are unaffected by line-of-sight depth effects to within their uncertainties.  

\begin{figure}
\gridline{\fig{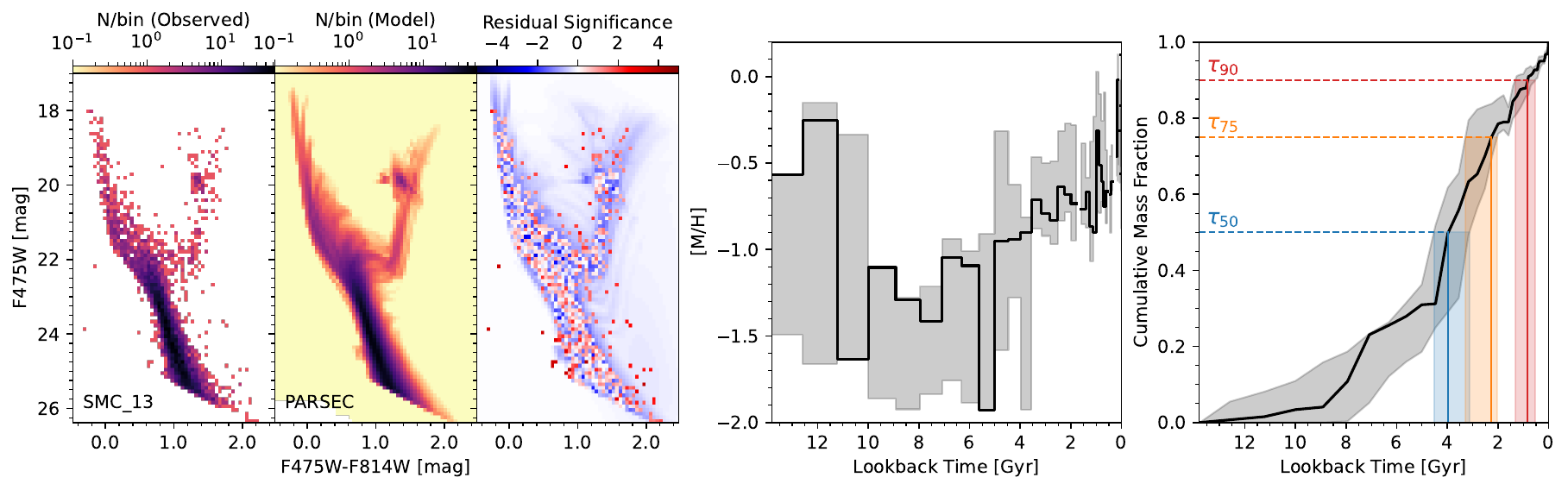}{0.99\textwidth}{}}
\gridline{\fig{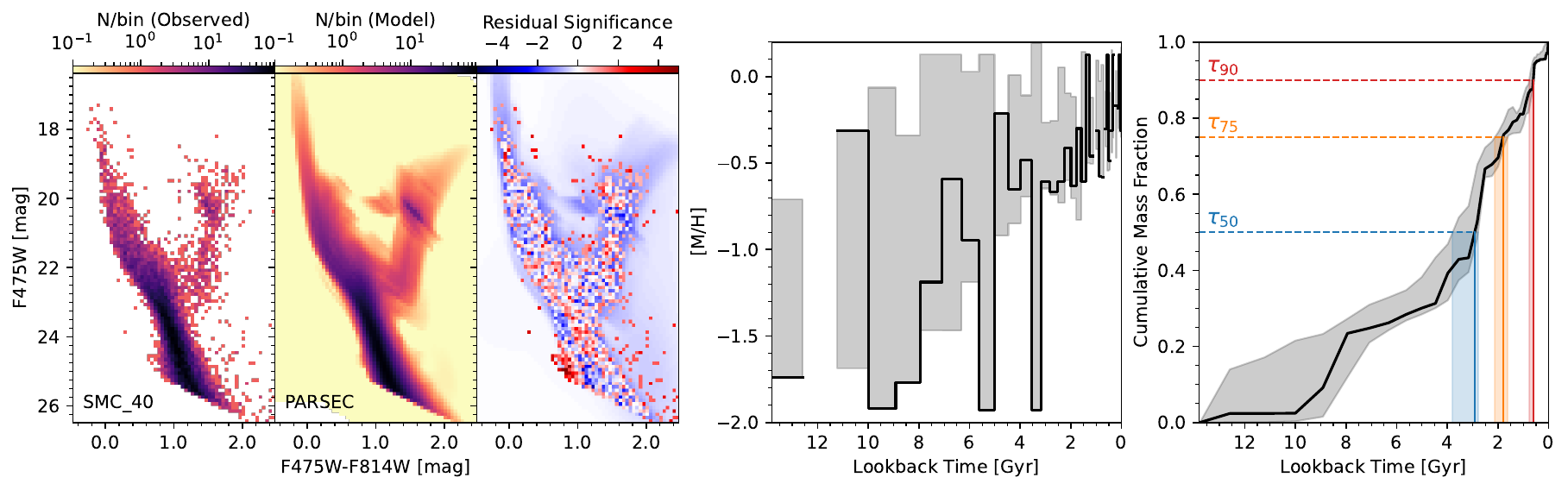}{0.99\textwidth}{}}
\caption{Example SFH fits for two fields (top row: SMC\_13; bottom row: SMC\_40) spanning the range of fit quality seen across our sample to illustrate our typical SFH and AMR precision on a per-field basis.  \textbf{Left:} Hess diagrams showing the observed CMD, the best-fit model, and the residuals in the sense (data-model).  \textbf{Center:} Best-fit age-metallicity relation, with uncertainties indicated by grey shading.  Despite large per-field uncertainties, both fields are consistent with an AMR that increases monotonically with time.  \textbf{Right:} Best-fit cumulative star formation history (CSFH), showing the fraction of total mass ever formed as a function of lookback time.  We quantify the CSFH using the metrics $\tau_{50}$, $\tau_{75}$ and $\tau_{90}$ to indicate the lookback times by which 50\%, 75\% and 90\% of the cumulative stellar mass had formed.  These values are overplotted on the CSFHs using blue, orange and red vertical lines respectively, with shading corresponding to 1-$\sigma$ uncertainties.  
\label{sfh_example_fig}}
\end{figure}

Examples of our per-field SFH fitting results are shown in Fig.~\ref{sfh_example_fig}, where each row presents SFH fits for a single field.  We have intentionally chosen two fields spanning the range of fit quality seen across our sample to illustrate the precision of our best-fit SFHs and AMRs on a per-field basis.  The top row is an example of a field with a high-quality SFH fit, where the SFH fitting residuals (third panel from left) show little to no structure across the CMD.  The results shown on the bottom row correspond to a field with more apparent structure in the CMD fit residuals.  Discrepancies are seen at the faint blue corner of the stellar sample as well as in the vicinity of the red clump and horizontal branch, and likely result from a combination of small-scale differential reddening, line-of-sight distance spread, and the known difficulty of evolutionary models to reproduce in detail the CMD morphology corresponding to late stages of stellar evolution (see \citealt{gallart05} for a review).  In the middle panel of Fig.~\ref{sfh_example_fig}, we show the best-fit AMR, which increases monotonically with age to within uncertainties.  This is an important validation of our SFH fitting procedure given that we placed no restriction on the allowed metallicity in each time bin.  The right-hand panel in each row illustrates the cumulative star formation history (CSFH) output by \texttt{MATCH}, which is the fraction of total stellar mass ever formed that had formed by a given lookback time.  We use the CSFH to quantify our per-field SFH fitting results, where $\tau_{50}$, $\tau_{75}$ and $\tau_{90}$ correspond to the lookback times by which 50\%, 75\% and 90\% of the cumulative mass had formed.  The values of $\tau_{50}$, $\tau_{75}$ and $\tau_{90}$ and their 1-$\sigma$ uncertainties are calculated via interpolation in the CSFH and its 1-$\sigma$ uncertainties, illustrated for the two example cases in the right-hand panels of Fig.~\ref{sfh_example_fig}.  All of the individual per-field values of $\tau_{50}$, $\tau_{75}$ and $\tau_{90}$ and their uncertainties for each of the three of the stellar evolutionary models we assume are provided in Table \ref{tautab} of Appendix \ref{indfieldsect}.

\section{Results \label{resultsect}}

Studies of the spatial variation in the SFH of the SMC have already revealed several trends that have ramifications for the interaction history of the SMC-LMC-Milky Way system.  Star formation in the SMC has generally been centrally concentrated, resulting in an outside-in age gradient such that the inner SMC is younger on average \citep{carrera08,cignoni13,rubele18}, due at least in part to a lack of recent star formation far from the SMC center \citep{dolphin01}.  However, in the direction of the SMC wing to the east, this trend does not hold, and star formation has continued over the most recent few Gyr \citep{mccumber05,noelsfh}.  To place quantitative constraints on these trends, we first combine SFH results for individual fields into subsets based on location (see the bottom panel of Fig.~\ref{map_fig}), allowing us to compare trends in the wing both to non-wing fields at similar distance from the SMC center as well as to the inner SMC (Sect.~\ref{globalsect}).  Subsequently, we use our field-by-field SFHs to examine radial SFH trends in detail, measuring their slopes and assessing their linearity (Sect.~\ref{radgradsect}).  

\subsection{Global Trends \label{globalsect}}

\begin{figure}
\gridline{\fig{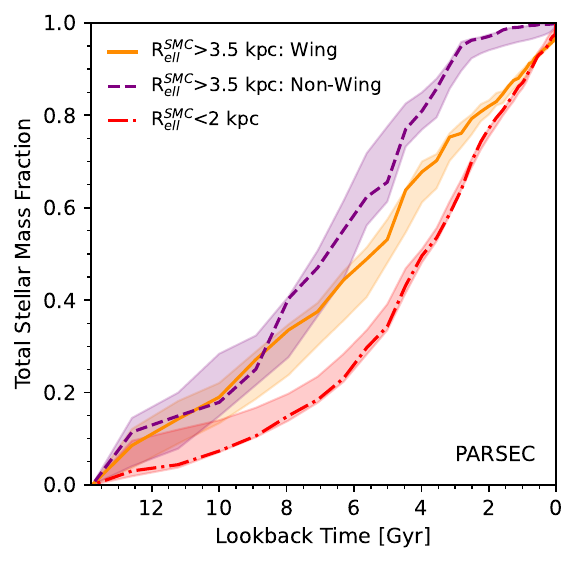}{0.485\textwidth}{}
          \fig{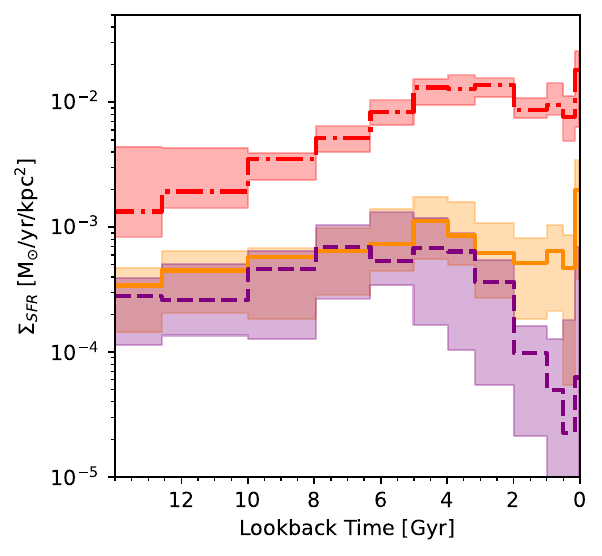}{0.505\textwidth}{}}
\caption{\textbf{Left: }Cumulative SFHs of fields in the wing (orange solid line), non-wing fields at similarly large radii of R$_{SMC,opt}^{ell}$$>$3.5 kpc (purple dashed line) and fields in the main body with R$_{SMC,opt}^{ell}$$<$2 kpc (red dash-dotted line).  The wing and non-wing fields assembled a similar fraction of their total mass at early times (lookback time $\geq$6 Gyr), but in the most recent $\sim$2 Gyr, the wing fields assembled a much larger fraction of their mass, similar to fields in the main body.  Conversely, the outer non-wing fields formed only $\sim$1\% of their mass within the last 1 Gyr.  \textbf{Right: }Average star formation rate per unit area in several illustrative time intervals.  Colors are as in the left panel.  Although the wing fields formed a similar mass fraction in the last $\sim$1$-$2 Gyr as the main body, their star formation rate per projected unit area $\Sigma_{\rm SFR}$ has remained an order of magnitude lower.  Meanwhile, the wing and non-wing fields, at similarly large distances from the SMC center, have similar star formation rates until $\sim$2 Gyr ago, when $\Sigma_{\rm SFR}$ in the non-wing outer fields decreased by an order of magnitude until the most recent LMC-SMC collision.} 
\label{combined_sfh_fig}
\end{figure}

The cumulative star formation histories (CSFHs) of the three positionally-selected subsets of fields (main body, wing, outer non-wing) are shown in the left panel of Fig.~\ref{combined_sfh_fig}, color-coded as in the bottom panel of Fig.~\ref{map_fig}.  These CSFHs were calculated by statistically combining the best-fitting SFHs and their uncertainties for each individual field.  Because systematic uncertainties cannot be assumed to be uncorrelated (unlike random uncertainties), the calculation of uncertainties when combining SFH results across multiple fields is non-trivial, and further details can be found in \citet[][see their Appendix C]{weisz11} and \citet{dolphin_syserr}.  Rather than simply fitting a single SFH to a combined CMD, statistically combining the best-fit SFHs and uncertainties from individual fields is necessary to account for per-field differences in photometric depth and incompleteness (and different observing setups).  

Two features are immediately apparent from the CSFH comparison in the left panel of Fig.~\ref{combined_sfh_fig}: First, an outside-in age gradient, such that the outer fields are older on average than the inner SMC (we quantify this gradient as a function of R$_{SMC}^{ell}$ below in Sect.~\ref{radgradsect}).  For example, the outer wing and non-wing fields (orange and purple in Fig.~\ref{combined_sfh_fig}) formed 50\% of their cumulative mass by 5.5$^{+0.3}_{-0.6}$ and 6.8 $^{+0.3}_{-0.8}$ Gyr ago, compared to 3.9$^{+0.2}_{-0.1}$ Gyr ago for the main body of the SMC (red in Fig.~\ref{combined_sfh_fig}).  Second, in the outer SMC, the wing and non-wing fields had formed similar cumulative mass fractions only until $\sim$6 Gyr ago.  This may be when the LMC and SMC began interacting \citep[e.g.,][]{besla12}, possibly driving continued star formation in the wing (and main body) but not the outer non-wing fields, as observed over the last few Gyr (see below). 

The nature, and timing, of the divergence in CSFH trends between the wing and outer non-wing fields is illustrated in the right-hand panel of Fig.~\ref{combined_sfh_fig}, where we plot the average star formation rate per unit area\footnote{To calculate $\Sigma_{\rm SFR}$, we use the non-overlapping spatial coverage of each field and assume a fixed SMC distance of 62 kpc.} $\Sigma_{\rm SFR}$ in several selected time intervals for the three subsets of fields.  For the wing fields, continued star formation resulted in 10.8\%$^{+1.5}_{-0.7}$ of their cumulative mass being formed in the last Gyr, calculated directly from interpolation in the CSFH and its uncertainties.  This cumulative mass fraction is in excellent agreement with the value of $\sim$7$-$12\% reported by \citet{noelsfh} for fields near the wing.  The outer non-wing fields, on the other hand, 
are nearly devoid of stars formed within the last several Gyr, reflected by a 
marked dropoff in star formation beginning $\sim$3$-$4 Gyr ago (also observed for a nearby field in the northwestern SMC by \citealt{dolphin01}).  In fact, the dropoff in star formation in the outer non-wing fields several Gyr ago was sufficiently dramatic that they formed only $\leq$1\% of their mass within the last Gyr, forming 90\% of their mass by 3.2$^{+0.1}_{-0.5}$ Gyr ago.  The wing and inner SMC fields, lacking such a dropoff and instead continuing to form stars, did not form 90\% of their mass until 0.90$^{+0.08}_{-0.13}$ and 0.77$^{+0.02}_{-0.07}$ Gyr ago respectively.  However, while the wing fields and the main body formed similar \textit{fractions} of their cumulative mass over the last 1$-$2 Gyr (left panel of Fig.~\ref{combined_sfh_fig}), the \textit{values} of $\Sigma_{\rm SFR}$ in the main body are consistently higher (by about an order of magnitude; right panel of Fig.~\ref{combined_sfh_fig}), in agreement with CMD-based SFH fits by \citet{cignoni13} and an SFH analysis of long-period variables \citep{rezaeikh14}.  However, the distribution of our fields is particularly concentrated, even within the main body of the SMC; our fields with R$_{ell}^{SMC}$$<$2 kpc have a median value of R$_{ell}^{SMC}$=1.0 kpc, compared to a value of 1.4 kpc that would result from a flat spatial distribution.  

We also find evidence for an enhancement in star formation following the most recent LMC-SMC collision $\sim$150 Myr ago \citep{choi22}, reported by \citet{rubele15,rubele18}.  Such a recent enhancement could be at least partially due to star formation from gas that was already tidally stripped \citep{elyoussoufi23}, and may be coeval with the formation of extended high-velocity cold gas outflows \citep{mccluregriffiths18}.  Although the saturation limit of our imaging prohibits robust constraints on star formation rates at such recent times (due to the higher luminosities of younger, more massive and short-lived stellar populations), the best-fit $\Sigma_{\rm SFR}$ values  we find in the inner SMC following the most recent LMC-SMC collision are $\Sigma_{\rm SFR}$=0.04$_{-0.01}^{+0.03}$ M$_{\odot}$/yr/kpc$^{2}$ when restricted to the most recent 25 Myr, nearly as high as H$\alpha$-based values seen in the SMC's most prominent star forming region N66 (NGC 346) over the last 5 Myr (\citealt{hony15}; also see \citealt{hagen17}).\footnote{When averaged over \textit{all} of our fields rather than just the main body of the SMC, we find $\Sigma_{\rm SFR}$=0.007$_{-0.002}^{+0.004}$ M$_{\odot}$/yr/kpc$^{2}$, in reasonable agreement with \citet[][see their Fig.~5]{jameson16} given the centrally concentrated spatial distribution of our fields.}  Our results also allow for the possibility that this most recent star formation enhancement is global in nature, seen also in the outer non-wing fields (see the right-hand panel of Fig.~\ref{combined_sfh_fig}).  While \citet{rubele18} and \citet{massana22} find that recent star formation in the SMC is instead centrally concentrated, at least in the former case uncertainties cannot rule out the global nature of an enhancement in $\Sigma_{\rm SFR}$ by a factor of a few.  Based on observations of an isolated pair of interacting dwarfs, \citet{privon17} suggested that such a global star formation enhancement could be caused by large-scale interaction-driven ISM compression, in this case triggered by the LMC-SMC collision.  While our focus here is on spatial trends in the lifetime assembly history of the SMC, a more detailed analysis of the timing, duration and location of star formation bursts in the LMC and SMC will be presented separately (C. Burhenne et al., in prep.).

The central concentration of star formation seen in the right-hand panel of Fig.~\ref{combined_sfh_fig} was noted in previous ground-based surveys \citep{hz04,rubele18,massana22} as well as studies using multiple HST pointings \citep{weisz13,cignoni13}.  In cases where star formation rates were provided, these values are in good agreement with our results.  
For example, prior to 5 Gyr ago before the star formation enhancements attributed to LMC-SMC interaction \citep{hz04,rubele18,massana22}, the VMC survey \citep{rubele18} gives an average star formation rate $\langle$$\Sigma_{\rm SFR}$$\rangle$$_{\rm t>5 Gyr}$ = 1.3$_{-0.1}^{+0.2}$$\times$10$^{-3}$M$_{\odot}$/yr/kpc$^{2}$ and SMASH \citep{massana22} finds $\langle$$\Sigma_{\rm SFR}$$\rangle$$_{\rm t>5 Gyr}$ = 0.7$_{-0.3}^{+0.3}$$\times$10$^{-3}$M$_{\odot}$/yr/kpc$^{2}$.  Over the most recent 3 Gyr, VMC results give $\langle$$\Sigma_{\rm SFR}$$\rangle$$_{\rm t<3 Gyr}$ = 2.3$_{-0.4}^{+0.5}$$\times$10$^{-3}$M$_{\odot}$/yr/kpc$^{2}$ and SMASH finds 1.2$_{-0.4}^{+0.4}$$\times$10$^{-3}$M$_{\odot}$/yr/kpc$^{2}$.  All of these values sit comfortably in between the values of $\Sigma_{\rm SFR}$ we find for the main body and the outer regions of the SMC (right-hand panel of Fig.~\ref{combined_sfh_fig}).  Survey-to-survey differences can be attributed at least in part to differences in spatial coverage, especially in light of the apparently strong radial gradient in $\langle$$\Sigma_{\rm SFR}$$\rangle$ apparent from the right-hand panel of Fig.~\ref{combined_sfh_fig}.  In particular, SMASH excluded the central region of the the SMC due to crowding, while VMC provided contiguous spatial coverage that included this central region but with differing coverage of the outer SMC.  Pre-dating these studies, MCPS also provided spatially resolved SFH fits \citep{hz04}, although their best-fit star formation rates were a factor of a few larger than VMC, SMASH, or the present work.  Furthermore, \citet{hz04} report an enhancement in star formation prior to 8.4 Gyr ago that is unconfirmed by any of these more recent studies (see Fig.~11 of \citealt{rubele18} and Fig.~2 of \citealt{massana22}).  As discussed by \citet[][see their Sects.~5.2$-$5.3]{rubele18}, such discrepancies with \citet{hz04} likely result from a combination of several factors, including their use of fixed discrete metallicity values, a different assumed stellar initial mass function, a fixed SMC distance of 60 kpc, and shallower imaging providing less leverage on the oldest stellar populations.

\subsection{Quantifying the Outside-In Radial Age Gradient \label{radgradsect}}

\begin{figure}
\gridline{\fig{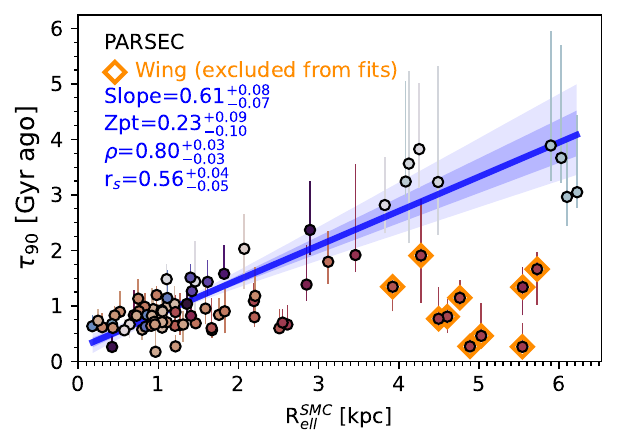}{0.314\textwidth}{}
          \fig{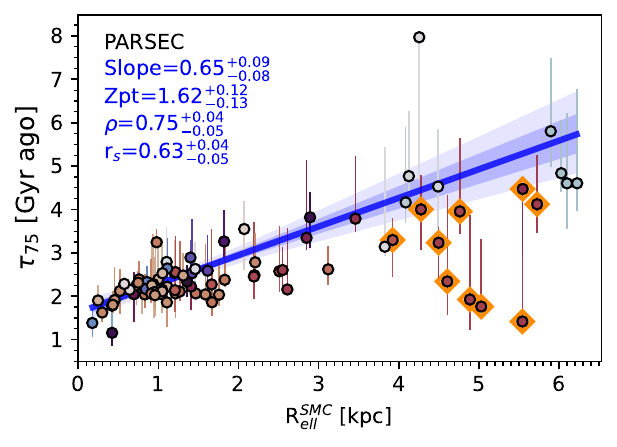}{0.314\textwidth}{}
          \fig{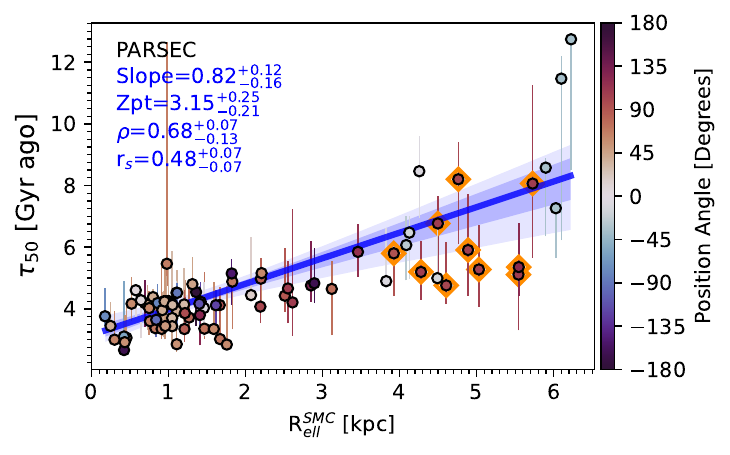}{0.363\textwidth}{}}
\vspace{-1.0cm}
\gridline{\fig{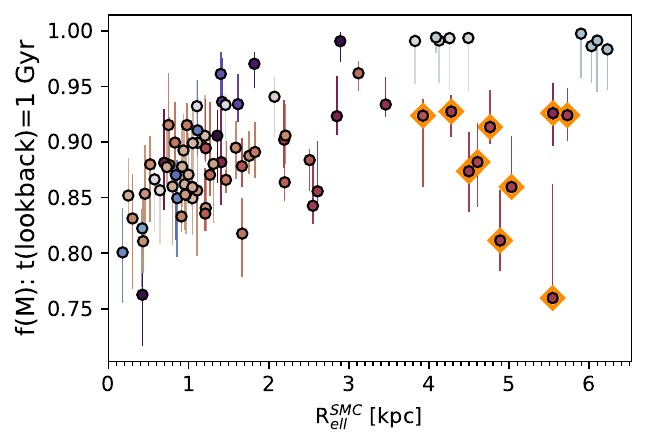}{0.314\textwidth}{}
          \fig{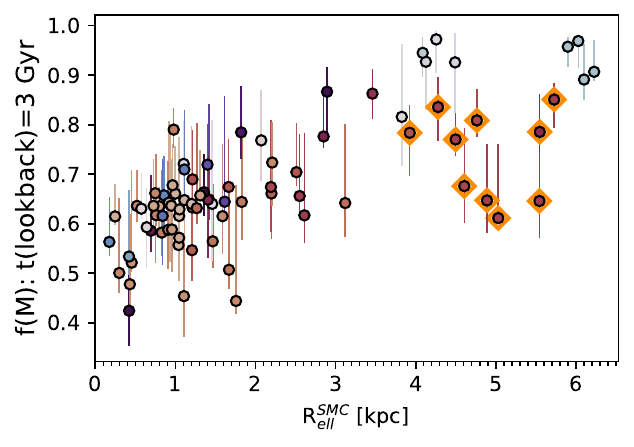}{0.314\textwidth}{}
          \fig{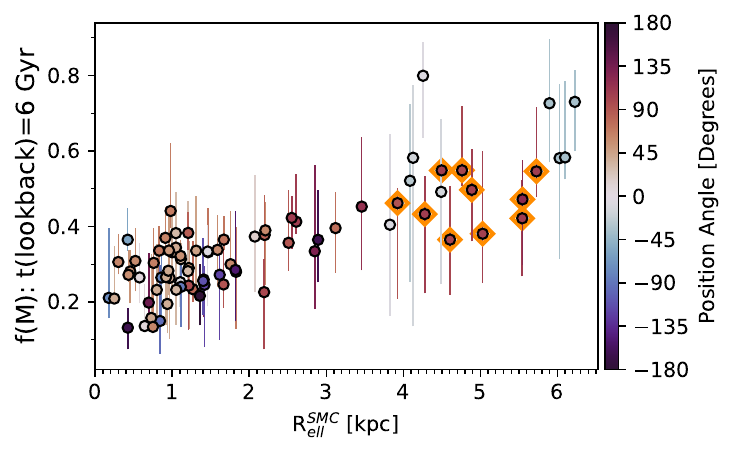}{0.363\textwidth}{}}
\vspace{-1.0cm}
\caption{Lifetime star formation history metrics versus semi-major axis equivalent distance from the optical center of the SMC R$_{ell}^{SMC}$.  \textbf{Top Row:} Lookback time to form (from left to right) 90\%, 75\% and 50\% of the cumulative stellar mass as a function of R$^{SMC}_{ell}$.  Fields are color-coded by their position angle east of north, and fields in the wing of the SMC are shown in orange, highlighting their increasing discrepancy with the remainder of fields at similarly large R$_{ell}^{SMC}$ at recent lookback times.  Linear fits excluding the wing fields are shown in blue, with dark (light) shading representing 1$\sigma$ (2$\sigma$) uncertainties.  The linear fit coefficients, Pearson linear correlation coefficient $\rho$ and Spearman rank correlation coefficient r$_{s}$ are given in each panel.  \textbf{Bottom Row:}  Fraction of cumulative stellar mass formed by lookback times of (from left to right) 1 Gyr, 3 Gyr and 6 Gyr, again versus R$_{ell}^{SMC}$, with symbols as in the top row.  Compared to other fields at similarly large distances from the center of the SMC (R$_{ell}^{SMC}$$>$3.5 kpc), fields in the wing assembled a larger fraction of their mass more recently than $\sim$6 Gyr ago.  Results shown assume PARSEC evolutionary models, and in Table \ref{coefftab} we give linear fit coefficients assuming other stellar evolutionary models for SFH fitting as well as a simple on-sky tangent plane (rather than elliptical) radius to characterize distance from the SMC center.} 
\label{csfh_rad_fig}
\end{figure}

A comparison of the lifetime SFHs in the inner (R$_{ell}^{SMC}$$<$2 kpc) versus outer (R$_{ell}^{SMC}$$>$3.5 kpc) SMC in Fig.~\ref{combined_sfh_fig} reveals an outside-in present-day age gradient, so we turn to our per-field SFH results to quantify the age gradient and its statistical significance.  We plot the CSFH metrics $\tau_{90}$, $\tau_{75}$ and $\tau_{50}$ 
versus semi-major axis equivalent distance R$_{ell}^{SMC}$ in the top row of Fig.~\ref{csfh_rad_fig}.  The individual fields are color-coded by position angle east of north, and the wing fields are indicated with orange diamonds.  The tendency for fields at large R$_{ell}^{SMC}$ to show larger uncertainties reflects the smaller number of stars available per field to constrain SFH fits far from the center of the SMC (see Table \ref{obstab}).  Overall, the CSFH metrics (i.e., lookback time to form a given cumulative mass fraction) show a linear trend with R$_{ell}^{SMC}$.  However, the wing fields deviate from this linear trend, increasingly so at more recent lookback times, with the wing fields assembling similar mass fractions as the inner SMC in the most recent few Gyr (as seen in Fig.~\ref{combined_sfh_fig}).  We therefore exclude the wing fields before performing linear fits to measure the slope of the radial age gradient, as well as the Pearson linear correlation coefficient $\rho$ and the Spearman rank correlation coefficient r$_{s}$, with uncertainties calculated using Monte Carlo draws from the asymmetric error distributions of the per-field CSFH metrics.  For all three of the CSFH metrics we examine, we find a moderate to high degree of linearity at very high statistical significance, with Pearson linear correlation coefficients $\rho$$>$0.68 and $p$-values $<$10$^{-3}$.  

In the bottom row of Fig.~\ref{csfh_rad_fig}, we reframe our results, choosing fixed lookback times of 1, 3 and 6 Gyr (left, middle and right panels) and plotting the cumulative mass fraction formed in each field by that lookback time (as a function of R$_{ell}^{SMC}$).  These lookback times were chosen to age-date the lookback time when the wing fields last formed a similar mass fraction as the rest of the outer SMC (i.e., the outer non-wing fields), $\sim$6 Gyr ago.  More recently, the bottom left panel of Fig.~\ref{csfh_rad_fig} illustrates on a per-field basis that \textit{all} of the outer non-wing fields (0$^{\circ}$$<$PA$<$90$^{\circ}$, R$_{ell}^{SMC}$$>$3.5 kpc) formed $\lesssim$1\% of their mass within the most recent 1 Gyr, while the bottom right panel of Fig.~\ref{csfh_rad_fig} illustrates the earliest lookback time at which our data quality allow the detection of a linear radial age gradient, 6 Gyr ago, for which we find the relationship between cumulative mass fraction formed versus R$_{ell}^{SMC}$ has a Pearson $\rho$=0.58$^{+0.08}_{-0.08}$ ($p$-value$<$10$^{-3}$).  At even earlier lookback times, observational uncertainties hamper the detection of the linear correlations between lookback time and R$_{ell}^{SMC}$ ($p$-values $\gtrsim$0.01).  

The fits shown in Fig.~\ref{csfh_rad_fig} parameterize distance from the SMC assuming an elliptical geometry frequently used in the literature \citep[e.g.,][]{piatti05,dias14,parisi22}, shown using grey ellipses in Fig.~\ref{map_fig}.
To test whether our results depend on this parameterization, we have reperformed our fits versus simple geometric on-sky distance rather than elliptical (semi-major axis equivalent) distance from the SMC center.  Because old stellar populations in the SMC show a nearly spherical geometry \citep{hz04,rubele18}, some previous literature studies do use projected on-sky distance without any correction for geometric effects \citep[e.g.,][]{nidever11,massana20,povick23}.  We found that while the slopes (and zeropoints) of the best-fit lines necessarily change, the linear correlation coefficients are unaffected to within their uncertainties ($\rho$$\geq$0.61) and retain their high statistical significance ($p$-values $<$10$^{-3}$).  We have also verified that our results are robust to the choice of stellar evolutionary model library used for our SFH fits, and the best-fit parameters for all combinations of evolutionary model and parameterization of radius are listed in Table \ref{coefftab}.

\begin{deluxetable}{cccccc}
\tablecaption{Linear Fit and Correlation Coefficients of CSFH Metrics Versus Radius \label{coefftab}}
\tablehead{\colhead{Metric} & \colhead{Model} & \colhead{Slope} & \colhead{Zeropoint} & \colhead {Pearson $\rho$} & \colhead{Spearman $r_{s}$} \\ \colhead{} & \colhead{} & \colhead{Gyr/kpc} & \colhead{Gyr} & \colhead{} & \colhead{}}
\startdata
\hline
\multicolumn{6}{c}{Versus Elliptical Distance R$_{ell}^{SMC}$} \\
\hline
$\tau_{90}$ & PARSEC & 0.61$^{+0.08}_{-0.07}$ & 0.23$^{+0.09}_{-0.10}$ & 0.80$^{+0.03}_{-0.03}$ & 0.56$^{+0.04}_{-0.05}$  \\
$\tau_{90}$ & MIST & 0.68$^{+0.11}_{-0.10}$ & 0.13$^{+0.11}_{-0.14}$ & 0.77$^{+0.04}_{-0.05}$ & 0.56$^{+0.04}_{-0.05}$  \\
$\tau_{90}$ & BaSTI18 & 0.62$^{+0.09}_{-0.08}$ & 0.24$^{+0.10}_{-0.11}$ & 0.76$^{+0.04}_{-0.05}$ & 0.53$^{+0.04}_{-0.05}$  \\
$\tau_{75}$ & PARSEC & 0.65$^{+0.09}_{-0.08}$ & 1.62$^{+0.12}_{-0.13}$ & 0.75$^{+0.04}_{-0.05}$ & 0.63$^{+0.04}_{-0.05}$  \\
$\tau_{75}$ & MIST & 0.71$^{+0.12}_{-0.10}$ & 1.69$^{+0.16}_{-0.17}$ & 0.69$^{+0.05}_{-0.05}$ & 0.55$^{+0.06}_{-0.06}$  \\
$\tau_{75}$ & BaSTI18 & 0.66$^{+0.11}_{-0.10}$ & 1.64$^{+0.15}_{-0.16}$ & 0.70$^{+0.05}_{-0.06}$ & 0.58$^{+0.05}_{-0.06}$  \\
$\tau_{50}$ & PARSEC & 0.82$^{+0.12}_{-0.16}$ & 3.15$^{+0.25}_{-0.21}$ & 0.68$^{+0.07}_{-0.13}$ & 0.48$^{+0.07}_{-0.07}$  \\
$\tau_{50}$ & MIST & 0.89$^{+0.13}_{-0.18}$ & 3.68$^{+0.28}_{-0.25}$ & 0.65$^{+0.07}_{-0.11}$ & 0.45$^{+0.07}_{-0.08}$  \\
$\tau_{50}$ & BaSTI18 & 0.96$^{+0.13}_{-0.16}$ & 3.21$^{+0.26}_{-0.22}$ & 0.69$^{+0.06}_{-0.10}$ & 0.49$^{+0.07}_{-0.07}$  \\
\hline
\multicolumn{6}{c}{Versus Projected On-Sky Distance } \\
\hline
$\tau_{90}$ & PARSEC & 1.14$^{+0.15}_{-0.13}$ & -0.05$^{+0.11}_{-0.13}$ & 0.74$^{+0.03}_{-0.03}$ & 0.52$^{+0.04}_{-0.04}$  \\
$\tau_{90}$ & MIST & 1.27$^{+0.21}_{-0.17}$ & -0.18$^{+0.15}_{-0.18}$ & 0.72$^{+0.04}_{-0.05}$ & 0.54$^{+0.04}_{-0.04}$  \\
$\tau_{90}$ & BaSTI18 & 1.16$^{+0.18}_{-0.15}$ & -0.05$^{+0.13}_{-0.16}$ & 0.71$^{+0.04}_{-0.04}$ & 0.51$^{+0.04}_{-0.05}$  \\
$\tau_{75}$ & PARSEC & 1.24$^{+0.16}_{-0.16}$ & 1.29$^{+0.16}_{-0.16}$ & 0.71$^{+0.04}_{-0.04}$ & 0.60$^{+0.05}_{-0.05}$  \\
$\tau_{75}$ & MIST & 1.34$^{+0.22}_{-0.19}$ & 1.33$^{+0.20}_{-0.21}$ & 0.65$^{+0.05}_{-0.05}$ & 0.52$^{+0.06}_{-0.06}$  \\
$\tau_{75}$ & BaSTI18 & 1.26$^{+0.21}_{-0.19}$ & 1.29$^{+0.19}_{-0.20}$ & 0.67$^{+0.05}_{-0.06}$ & 0.56$^{+0.05}_{-0.06}$  \\
$\tau_{50}$ & PARSEC & 1.52$^{+0.21}_{-0.28}$ & 2.78$^{+0.30}_{-0.25}$ & 0.62$^{+0.07}_{-0.12}$ & 0.48$^{+0.07}_{-0.07}$  \\
$\tau_{50}$ & MIST & 1.67$^{+0.25}_{-0.32}$ & 3.24$^{+0.34}_{-0.30}$ & 0.61$^{+0.07}_{-0.10}$ & 0.45$^{+0.07}_{-0.08}$  \\
$\tau_{50}$ & BaSTI18 & 1.81$^{+0.24}_{-0.31}$ & 2.74$^{+0.32}_{-0.28}$ & 0.65$^{+0.06}_{-0.09}$ & 0.48$^{+0.06}_{-0.07}$  \\
\enddata
\end{deluxetable}

Our spatial sampling of the SMC is not unbiased, and the spatial distribution of our fields is highly centrally concentrated, with a median R$^{SMC}_{ell}$=1.4 (1.2) kpc including (excluding) the wing fields.  To test the extent to which our finding of a linear age gradient hangs on the results for our nine outer non-wing fields with R$_{ell}^{SMC}$$>$3.5 kpc, we have reperformed our fits excluding these fields (i.e., including only fields with R$_{ell}^{SMC}$$<$3.5 kpc), finding moderate values of the linear correlation coefficient ($\rho$$\geq$0.49) for the more recent CSFH metrics $\tau_{90}$ and $\tau_{75}$ (still at high significance, $p$-value$<$10$^{-3}$), and less evidence of linearity for $\tau_{50}$ ($\rho$=0.30, $p$-value=0.013).  This is at least partially reflective of real field-to-field SFH variations, some of which are apparent in Fig.~\ref{csfh_rad_fig}.  While our pointings do not coincide with any of the known substructure in the SMC other than the wing \citep[e.g.,][]{pieres17,mackey18,martinezdelgado19,elyoussoufi21,cullinane23,massana24}, the fields on the eastern side of the SMC (in the direction of the LMC) with 2$\lesssim$R$_{ell}^{SMC}$$\lesssim$3 kpc tend to show indications of late-time star formation similar to the wing fields located even farther to the east.  The preferentially higher mass fraction formed in the eastern fields over the last couple of Gyr was also detected by \citet{massana22}, and may explain the eastward offset found in the spatial distribution of intermediate-age stellar populations by \citet{mackey18}.  Furthermore, \citet{noelsfh} noted that among their fields located throughout the outer SMC in various directions, only their eastern fields showed ongoing star formation over the last $\sim$2 Gyr.  Specifically, with the exception of their wing fields, all of the \citet{noelsfh} fields show similar SFH trends as our outer non-wing fields, supporting the idea that our outer non-wing fields are typical for their R$_{ell}^{SMC}$.  Additional support comes from our lone field to the south of the SMC main body (SMC\_A25), which has a mass assembly history essentially identical to the outer non-wing fields, completing nearly all of its star formation (99.1\%$^{+0.8}_{-1.8}$ of its cumulative stellar mass formed) by 1 Gyr ago (seen at R$_{ell}^{SMC}$$\sim$2.9 kpc and position angle $\approx$$-$170$^{\circ}$ in Fig.~\ref{csfh_rad_fig}).  

We have provided the first quantitative measurements of the radial stellar age gradient in the SMC and its statistical significance.  However, two previous studies have provided spatially resolved stellar age information over a large number of sightlines (N$>$100), allowing us to calculate radial age gradients in the SMC for comparison with our results.  First, stellar ages were provided for 10 non-contiguous ground-based imaging fields by \citet[][see their Fig.~1]{piatti12}, who spatially divided each of their fields into a grid of 16 subregions for a total of 160 measurements.  Calculating R$_{ell}^{SMC}$ for each of their fields identically as for ours, assuming the same SMC center location, axis ratio and position angle, their measurements yield a radial age gradient of $\delta$(age)/$\delta$(R$_{ell}^{SMC}$) = 1.56$^{+0.39}_{-0.36}$ Gyr/kpc over 0.2$\gtrsim$R$_{ell}^{SMC}$$\gtrsim$5.5 kpc, substantially steeper than any of the values we find in Fig.~\ref{csfh_rad_fig}.  This difference is likely due at least in part to their methodology.  In contrast to the CMD synthesis technique used here and in previous studies of the spatially resolved SFH in the SMC \citep{hz04,noelsfh,rubele15,rubele18,massana22}, the age measurement technique used by \citet{piatti12} was designed and calibrated using star clusters, and measures the age of the dominant stellar population without any regard for the age \textit{distribution}.  Furthermore, the age calibration used by \citet{piatti12}, detailed in \citet{geisler97}, is based on a 3rd-order polynomial fit to ages for only six clusters, each with assumed ages nearly three decades old or more.  \citet{piatti12} also point out that their method for measuring cluster ages would in fact fail in the case of a constant SFH since their technique assumes the detection of a peak in the stellar age distribution.  While they find that this is not the case for the fields they analyze, a constant SFH seems to be a non-negligible possibility in the case of fields towards the SMC based on Fig.~\ref{combined_sfh_fig}.     

The other study providing the necessary information to calculate age gradients is \citet{rubele18}, based on contiguous near-infrared imaging from the VMC survey.  A methodological issue with the \citet{piatti12} ages is supported by results based on \citet{rubele18}, yielding radial age gradients statistically consistent with our results in Fig.~\ref{csfh_rad_fig}.  Using SFH results for individual VMC tile subregions from \citet{rubele18} and again performing linear fits versus R$_{ell}^{SMC}$ (excluding fields towards the wing\footnote{We have excluded VMC tile 3\_5 and subregions 1$-$3, 5$-$7 and 9 of tile 4\_5.} as in our analysis), we find radial gradient slopes of ($\delta$$\tau_{90}$, $\delta$$\tau_{75}$, $\delta$$\tau_{50}$)/$\delta($R$_{ell}^{SMC}$) = (0.64$^{+0.02}_{-0.02}$, 0.59$^{+0.03}_{-0.03}$, 0.65$^{+0.04}_{-0.04}$) Gyr/kpc, in agreement with our results to within 1-$\sigma$ uncertainties.  The agreement is particularly good at more recent lookback times ($\tau_{90}$, $\tau_{75}$), likely because of the shallower photometric depth of the VMC imaging (see Table 1 of \citealt{rubele18}).  Furthermore, the agreement we find between our radial gradient slopes and those from \citet{rubele18}, which are based on contiguous ground-based imaging, argues against spatially biased sampling affecting our radial age gradient measurements at a statistically significant level.

\subsection{Comparison to Previous Studies \label{litsect}}

\begin{figure}
\gridline{\fig{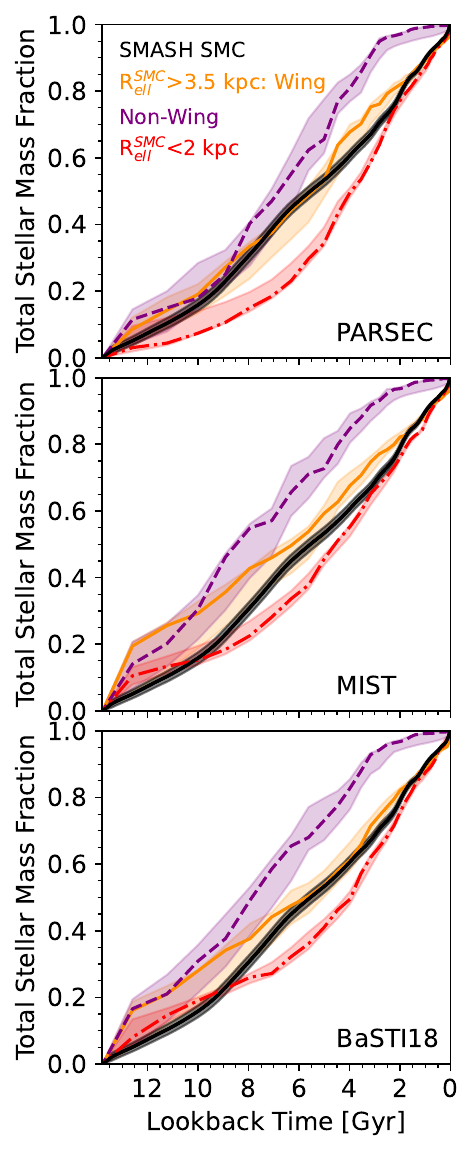}{0.3\textwidth}{}
          \fig{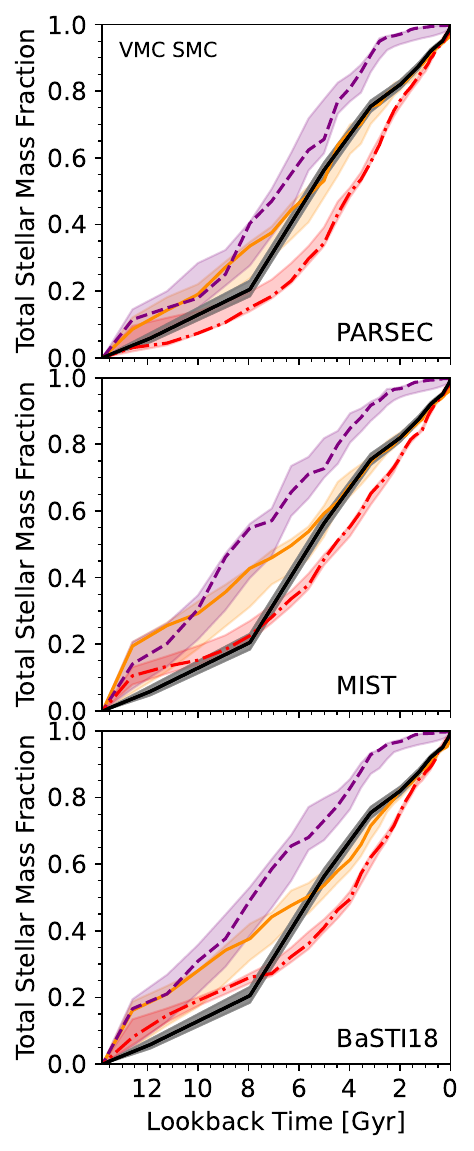}{0.3\textwidth}{}
          \fig{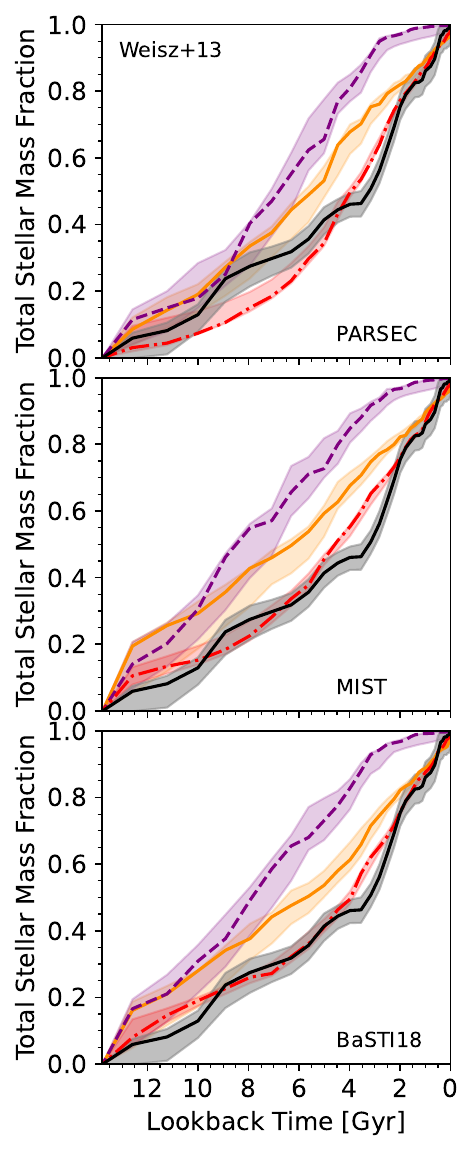}{0.3\textwidth}{}}    
\caption{Comparison between our combined CSFHs for different spatial regions in the SMC (shown as in Fig.~\ref{combined_sfh_fig}) and SMC CSFHs from \citet{massana22} (left column), \citet{rubele18} (middle column) and \citet{weisz13} (right column).  Each row corresponds to a different stellar evolutionary model that we assumed when calculating our SFH results, while each of the literature SFH studies shown used a single, different evolutionary model.  Agreement is generally good, especially given differences in spatial sampling of the SMC.}
\label{complitfig}
\end{figure}

While only \citet{piatti12} and \citet{rubele18} provide sufficient spatial sampling of the SMC to quantify radial stellar age gradients for comparison with our results, more general comparisons with recent literature studies provide further validation of the SFH fits on which our radial age gradient measurements are based.  In Fig.~\ref{complitfig}, we compare our CSFHs from three spatially selected regions of the SMC, as in Fig.~\ref{combined_sfh_fig}, to CSFHs available in the literature.  Each row of Fig.~\ref{complitfig} corresponds to a different assumed stellar evolutionary model in our SFH fitting, and each column corresponds to a different literature study for comparison, including (from left to right) SMASH \citep{massana22}, VMC \citep{rubele18}, and the WFPC2 fields analyzed by \citet{weisz13}.  Previous surveys tend to find CSFHs intermediate between our results for the inner and outer SMC, which is to be expected based on differences in spatial sampling.  In particular, our fields are biased towards the inner SMC, with a median projected  R$_{ell}^{SMC}$=1.3 kpc, whereas SMASH excluded the innermost SMC due to crowding and VMC coverage was contiguous.  Slight discrepancies are seen at early ($\gtrsim$7 Gyr) lookback times compared to the VMC results, and at intermediate lookback times ($\sim$2$-$4 Gyr) compared to the \citet{weisz13} results, although additional causes include differences in assumed stellar evolutionary model, SMC distance, and details of the SFH fitting methodology used in each study.  The minor ($<$2-$\sigma$) differences seen between our CSFHs and those from the VMC survey at early lookback times are not present in the comparison to SMASH, implying that the shallower imaging and/or near-infrared wavelength regime used by VMC may be a contributing factor.  Furthermore, unlike our approach, each of the literature studies assumed a single set of evolutionary models for SFH fitting, rendering it impossible to assess the impact of this choice on their results.  

Regarding the comparison to \citet{weisz13} in the right-hand column of Fig.~\ref{complitfig}, discrepancies in the CSFH at intermediate ages (2$-$6 Gyr) can also be traced to differences in spatial coverage.  The combined CSFH from \citet{weisz13}, based on only seven WFPC2 fields, is dominated by the innermost three of these fields at R$_{ell}^{SMC}$$<$1 kpc, which are much more well-populated.  Accordingly, viewed on a per-field basis, the \citet{weisz13} results are in reasonable agreement with our measurements since their three innermost fields have $\tau_{50}$$\sim$3 Gyr, while the outer four fields, with R$_{ell}^{SMC}$$\sim$4 kpc, have $\tau_{50}$$\sim$7 Gyr (see Fig.~4 of \citealt{weisz13}), quite similar to our results in the upper left-hand panel of Fig.~\ref{csfh_rad_fig} given different assumptions on distance, extinction and input stellar evolutionary models between our study and \citet{weisz13}.  In addition, detailed per-field CSFH comparisons can function as checks of both internal and external consistency, so we provide further comparisons along with our results for all individual fields in Appendix \ref{indfieldsect}.

\begin{figure}
\gridline{\fig{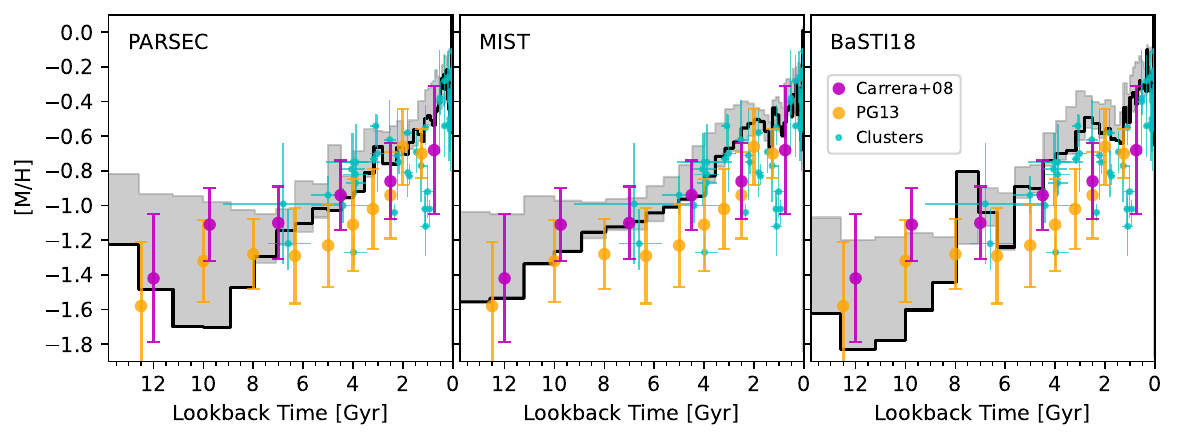}{0.99\textwidth}{}}    
\caption{Combined SMC AMR for each of our three assumed evolutionary models.  The SMC field AMRs from \citet{carrera08} and \citet{pg13} are overplotted in magenta and orange respectively, and SMC clusters are overplotted in cyan, with ages and metallicites from \citet{viscacha1,viscacha3,viscacha4,oliveira23}.}  
\label{amrfig}
\end{figure}

In Fig.~\ref{amrfig}, we show the AMR obtained by statistically combining all of our SMC fields as a black line, with uncertainties indicated by grey shading.  This AMR is compared to results for SMC field stars from CaII triplet spectroscopy by \citet{carrera08} and isochrone-based field star ages and metallicities from \citet{pg13} as well as isochrone-fitting values for clusters from \citet{viscacha1}, \citet{viscacha3,viscacha4} and \citet{oliveira23}.  Given that we imposed no restrictions whatsoever on the allowed AMR in any of our fields, agreement is very good (generally within uncertainties), providing an important validation of our SFH fitting procedure.  Again, minor discrepancies are due to differences in spatial sampling of the SMC, with lower literature metallicities resulting from more radially distributed targeting of the SMC field in literature studies \citep{carrera08,pg13} given its negative radial metallicity gradient \citep[e.g.,][]{choudhury20,grady21,massana24}.

\section{Discussion \label{discusssect}}
 
\begin{figure}
\gridline{\fig{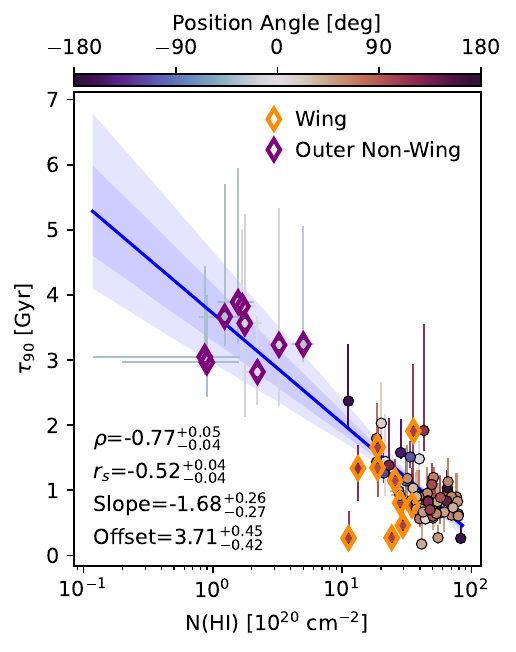}{0.327\textwidth}{}
          \fig{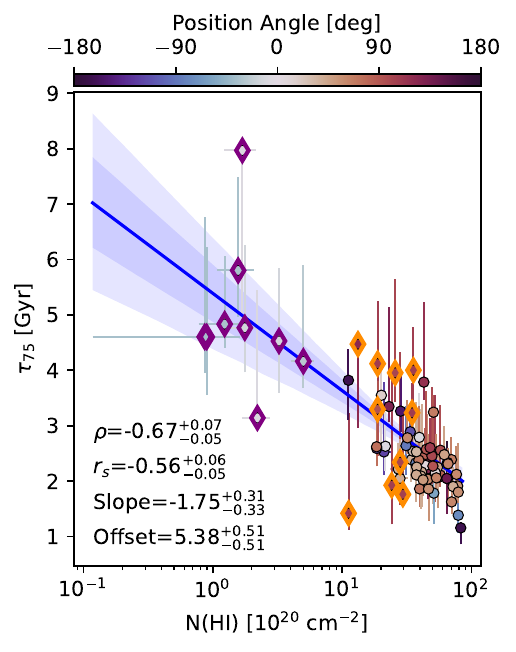}{0.327\textwidth}{}
          \fig{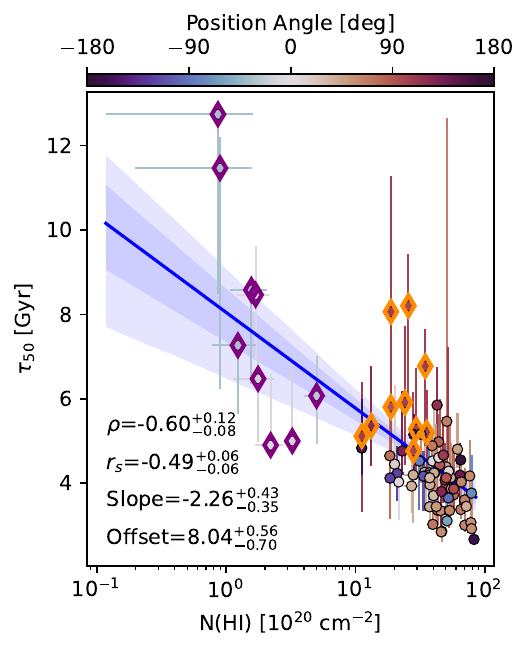}{0.337\textwidth}{}}
\caption{Relationship between lookback time to form a given mass fraction (from Fig.~\ref{csfh_rad_fig}) versus HI column density \citep{gaskap}.  The Pearson linear correlation coefficient $\rho$ and the Spearman rank correlation coefficient $r_{s}$ are given in the lower left corner of each panel versus Log$_{\rm 10}$ N(HI) \textbf{along with the coefficients of the best-fit line (shown in blue).}  The mass assembly histories of our fields correlate loosely with present-day N(HI) over a broad range of lookback times, although this may simply reflect the fairly concentrated present-day HI morphology of the SMC.}
\label{nhivtau_fig}
\end{figure}

Overall, our SFH results imply that in the SMC, 
the spatial distribution of its 
stellar populations is characterized globally by an outside-in present-day age gradient, but modified by the LMC-SMC interaction in a location-dependent way.  The global, long-term nature of the outside-in age gradient suggests that it is driven primarily by in situ (rather than external) processes, which consist of at least two non-exclusive possibilities.  
One potential explanation for the lack of recent star formation in the outer SMC could be outside-in quenching due to a shrinking gas reservoir \citep[e.g.,][]{stinson09}.  This idea is supported by an outside-in cessation of star cluster formation over the last Gyr \citep{bitsakis18,bica20}, as well as a correlation between the HI column density N(HI) and our CSFH metrics ($\tau_{90}$, $\tau_{75}$, $\tau_{50}$) over a broad range of lookback times, shown in Fig.~\ref{nhivtau_fig}.  However, given the turbulent interaction history of the SMC and its complex present-day HI position and velocity structure, such a correlation may simply reflect the fact that N(HI), in a gross sense, tends to decrease moving away from the SMC center.  Indeed, at sufficiently recent lookback times, the correlation breaks down -- the correlation between $\tau_{99}$ (sampling the most recent 0.5 Gyr to within uncertainties) and N(HI) has 
Spearman $r_{s}$=$-$0.14$^{+0.10}_{-0.08}$ at low statistical significance ($p$-value $>$0.1).  

Another potential explanation for the present-day outside-in age gradient is that older stellar populations have been driven to preferentially larger radii over time by star formation feedback (see \citealt{graus} and references therein).  Observationally, this idea is supported by the ubiquity of an extended, round stellar component seen for SMC-mass dwarfs regardless of environment or color \citep{kadofong20}, while simulations predict that feedback-driven stellar radial migration is most efficient at SMC-like masses \citep{elbadry16}.  However, in addition to these two internally-driven processes, the extended LMC-SMC interaction may have also contributed to the present-day observed radial age gradient.  Simulations predict that galaxy-galaxy interactions can dynamically heat stars in the outer regions of disks, temporarily inducing large radial excursions on timescales of a few hundred Myr \citep{carr22}.  Such interaction-driven dynamical heating is directly observed in the LMC \citep{choi22}, and further modeling work is needed to address whether this phenomenon is relevant to the SMC.

Our results also suggest that infall into the Milky Way may not have played an important role in regulating star formation in the SMC. The discrepancy between the star formation rates towards the wing compared to the outer non-wing fields begins $\sim$3$-$6 Gyr ago (see Fig.~\ref{csfh_rad_fig} and the right-hand panel of Fig.~\ref{combined_sfh_fig}). 
This timing argues against ram pressure stripping as the 
cause of a lack of younger ($\lesssim$3$-$4 Gyr) stars in the outer SMC since the LMC and SMC likely did not enter the Milky Way's virial radius until $\sim$1$-$2 Gyr ago \citep{kallivayalil06a,besla12,patel17}.  More generally, our results highlight the idea that the constraining power of our spatially resolved lifetime SFHs is maximized when leveraged together with the wealth of additional information available for sightlines towards the SMC, so with this in mind, we now examine our SFH fits in the context of available stellar kinematics and abundances.  We then place our SFH results in context with our companion study of the LMC, compare them with other nearby dwarf galaxies, and finally describe some plans for future studies enabled by this work.  

\subsection{Constraints From Stellar Kinematics and Abundances \label{chemodynsect}}

In the wing, in contrast to the rest of the outer SMC, we find a best-fit SFH consisting of nearly constant star formation over the last $\sim$3$-$4 Gyr (in fact, over its entire lifetime; see Fig.~\ref{combined_sfh_fig}), implying that multiple previous LMC-SMC interactions played a role in sustaining star formation there.  On one hand, the most recent LMC-SMC collision $\sim$150 Myr ago \citep{zivick18,choi22} has been put forth as a contender for generating foreground tidal debris along sightlines towards the wing, in addition to the Magellanic Bridge, based largely on a comparison to simulations \citep[e.g.,][]{diazbekki12,zivick19}.  Accordingly, the constant SFH we observe is at least partially a consequence of composite stellar populations viewed along the line of sight, evidenced by bimodalities in stellar kinematics and distances \citep[e.g.,][]{dobbie14,elyoussoufi21,james21,elyoussoufi23,mucciarelli23}.  In addition, star clusters towards the wing show a broad range of star cluster distances, ages and metallicites, including both a young ($\lesssim$200 Myr) metal-rich \textit{in situ} component and a metal-poor component with a broad range of ages \citep[0.5$-$7 Gyr;][]{oliveira23}.

Stellar kinematics and chemical abundances \citep{omkumar21,almeida23a} have both been used to argue that the recent LMC-SMC collision was responsible for transporting stars outward from the inner SMC (and, in one case, towards a field as far as $\gtrsim$9 kpc from the optical center of the SMC; \citealt{cullinane23}).  Proper motion measurements in the eastern SMC and the wing are reasonably consistent with this hypothesis, given residual mean motions towards the Magellanic Bridge of $\sim$0.25 mas/yr ($\sim$75 km/s) for RGB stars \citep{zivick18,deleo20,zivick21}, naively translating to a travel time of $\sim$13 Myr/kpc.  The link between the recent LMC-SMC collision and the bimodal stellar distance distribution is also consistent with the  locations of the peaks in the bimodal radial velocity distribution of red giants towards the eastern SMC \citep{james21}.

However, additional LMC-SMC interactions previous to the most recent collision must have also contributed to star formation in the wing given its near-constant star formation rate.  A previous interaction $\sim$1.7$-$2 Gyr ago (that likely created the Magellanic Stream; e.g., \citealt{nidever08,besla12,diazbekki12}) could have played at least two roles: First, this interaction may also have been responsible for sourcing stars from the inner SMC, explaining the quite mild ($\lesssim$0.2 dex) metal-rich offset in the metallicity distribution of the less distant of the two line-of-sight components reported by \citet{almeida23a}.  Second, this interaction almost certainly triggered some \textit{in situ} star formation in the wing as well, evidenced by the spike in stellar cluster production in the outer SMC $\sim$2 Gyr ago, possibly due to interaction-enhanced massive giant molecular cloud formation \citep{bitsakis18,williams22}.  An even earlier LMC-SMC interaction also played an important role of course, since the peak of star formation in the wing was $\sim$4 Gyr ago.  This is roughly contemporaneous with the increase in star formation rate in the LMC and SMC more generally \citep{weisz13,massana22}, as well as a change in chemical abundance trends in the SMC \citep{povick23}.  This earlier interaction $\sim$3$-$4 Gyr ago could also be responsible for the lack of stellar clusters older than $\sim$3 Gyr in the northeastern SMC \citep{parisi24}, corroborated by a dropoff in the star cluster age distribution in the outer LMC as well \citep{gatto24}.  Furthermore, a correspondence between SFH features and LMC-SMC interactions extending to at least $\sim$2$-$3.5 Gyr ago was reported by \citet{sakowska23} for the shell-like feature $\sim$2 deg to the northeast of the SMC optical center (they report star formation rates quite similar to our wing fields, $\Sigma_{\rm SFR}$$\sim$0.5$-$1$\times$10$^{-3}$ M$_{\odot}$/yr/kpc$^{2}$, peaking $\sim$200 Myr ago at $\sim$3$\times$10$^{-3}$ M$_{\odot}$/yr/kpc$^{2}$).  

\subsection{Comparison to the LMC}

The multitude of previous LMC-SMC interactions has left detectable imprints on the lifetime SFH of each Cloud when viewed in a spatially resolved way.  In the SMC, this is exemplified by the recent star formation towards the wing, atypical compared to other fields at similar R$_{ell}^{SMC}$ (Figs.~\ref{combined_sfh_fig}$-$\ref{csfh_rad_fig}).  In the LMC, interactions have likely played a similarly important but somewhat indirect role, boosting radial migration to create a \enquote{V-shaped} radial age gradient as seen in some other similarly massive (M$_{\star}$$\gtrsim$10$^{9}$ M$_{\odot}$) late-type galaxies (discussed in Cohen et al.~2024 and references therein).  In addition, while the gaseous content of both Clouds is clearly indicative of previous disturbances, repeated LMC-SMC interactions have wreaked havoc on the HI morphology and kinematics of the less massive SMC.  
For this reason, unlike in our analysis of the LMC (Cohen et al.~2024), we have not presented radial age gradients under the assumption of an SMC center location based on HI kinematics; given the finding of two (or possibly more) components with a complex interplay discussed by \citet{murray23}, the utility of attempting to define such a center seems somewhat unclear.

\begin{figure}
\gridline{\fig{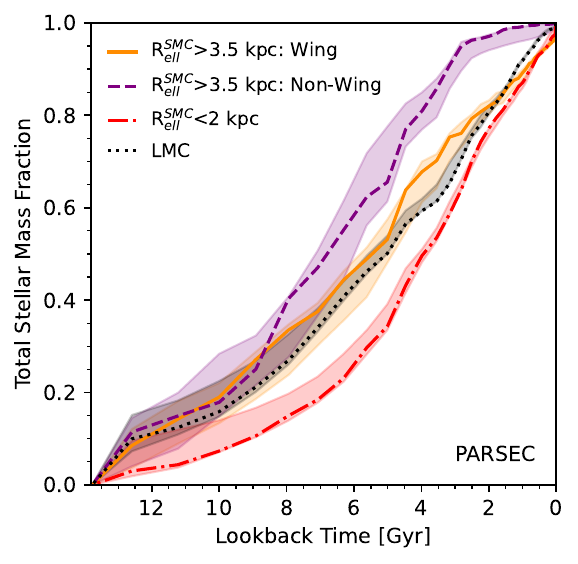}{0.49\textwidth}{}}    
\caption{CSFHs for different SMC regions as in the left-hand panel of Fig.~\ref{combined_sfh_fig}, but now with our combined CSFH for the LMC (Cohen et al.~2024) overplotted as a black dotted line.} 
\label{csfh_combined_lit_fig}
\end{figure}

While the LMC and SMC have clearly distinct \textit{intra-}galaxy stellar age trends, trends in their \textit{global} (i.e., galaxy-wide) mass assembly histories over cosmic time (i.e., CSFHs) are quite similar, with both Clouds having lifetime SFHs approximating constant star formation.  This is illustrated in Fig.~\ref{csfh_combined_lit_fig}, showing that our combined LMC CSFH from Cohen et al.~(2024) falls intermediate between the CSFHs of the different SMC regions we analyze.  Simulations predict that such constant SFHs, and some detectable galaxy-to-galaxy scatter, are fairly generic features of SMC- to LMC-mass galaxies in the vicinity of massive hosts \citep{engler23}.  These simulations further predict that at LMC/SMC-like galaxy masses, these constant SFHs are quite insensitive to time of infall, reinforcing independent predictions that quenching times are maximized at LMC/SMC-like satellite masses \citep{wetzel15b}.  These predictions are generally supported by observed SFHs for Local Group dwarfs, especially given the near-constant SFHs observed for some relatively more isolated, massive, gas-rich Local Group dwarf irregulars \citep[e.g.,][]{skillman14,weisz14,weisz15}.  Hence, current evidence, although sparse, seems to suggest that prolonged dwarf-dwarf interactions are not a prerequisite for constant lifetime SFHs in the Local Group, although stringent comparisons to simulations will require larger samples in more diverse (i.e., isolated) environments.  

\subsection{The Outside-In Present-Day Age Gradient in Context}

Our results reveal a dichotomy where the LMC and SMC have nearly identical trends in their CSFHs when viewed globally, but have schematically different \textit{internal} radial age trends when viewed in a spatially resolved way.  In fact, the SMC effectively sets an upper limit on the stellar mass of dwarf galaxies observed to have outside-in age gradients. An increase in average age with radius 
has been observed (or inferred from stellar metallicity gradients) in numerous Local Group dwarfs 
\citep{faria07,hidalgo13,beccari14,martinezvasquez15,santana16,delpino17,albers19,bettinelli19,pace20,ruizlara21,rusakov21,abdollahi23} but all of these dwarfs are at least an order of magnitude less massive than the SMC, with the next most massive being the gas-rich dwarf irregular WLM (M$_{\star}$$\approx$ 4.7$\times$10$^{7}$M$_{\odot}$; \citealt{mcconnachie12}).  For comparison, estimates of the stellar mass of the SMC range from  $\sim$3.0$-$4.8$\times$10$^{8}$ M$_{\odot}$ within the inner $\sim$3$-$4 kpc \citep{hz04,skibba12,rubele18,diteodoro19,deleo23}.  Furthermore, these values are likely lower limits, since the old (red giant and red clump) stellar population of the SMC extends to $\gtrsim$8 kpc \citep{massana20}, and the SMC hosts a variety of additional stellar substructure in its periphery \citep[e.g.,][]{pieres17,mackey18,elyoussoufi21,chandra23,cullinane23}.  It bears mention that outside-in age gradients have been reported for a few galaxies much more massive than the SMC (Log$_{\rm 10}$ M$_{\star}$/M$_{\odot}$$\gtrsim$10; \citealt{staudaher19}) based on fits to spectral energy distributions from broadband imaging \citep{dale20}, but these age measurements assume a parametric (delayed-$\tau$) star formation history (discussed by, e.g., \citealt{carnall19}), inconsistent with observed SFHs for dwarfs showing outside-in age gradients in at least some cases (e.g., \citealt{rusakov21,ruizlara21}; see the discussion of \enquote{fast} vs.~\enquote{slow} dwarfs by \citealt{gallart15}).

While all of the Local Group dwarfs with outside-in age gradients are less massive than the SMC, the converse is not true -- not all dwarfs less massive than the SMC have outside-in age gradients.  For example, NGC 6822 has a flat age gradient out to $\sim$2.5 disk scalelengths, although it may have undergone previous interactions based on the flattened, twisted stellar morphology in its outskirts \citep{cannon12,fusco14,zhang21}.  More generally, a variety of age gradients are implied in late-type Local Volume dwarfs based on radial distributions of both blue helium burning stars and asymptotic giant branch stars relative to red giants \citep{mcquinn12,mcquinn17}.  Similarly, radial age gradients ranging from outside-in to flat are predicted in dwarfs by at least one set of zoom-in hydrodynamical cosmological simulations \citep{graus}.  In these simulations, feedback-driven stellar radial migration creates the outside-in gradient, which can be counteracted by late-time star formation at large radii, flattening the outside-in gradient.  The \citet{graus} simulations predict, as a result of these competing processes, a correlation between global mass assembly history (quantified by galaxy-wide $\tau_{50}$ and $\tau_{90}$) and radial age gradient slope, 
such that globally older galaxies show steeper outside-in age gradients (normalized to their half-mass radius).  Observationally, such a trend has yet to be confirmed or refuted, as the sample of dwarfs with directly measured radial age gradients is both miniscule and severely biased, limited to the Local Group by observational capabilities.  Fortunately, radial age gradient measurements of more distant targets are now within the reach of modern observational facilities, and radial age gradient measurements of a larger sample will be presented elsewhere (R.~E.~Cohen, in prep.).  

\subsection{Caveats and Future Work}

It is evident (e.g., from the discussion in Sect.~\ref{chemodynsect}) that high-dimensional information, combining ages, distances, kinematics and chemical abundances of resolved stars (and/or stellar clusters) provides additional leverage towards disentangling the history of the LMC-SMC-Milky Way system; it is equally evident that this task is far from complete.  Our SFH fitting technique assumes a single line-of-sight distance, and in Appendix \ref{simsect}, we demonstrate that this assumption does not impact our SFH results beyond their uncertainties.  However, the ability to simultaneously solve for the ages and distances of individual resolved stars could provide the diagnostic power needed to map the contributions of particular LMC-SMC interactions to the present-day stellar content of the SMC and its substructures, posing strong constraints for models of the LMC-SMC interaction.  Indeed, SFH fitting methodologies that incorporate non-negligible stellar distance distributions have already been applied to substructure in the SMC in at least one instance \citep{sakowska23}.  As another example, \citet{almeida23a} used \textit{Gaia}-based distances and kinematics together with abundances of resolved stars to characterize the two line-of-sight components towards the SMC's wing.  However, their stellar sample was comprised exclusively of red giants with available spectroscopic abundances, preventing precise individual stellar age and distance measurements.  

In forthcoming studies of this series, we will apply a technique that fits the spectral energy distribution (SED) of each star to simultaneously solve for stellar and dust properties on a star-by-star basis \citep{beast}.  In contrast to our approach, which uses the well-established CMD synthesis technique for each field, the star-by-star SED-fitting approach can provide, among other things, individual stellar ages and distances and their uncertainty distributions (C.~E.~Murray, in prep.).  This approach has already been used to characterize differences in the three-dimensional structure of stars and dust towards the southwestern bar of the SMC \citep{petia21} and will be applied to new imaging towards the wing (C. Lindberg, in prep.), offering tantalizing possibilities for understanding the evolution of the SMC.  As one example of these possibilities, \citet{almeida23a} point out that disentangling the stellar age, distance and chemical abundance distributions of the two line-of-sight components towards the wing can confirm the hypothesis that the nearby component was stripped from the inner SMC by the most recent LMC-SMC interaction.  Another intriguing possibility is to definitively refute (or confirm) a scenario where the LMC and SMC are not on first infall and have instead made multiple previous pericentric passages of the Milky Way, suggested by \citet{vasiliev24}.  Our SFHs are not incompatible with such a scenario, and include subtle hints of a star formation enhancement $\sim$7$-$8 Gyr ago in both the SMC (right-hand panel of Fig.~\ref{combined_sfh_fig}) and LMC (Fig.~14 of Cohen et al.~2024).  \citet{vasiliev24} detail that current data cannot completely rule out the multiple passage scenario, and high-dimensional, high fidelity information for large stellar samples along many sightlines may provide critical clues.  Fortunately, in addition to deep multiwavelength imaging, steps in this direction are already being taken with massively multiplexed spectroscopic campaigns targeting the LMC and SMC \citep[e.g.,][]{cioni19,almeida23b}.

\section{Conclusions}

We have harnessed together 83 individual HST imaging fields towards the SMC and fit lifetime SFHs to each individually.  We have combined fields by location within the SMC to quantify differences in their assembly histories (Sect.~\ref{globalsect}), and used per-field results to measure radial stellar age gradients in the SMC (Sect.~\ref{radgradsect}), using simulations to verify that our SFH fits are robust to various line-of-sight distance distributions (Appendix \ref{simsect}).  Our main results are as follows:

\begin{enumerate}
\item We observe an outside-in age gradient in the SMC.  Assuming PARSEC evolutionary models and an elliptical geometry for the SMC, we find radial age gradient slopes of $\delta$($\tau_{90}$, $\tau_{75}$, $\tau_{50}$)/$\delta$R  
= (0.61$^{+0.08}_{-0.07}$, 0.65$^{+0.09}_{-0.08}$, 0.82$^{+0.12}_{-0.16}$) Gyr/kpc.  The correlation between lookback time to form a given cumulative mass fraction (i.e., $\tau_{90}$, $\tau_{75}$, $\tau_{50}$) and distance from the optical center of the SMC remains linear (Pearson $\rho$$>$0.6) at high statistical significance ($p$-values$<$10$^{-3}$) regardless of choice of stellar evolutionary model for SFH fits or the parameterization of distance from the SMC center as linear or elliptical (semi-major axis equivalent).  We argue that the observed age gradient is driven primarily by in situ processes, supported by timing arguments, including a similar trend seen in star cluster ages.  

\item Fields in the inner SMC (R$_{ell}^{SMC}$$<$2 kpc) have been forming stars throughout its lifetime at a rate that is approximately constant, and an order of magnitude higher than stars farther from the SMC center (R$_{ell}^{SMC}$$>$3.5 kpc).  In accord with previous studies, we also find evidence for a further enhancement in star formation $<$150 Myr ago, following the most recent LMC-SMC collision (right-hand panel of Fig.~\ref{combined_sfh_fig}).  However, our data are insufficient to more precisely age-date this star formation enhancement or assess whether it is global in nature.

\item In contrast to the remainder of the outer SMC (3.5$<$R$_{ell}^{SMC}$$<$6.5 kpc), fields in the vicinity of the wing show near-constant best-fit SFHs even over the most recent $\sim$3$-$4 Gyr.  This is likely due to a superposition of multiple stellar components along the line of sight, including tidal debris.  Kinematics of RGB stars are consistent with the hypothesis of such tidal debris being stripped from the inner SMC during the most recent LMC-SMC collision, but earlier interactions likely also played a role, supported by both the field star metallicity distribution and the age distribution of star clusters in the outer SMC.   

\end{enumerate}

\begin{acknowledgements}

It is a pleasure to thank Michele Cignoni, Pol Massana and Dan Weisz for sharing SFH results for comparison, and the anonymous referee for their detailed and constructive comments.  Some of the data presented in this article were obtained from the Mikulski Archive for Space Telescopes (MAST) at the Space Telescope Science Institute. The specific observations analyzed can be accessed via\dataset[doi: 10.17909/8ads-wn75]{https://doi.org/10.17909/8ads-wn75}.  Based on observations with the NASA/ESA Hubble Space Telescope obtained at the Space Telescope Science Institute, which is operated by the Association of Universities for Research in Astronomy, Incorporated, under NASA contract NAS 5-26555.  
Support for programs HST GO-15891, GO-16235 and GO-16786 were provided by NASA through a grant from the Space Telescope Science Institute, which is operated by the Association of Universities for Research in Astronomy, Inc., under NASA contract NAS 5-26555.  R.~E.~C. acknowledges support from Rutgers the State University of New Jersey.  

\end{acknowledgements}

\facilities{HST (ACS/WFC, WFC3/UVIS, WFPC2)}
\software{astropy \citep{astropy}, matplotlib \citep{matplotlib}, numpy \citep{numpy}, emcee \citep{emcee}, Dolphot \citep{dolphot}, MATCH \citep{match}}

\appendix

\section{Clusters Removed From Fields \label{clustersect}}

\begin{figure}[hb!]
\gridline{\fig{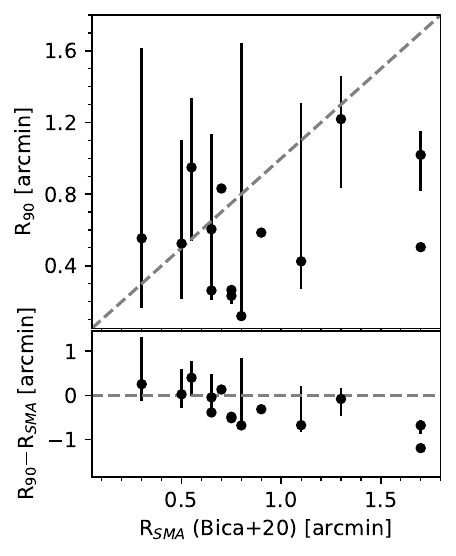}{0.35\textwidth}{}}
\caption{Comparison of R$_{\rm 90}$, the radius enclosing 90\% of the cluster light from \citet{hz06} against R$_{\rm SMA}$, the cluster semi-major axis from the \citet{bica20} catalog inside which we remove all stars from our catalogs before fitting SFHs.  Among the 15 clusters in common, all but one have R$_{\rm SMA}$$\geq$R$_{\rm 90}$ to within uncertainties.}
\label{clusterfig}
\end{figure}

Using the \citet{bica20} star cluster catalog, we have removed from our photometric catalogs all sources within their listed semi-major axis distance R$_{\rm SMA}$ from their cluster centers 
before fitting SFHs.  The relevant clusters and affected fields are listed along with the cluster centers and semi-major axes from \citet{bica20} in Table \ref{clustab}.  The cluster name given in the first column corresponds to the name used in the SIMBAD database, with any alternate names given in the following column.  To gauge how much of the cluster population is removed by our spatial cut using R$_{\rm SMA}$ from \citet{bica20}, in Fig.~\ref{clusterfig} we compare R$_{\rm SMA}$ against R$_{\rm 90}$ from \citet{hz06}, the radius enclosing 90\% of the cluster light based on surface brightness profile fits.  Among the 15 clusters matched to the sample analyzed by \citet{hz06}, the \citet{bica20} semi-major axis values we use are equal to or greater than the radius enclosing 90\% of the cluster light to within uncertainties in all cases except one cluster (OGLE-CL SMC 35) for which uncertainties on the surface brightness profile fits were not given by \citet{hz06}.  

\begin{figure}[ht!]
\gridline{\fig{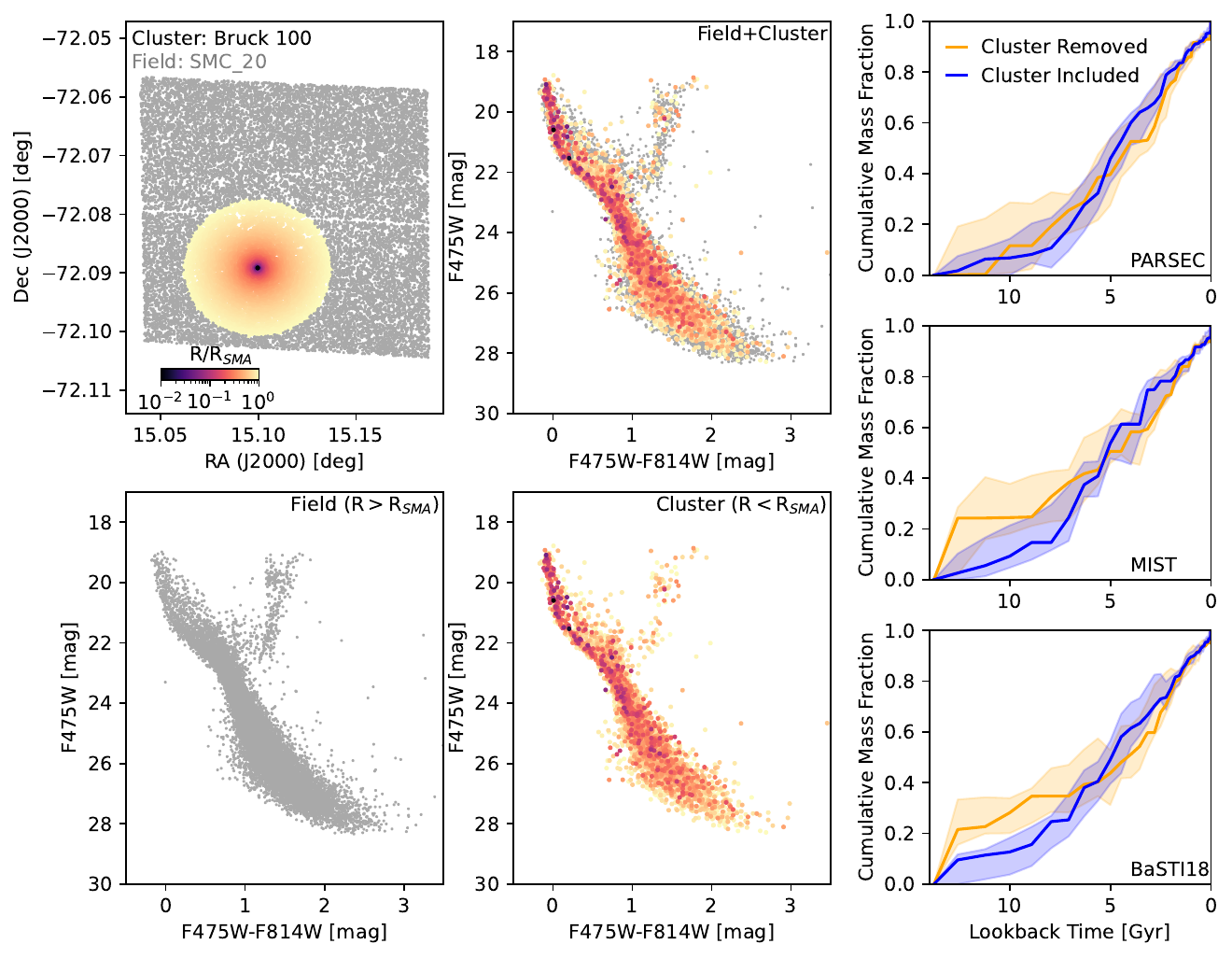}{0.99\textwidth}{}}
\caption{An example of our removal of known clusters from our fields and its impact on the resulting CSFH.  This example corresponds to the most pessimistic case, in  which the cluster (Bruck 100) occupies the largest projected area (1.54 arcmin$^{2}$, or 21\% of the field of view) in any of our target fields.  In the left four panels, stars removed from our catalog are color-coded by distance from the center of the cluster R$_{\rm SMA}$ listed by \citet{bica20}, and stars beyond R$_{\rm SMA}$, retained in our catalog, are shown in grey.  In the right-hand column, we compare the CSFH obtained after removing sources with R$<$R$_{\rm SMA}$ (orange) to the CSFH obtained without removing any sources over the entire field of view (blue).  The values of $\tau_{50}$, $\tau_{75}$ and $\tau_{90}$ are not impacted beyond their uncertainties even if no effort is made to remove the cluster from our photometric catalog.}   
\label{clusterexamplefig}
\end{figure}

As an additional check, we intentionally select the field with the worst cluster contamination by projected area, which is SMC\_20, with the cluster Bruck 100 occupying 1.54 arcmin$^{2}$, or 21\% of the field of view based on its R$_{\rm SMA}$=0.7$\arcmin$ \citep{bica20}.  This is shown in the top left panel of Fig.~\ref{clusterexamplefig}, where stars within R$_{\rm SMA}$ are color-coded by their distance from the center of the cluster and stars outside R$_{\rm SMA}$, retained in our catalog, are shown in grey.  The remaining three left-hand panels of Fig.~\ref{clusterexamplefig} compare the CMD loci of stars using this spatial selection.  To assess the impact of the cluster on the CSFH in the limiting case where no cluster stars were removed, we have recalculated the CSFH of the entire field of view without removing any stars.  The results are shown in the three vertical right-hand panels in Fig.~\ref{clusterexamplefig}, with each panel corresponding to a different assumed stellar evolutionary model.  Comparing the CSFHs obtained removing all stars within R$_{\rm SMA}$ (orange) to the CSFHs obtained without removing any stars (blue), we find that the values of $\tau_{50}$, $\tau_{75}$ and $\tau_{90}$ are not impacted beyond their uncertainties.

\begin{deluxetable}{lccccc}
\tablecaption{Known Clusters Excluded from Photometric Catalogs \label{clustab}}
\tablehead{
\colhead{Cluster} & \colhead{Other Names} & \colhead{RA (J2000)} & \colhead{Dec (J2000)} & \colhead{Radius} & \colhead{Field} \\ \colhead{} & \colhead{} & \colhead{$^{\circ}$} & \colhead{$^{\circ}$} & \colhead{$\arcmin$} & \colhead{}}
\startdata
OGLE-CL SMC 131 & & 15.515833  & -72.521111 & 0.70 & SMC\_2 \\ 
OGLE-CL SMC 144 & OGLE-CL SMC 236 & 16.021667 & -72.120556 & 0.60 & SMC\_8 \\  
BS95 11 & SBica7 & 9.580000  &  -73.378611 & 0.30 & SMC\_15 \\ 
OGLE-CL SMC 137 & B115, RZ144 & 15.846667 & -72.650833 & 0.90 & SMC\_18 \\ 
OGLE-CL SMC 140 & BS118 & 16.057083  &  -72.646111  & 1.30 & SMC\_18 \\ 
HGH2008 Cluster 1 & SSN16 & 14.879167 & -72.153056 & 0.20 & SMC\_19 \\  
Bruck 100 & & 15.099583 & -72.089167 & 0.70 & SMC\_20 \\ 
OGLE-CL SMC 221 & H86-164 & 13.937917 & -72.704444 & 0.45 & SMC\_25 \\
Cl Kron 28 & L43,RZ69 & 12.923750  &  -71.998333 & 1.70 & SMC\_35 \\ 
OGLE-CL SMC 73 & B59, RZ70 & 12.932083 & -72.840833 & 0.80 & SMC\_41 \\ 
H86 173 & & 14.338333   & -72.576111 & 0.50 & SMC\_42 \\ 
Bruck 135 & & 17.341250  &  -73.188611 & 0.70 & SMC\_44 \\ 
OGLE-CL SMC 67 & L41, RZ67, BMS231 & 12.732500 & -72.726389 & 0.65 & SMC\_45 \\ 
NGC 416 & K59, L83, ESO29-SC32, OGLE-CL SMC 158  & 16.996250  &  -72.355000  &  1.70  & SMC\_46  \\ 
BS95 138 & RZ164   & 16.816250  &  -72.102500  & 1.30  & SMC\_49 \\ 
OGLE-CL SMC 248 & & 22.455000  &   -73.430278 & 1.50 & SMC\_52 \\ 
OGLE-CL SMC 286 & & 22.639167  &   -73.422500 & 1.50 & SMC\_52 \\ 
OGLE-CL SMC 134 & RZ143,K47,L70  & 15.797500  &  -72.272222 & 0.75 & SMC\_54 \\ 
HW 63 & RZ182   &  17.551250  &  -73.208889 &  0.75  & SMC\_53 \\ 
OGLE-CL SMC 204 & H86-121 & 12.840000 & -73.138333 & 0.55 & SMC\_A4 \\ 
OGLE-CL SMC 75 & H86-126 & 12.973333 & -73.098333 & 0.50 & SMC\_A4 \\ 
H86 146 & & 13.413333 & -72.392778 & 0.90 & SMC\_A10 \\ 
H86 142 & & 13.394583 & -72.490278 & 0.90 & SMC\_A11 \\ 
Bruck 143 & RZ192 & 18.183333 & -72.750000 & 0.50 & SMC\_A12 \\ 
RZ2005 38 & & 11.121250 & -72.815833 & 0.80 & SMC\_A13 \\ 
OGLE-CL SMC 151 & & 16.554583 & -72.794167 & 1.20 & SMC\_A16 \\ 
OGLE-CL SMC 152 & & 16.587500 & -72.830000 & 0.65 & SMC\_A16 \\ 
OGLE-CL SMC 6 & & 9.889167 & -73.176944 & 0.80 & SMC\_A17 \\ 
RZ2005 180 & HW61 & 17.426250 & -72.295278 & 0.65 & SMC\_A21 \\ 
OGLE-CL SMC 103 & RZ95,B83 & 13.876667  & -73.072222 & 0.55 & SMC\_W1 \\ 
OGLE-CL SMC 185 & H86-84 & 11.641667  &  -72.765556 & 0.40 & SMC\_W6 \\ 
OGLE-CL SMC 35 & RZ48,H86-83 & 11.641667  &  -72.773889 & 0.70 & SMC\_W6 \\ 
OGLE-CL SMC 36 & RZ47,L31 & 11.648750  &  -72.741111 & 1.10 & SMC\_W6 \\ 
OGLE-CL SMC 181 & & 11.366667  &  -72.885833 & 0.70 & SMC\_W7 \\ 
RZ2005 43 & & 11.414583 &   -72.861944 & 0.30 & SMC\_W7 \\ 
Bruck 36 & & 11.424583  &  -72.841389 & 0.70 & SMC\_W7 \\ 
\enddata
\end{deluxetable}

\section{Simulations \label{simsect}}

We used a series of simulations to assess the impact of various line-of-sight stellar distance distributions on our recovered CSFHs.  Several previous SMC SFH studies have tested the impact of line-of-sight depth on SFH recovery, finding that their results were generally unaffected beyond their uncertainties.  However, each of these earlier studies tested only a single analytical form for the stellar distance distribution.  \citet{massana22} tested \enquote{gaussian-like} depths with standard deviations up to 5.5 kpc when fitting SMASH photometry, while \citet{rubele18} tested a Cauchy distribution based on RR Lyrae distances from \citet{muraveva18} for fitting VMC photometry, and \citet{hz04} tested a \enquote{characteristic depth} of 12 kpc in their fits to MCPS photometry.  

There is now a substantial body of observational evidence revealing that the line-of-sight distribution of the SMC's stellar populations is position-dependent, in terms of not only its \textit{extent}, but also in terms of its \textit{shape}.  For example, \citet[][see their Fig.~13]{nidever13}, using the red clump as a tracer, find distance distributions varying from flat to single-peaked to double-peaked with changing relative amplitudes of the two peaks.  Exploiting the contiguous coverage of the VMC survey, \citet{sub17} and \citet{elyoussoufi21} expanded these results by fitting single- and double Gaussians to the red clump distribution along numerous sightlines, finding standard deviations of 3$\lesssim$$\sigma$$\lesssim$11 kpc in the inner $\sim$5$^{\circ}$ of the SMC with values as large as $\sigma$$\sim$16 kpc in the outer regions.  The amplitude ratios of the double-peaked distance distributions are also positionally dependent, with the fainter component being equally or more well-populated than the brighter component in the inner $\sim$3$^{\circ}$ of the SMC where our fields lie \citep[][their Fig.~13]{sub17}.  In addition to results based on Gaussian fits, \citet{tatton21} use the red clump magnitude distribution from VMC photometry to map the full and interquartile ranges of the line-of-sight depth across the SMC, finding full depths from $\sim$10$-$26 kpc over the area covered by our fields.  Using VMC RR Lyrae stars as a tracer, \citet{muraveva18} find a line-of-sight depth of $\sim$9 kpc, in good agreement with results from optical imaging \citep{jd17}.  

\begin{figure}
\gridline{\fig{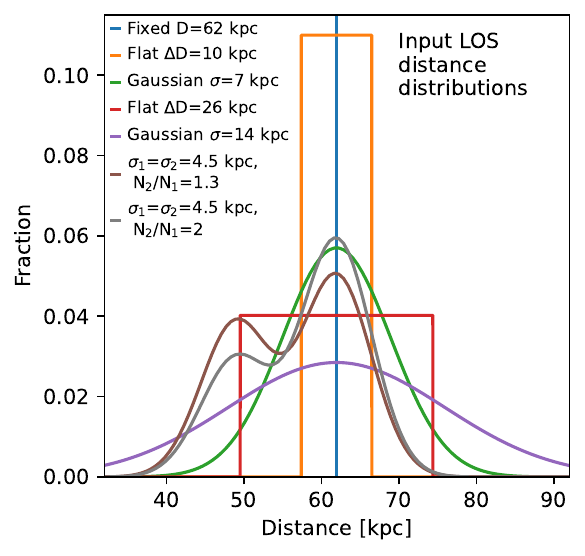}{0.39\textwidth}{}
          \fig{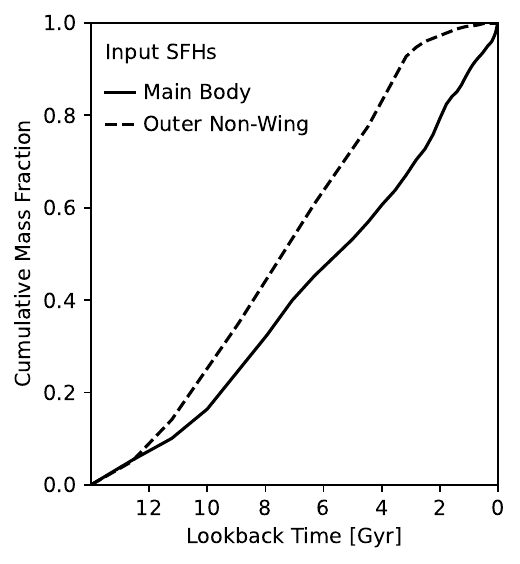}{0.35\textwidth}{}
          \fig{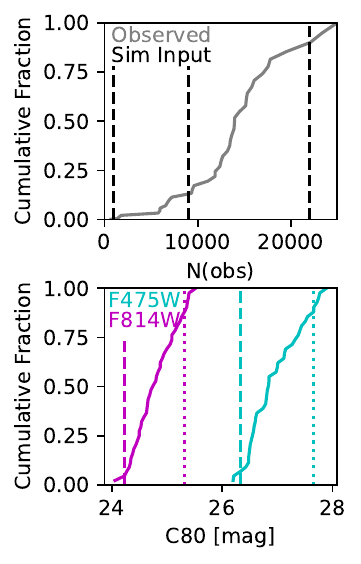}{0.25\textwidth}{}}
\vspace{-0.75cm}
\caption{\textbf{Left:} The seven input line-of-sight distance distributions applied to our synthetic CMDs before SFH recovery.  \textbf{Middle:} The two input SFHs we used in our simulations, one based on the results of \citet{massana22} representing the main body of the SMC and another representing our outer SMC fields, which show a dropoff in star formation in the last $\sim$3 Gyrs (Sect.~\ref{globalsect}.  \textbf{Top Right:} The three values used in our simulations for the number of observed stars per field, overplotted on the cumulative distribution of the number of stars per field across our sample.  \textbf{Bottom Right:} Cumulative distributions of the 80\% completeness limit across our Scylla fields, shown for the F475W filter in cyan and F814W in magenta.  Using vertical lines, we overplot the 80\% completeness limits in the \enquote{deep} (SMC\_23; dotted lines) and \enquote{shallow} (SMC\_35; dashed lines) fields from which we draw our observational noise model based on artificial star tests.  The input values we use in our simulations essentially bracket the range of values across our sample of observed fields.}  
\label{siminputfig}
\end{figure}

In light of the variety of line-of-sight distance distributions seen across the SMC, we simulate (for the first time as far as we are aware) the impact of multiple analytical forms for the distance distribution of individual stars.  These include a single fixed distance as a control sample, and also include flat and Gaussian distance distributions, each with two different widths, based on the above observational evidence.  Based on fits by \citet{sub17} and \citet{elyoussoufi21}, we also simulate double Gaussians where we have varied the relative amplitude of the two components.  We test seven input distance distributions in total, shown using color-coding in the left panel of Fig.~\ref{siminputfig}:

\begin{itemize}
\item A single, fixed distance
\item A flat distance distribution of width 10 kpc
\item A Gaussian distance distribution with a standard deviation $\sigma$=7 kpc
\item A broader flat distance distribution of width 26 kpc
\item A broader Gaussian distance distribution with $\sigma$=14 kpc
\item A distance distribution comprised of two Gaussians (each with $\sigma$=4.5 kpc), where the more nearby Gaussian has a mean distance 13 kpc in front of the SMC and has an amplitude 1.3 times smaller than the more distant Gaussian.
\item A double-Gaussian distribution as above, but with the nearer (i.e., foreground) distribution having an amplitude 2 times (rather than 1.3 times) smaller than the farther distribution.
\end{itemize}

\begin{figure}[h]
\gridline{\fig{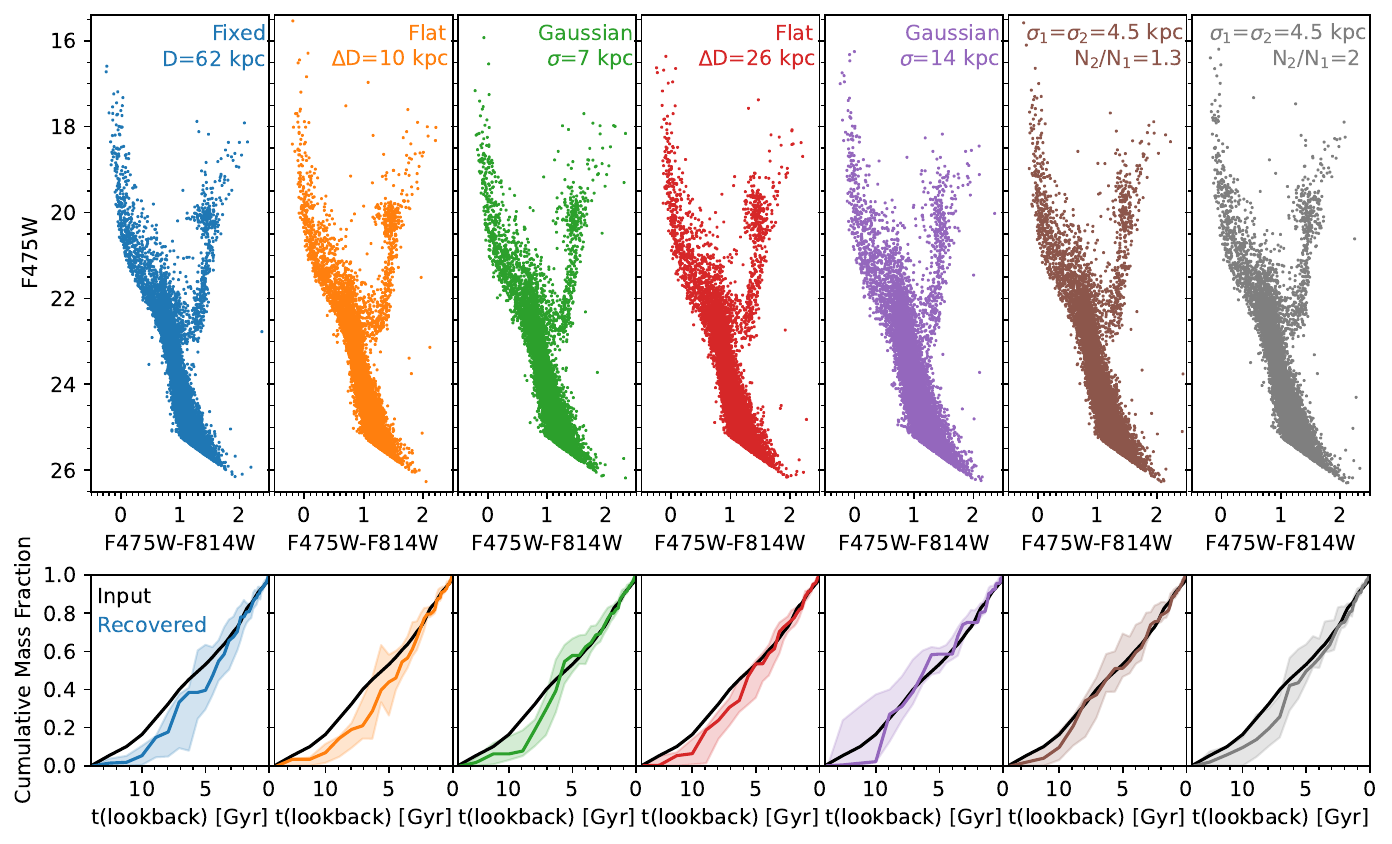}{0.99\textwidth}{}}
\vspace{-0.75cm}
\caption{\textbf{Top row:} Example synthetic CMDs generated assuming the different line-of-sight stellar distance distributions we sample, color-coded and labeled as in the left panel of Fig.~\ref{siminputfig}.  To allow a direct visual comparison, all synthetic CMDs were generated assuming the same PARSEC evolutionary models, input SFH, observational noise model, and number of observed stars.  \textbf{Bottom row:} The recovered star formation history for each CMD (assuming PARSEC evolutionary models also for SFH fitting) is overplotted on the input SFH, which is the same \enquote{Main Body} SFH (middle panel of Fig.~\ref{siminputfig}) in all cases.  While results vary slightly beyond 1-$\sigma$ uncertainties in some cases at early lookback times $\gtrsim$6 Gyr, input values of $\tau_{50}$, $\tau_{75}$ and $\tau_{90}$ (sampling more recent lookback times) are recovered to within 1-$\sigma$ uncertainties for all stellar distance distributions.}   
\label{simcmdfig}
\end{figure}

An example of simulated CMDs corresponding to each assumed distance distribution is shown in Fig.~\ref{simcmdfig}, with color-coding corresponding to the distance distributions in the left panel of Fig.~\ref{siminputfig}.  To isolate the impact of the different distance distributions on the CMD for illustrative purposes, all other input parameters (assumed stellar evolutionary model library, input SFH, observational noise model, number of stars; see below) were held fixed in Fig.~\ref{simcmdfig}.  The morphology of the red clump, at (F475W-F814W)$\sim$1.4 and F475W$\sim$20.1, is most visibly affected by the various distance distributions, although changes in the widths of other stellar evolutionary sequences including the subgiant branch and main sequence are also apparent in some cases.

For each assumed distance distribution, we also varied several additional input parameters, which did not ultimately correlate with the results.  These include:

\noindent \textit{Assumed star formation history:}  We tested two input SFHs, shown in the middle panel of Fig.~\ref{siminputfig}, one based on the results of \citet{massana22} corresponding to the main body of the SMC, and another based on our results for the outer fields of the SMC, with a sharp dropoff in star formation in the most recent few Gyrs (see Fig.~\ref{combined_sfh_fig}).

\noindent \textit{Number of observed stars:}  To test the impact of sample size on our results, we chose three sample sizes (22k, 9k, and 1k stars, with the latter being representative of the fields in the SMC wing).  In the upper right panel of Fig.~\ref{siminputfig}, we overplot these three values on the cumulative distribution of stars per field across our Scylla fields, showing that they reasonably bracket the range of observational sample sizes.

\noindent \textit{Observational Noise Model:} To simulate an observed field, \texttt{MATCH} applies observational noise (photometric error, bias, incompleteness) by drawing from a user-supplied ensemble of artificial star tests.  Field-to-field differences in photometric precision and completeness limits can be caused by differences in total exposure time, but also other effects including stellar crowding that may vary from field to field.  We explored the impact of such differences in photometric precision by supplying artificial stars from each of two fields (\enquote{deep} and \enquote{shallow}) encompassing the range of completeness limits seen across our fields.  This is illustrated in the lower right panel of Fig.~\ref{siminputfig}, where we show cumulative distributions of the 80\% completeness limits across our fields in each filter and overplot the 80\% completeness limits for the \enquote{deep} and \enquote{shallow} fields using dashed (dotted) lines corresponding to the deep (shallow) field.

\noindent \textit{Assumed Stellar Evolutionary Model:} For each simulation run (i.e., given an assumed input SFH, number of stars, and observational noise model), we use PARSEC evolutionary models to generate synthetic CMDs.  We then recover the SFH three times separately, each time assuming a different evolutionary model for SFH fitting (PARSEC, MIST, BaSTI18) identically as for our observed fields (see Sect.~\ref{sfhsect}).

Lastly, we assume fixed input values for both foreground Galactic extinction and differential extinction internal to the SMC when generating synthetic CMDs.  Our assumed foreground extinction A$_{V,fg}$=0.1 mag is in good agreement with dust emission maps in the outer SMC where they remain reliable (\citealt{sf11}, also see Fig.~14 of \citealt{skowron}).  Similarly, we assume additional internal differential extinction $\delta$A$_{V}$=0.1 mag, in good agreement with the values from \citet{skowron}.  Specifically, they quantified internal differential extinction by fitting half-Gaussians to the CMD distribution of the red clump along the reddening vector, measuring standard deviations $\sigma_{1}$ and $\sigma_{2}$ corresponding to extinction in the near and far sides of the SMC.  Averaging their reported $\sigma_{1}$ and $\sigma_{2}$ values across our observed sightlines and converting to $V$-band extinction using the coefficients of \citet{sf11}, we find a mean and standard deviation of 0.12$\pm$0.04 mag.   

\begin{figure}
\gridline{\fig{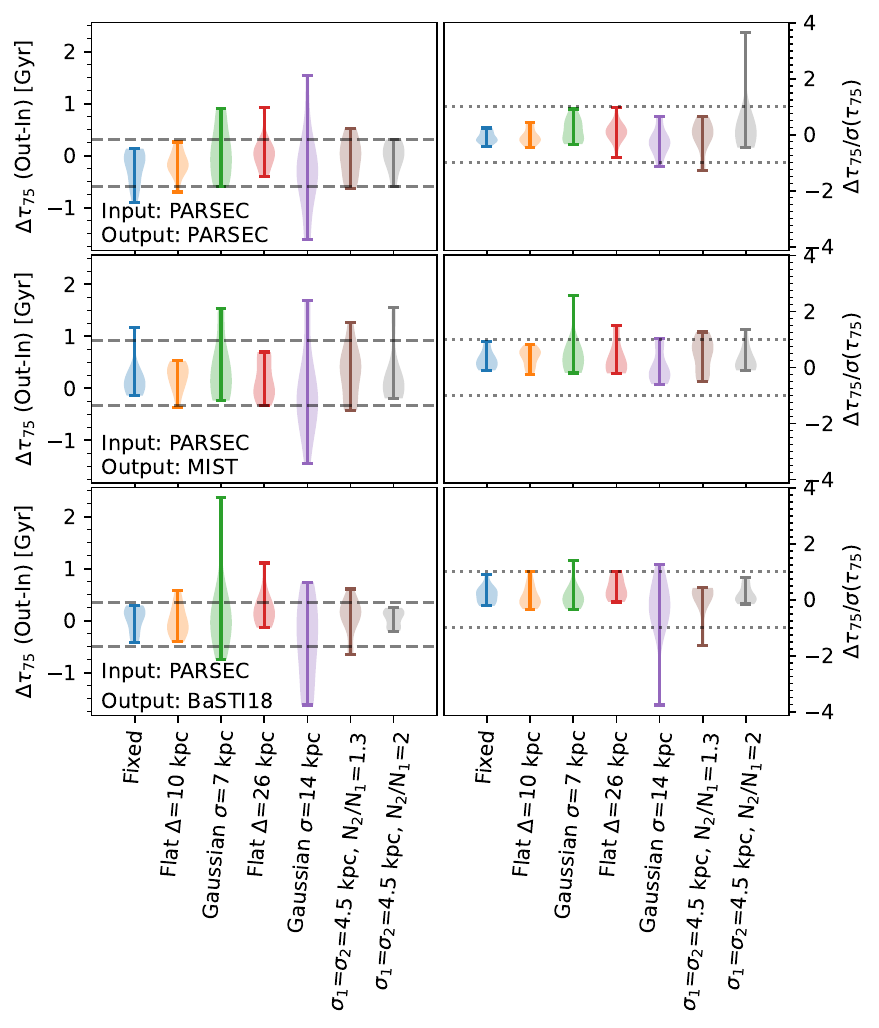}{0.49\textwidth}{}
          \fig{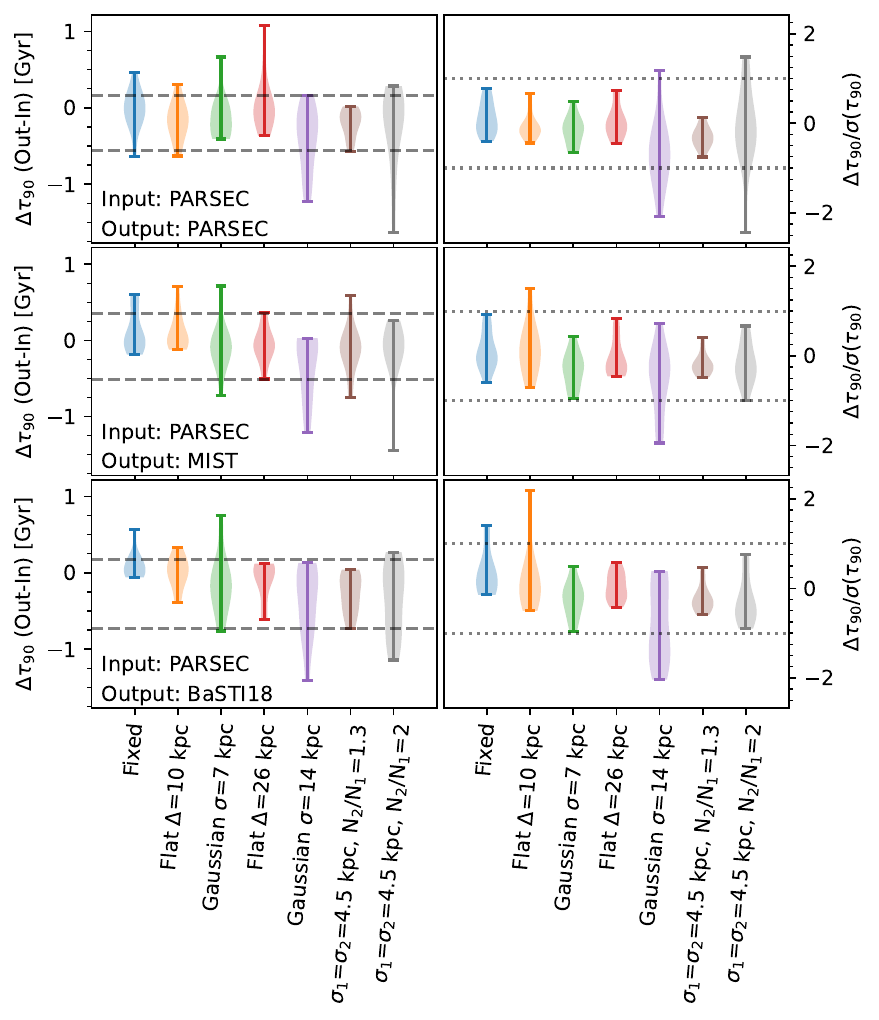}{0.49\textwidth}{}}
\caption{\textbf{Left:} Difference between recovered and input values of $\tau_{75}$ for input stellar populations with different line-of-sight distance distributions, labeled along the horizontal axis and color-coded as in Figs.~\ref{siminputfig}-\ref{simcmdfig}.  Synthetic CMDs were generated using PARSEC stellar evolutionary models and then recovered assuming three different assumed stellar evolutionary libraries (one per row), identically as for our observational sample.  The leftmost column shows the difference between recovered and input values of $\tau_{75}$ in Gyr (with the 10th and 90th percentiles over all simulation runs indicated by dashed grey lines), and the second column from the left shows these differences normalized by their 1-$\sigma$ uncertainties (with $\pm$1-$\sigma$ deviations indicated by dotted grey lines).  \textbf{Right:} Same, but for $\tau_{90}$.}  
\label{simfig}
\end{figure}

The results of our simulations are summarized in Fig.~\ref{simfig}.  The left-hand two panels illustrate our ability to recover $\tau_{75}$, and the right-hand two panels illustrate our ability to recover $\tau_{90}$.  For each of these two CSFH metrics, the left-hand column illustrates the difference in lookback times in Gyr in the sense (recovered-input).  The right-hand column in each panel illustrates the (recovered-input) difference normalized to its 1$\sigma$ uncertainty, illustrating that the CSFH metrics are nearly always recovered to within their uncertainties, exceeding them in $<$5\% of cases for $\tau_{90}$ and $\tau_{75}$ (and, although not shown, $<$10\% of cases for $\tau_{50}$).  The input CMDs were generated assuming PARSEC evolutionary models, and in each row we assume a different set of assumed stellar evolutionary models for SFH recovery, further illustrating that our results are robust to the choice of stellar evolutionary model used for SFH fitting.  Our simulations also demonstrate that the discrepancy between the values of $\tau_{75}$ and $\tau_{90}$ observed in the wing compared to the radial trend seen for the rest of our fields (see Fig.~\ref{csfh_rad_fig}) is real, since differences between recovered and input values of $\tau_{75}$ and $\tau_{90}$ are insufficient to reconcile this discrepancy, even in the presence of a distance spread of as much as 26 kpc or a bimodal distance distribution in the wing \citep[e.g.,][]{nidever13,sub17,tatton21}.  

\section{Results and Comparisons For Individual Fields \label{indfieldsect}}

\begin{figure}
\gridline{\fig{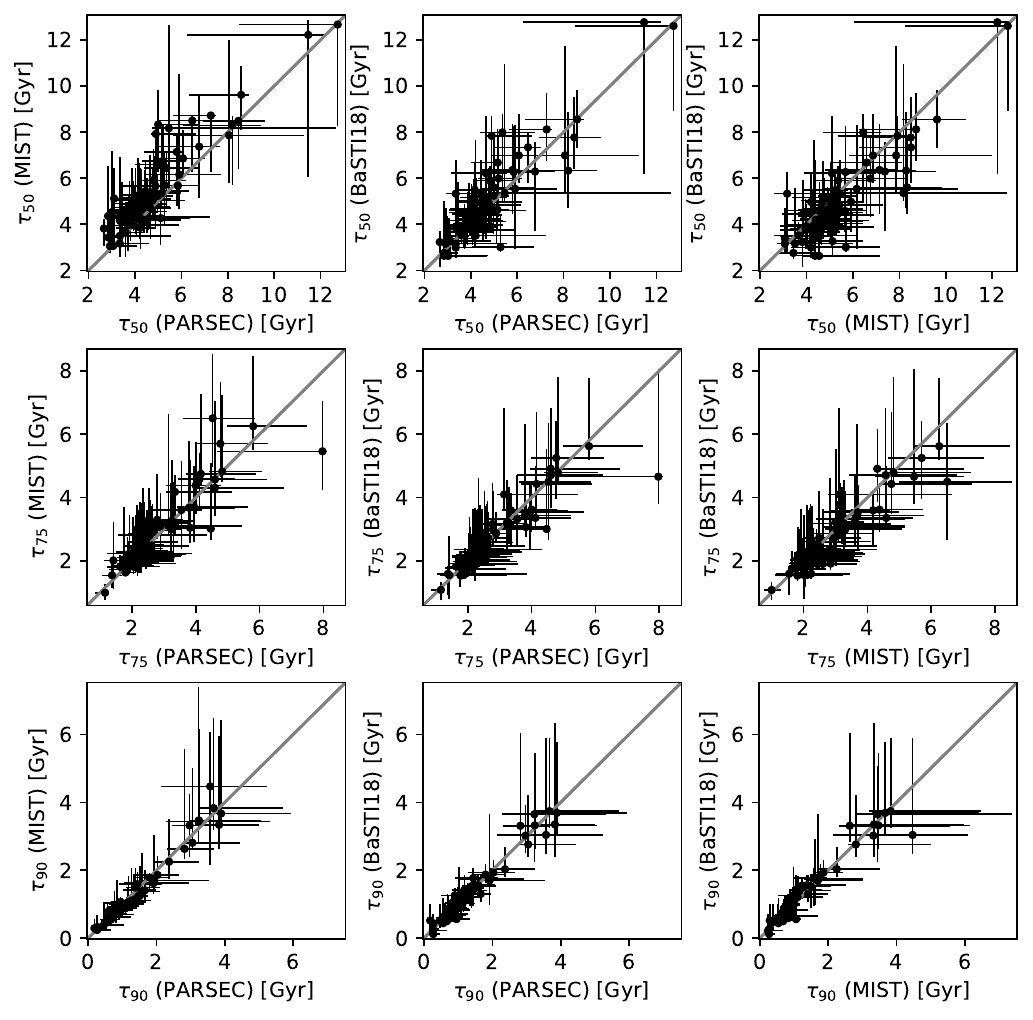}{0.8\textwidth}{}}
\caption{Comparison of our best-fit values for $\tau_{50}$ (top row), $\tau_{75}$ (middle row) and $\tau_{90}$ (bottom row) assuming different stellar evolutionary models for SFH fitting.  Results are in agreement across different evolutionary models in nearly all ($>$97\%) cases.  
\label{comptaufig}}
\end{figure}

\begin{figure}
\gridline{\fig{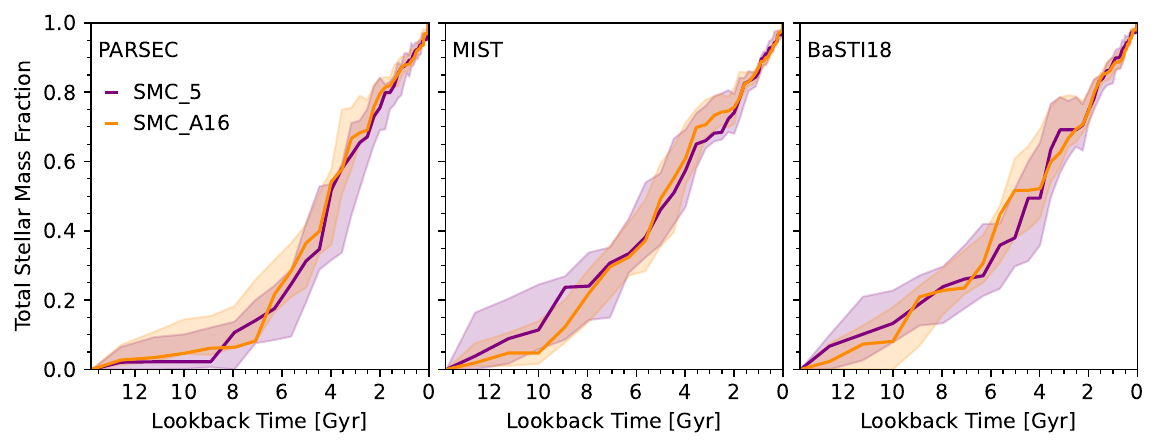}{0.9\textwidth}{}}
\vspace{-0.5cm}
\caption{Comparison between CSFHs for two co-spatial fields observed with different instrument and filter combinations: SMC\_5 was observed with WFC3/UVIS F475, F814W filters, and SMC\_A16 was observed with ACS/WFC F606W, F814W filters.  Each panel presents a comparison assuming a different stellar evolutionary library for SFH fitting.  The best-fit CSFHs are in excellent agreement, showing no evidence of observational systematics.}
\label{comp_overlap_fig}
\end{figure}

\begin{figure}
\gridline{\fig{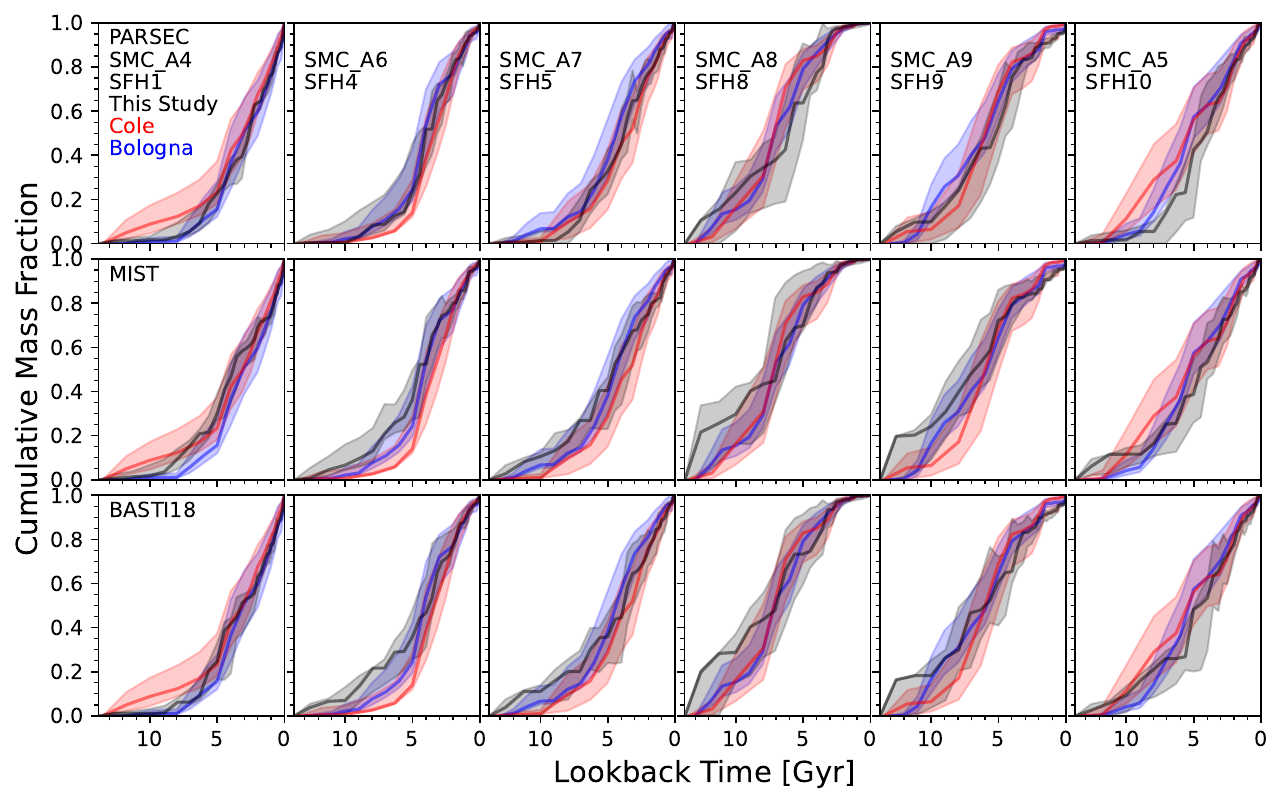}{0.99\textwidth}{}} 
\vspace{-0.5cm}
\caption{Per-field comparison between our CSFHs from \texttt{MATCH} (black) and those of \citet{cignoni12,cignoni13}, who used both the Cole (red) and Bologna (blue) SFH fitting codes and then-recent Padova evolutionary models \citep{padova1,padova2}.  Each row shows results for all six fields assuming a different set of evolutionary models for our SFH fitting.  Each column is labeled with the corresponding field name used here as well as the field name used by \citet{cignoni13}.}
\label{cignonifig}
\end{figure}

In Table \ref{tautab} we list the best-fit values of $\tau_{50}$, $\tau_{75}$ and $\tau_{90}$ and uncertainties for each field assuming three different stellar evolutionary models for SFH fitting.  Using these values, we have performed several per-field checks of internal and external consistency in addition to the sample-wide comparisons with literature results in Sect.~\ref{litsect}:
\begin{itemize}
    \item In Fig.~\ref{comptaufig}, we compare our per-field values of $\tau_{50}$, $\tau_{75}$ and $\tau_{90}$ across the three different sets of evolutionary models we assumed for SFH fitting, finding that they agree to within 1-$\sigma$ in 97\% of cases.  
    \item We exploit serendipitous imaging of two fields with near-complete ($>$90\%) spatial overlap that were observed with different observing setups.  A Scylla field (SMC\_5) was observed with WFC3/UVIS F475W and F814W filters and an archival ACS/WFC field (SMC\_A16) was observed with ACS/WFC F606W and F814W filters.  The two CSFHs, shown in Fig.~\ref{comp_overlap_fig}, are in excellent agreement, arguing strongly against any systematic impact of differences in observing setup across our fields.  The shallower of the two fields (SMC\_5) was excluded from our analysis in order to keep our stellar samples statistically independent.  
    \item Our results include a reanalysis of the six fields studied by \citet{cignoni12,cignoni13}, who used two different SFH fitting codes.  In Fig.~\ref{cignonifig}, we perform a direct comparison between our CSFHs and those obtained by \citet{cignoni12,cignoni13} for each field individually, again showing excellent agreement.  
\end{itemize}

\startlongtable
\begin{deluxetable}{lccccccccc}
\tablecaption{Lookback Times to Form 50\%, 75\% and 90\% of Cumulative Stellar Mass for Three Evolutionary Models\label{tautab}}
\tablehead{
\colhead{Model:} & \multicolumn{3}{c}{PARSEC} & \multicolumn{3}{c}{MIST} & \multicolumn{3}{c}{BaSTI18} \\
\colhead{Field} & \colhead{$\tau_{50}$} & \colhead{$\tau_{75}$} & \colhead{$\tau_{90}$} & \colhead{$\tau_{50}$} & \colhead{$\tau_{75}$} & \colhead{$\tau_{90}$} & \colhead{$\tau_{50}$} & \colhead{$\tau_{75}$} & \colhead{$\tau_{90}$} \\ \colhead{} & \colhead{Gyr} & \colhead{Gyr} & \colhead{Gyr} & \colhead{Gyr} & \colhead{Gyr} & \colhead{Gyr} & \colhead{Gyr} & \colhead{Gyr} & \colhead{Gyr} }
\startdata
SMC\_1 & 4.25$^{+1.25}_{-0.56}$ & 2.38$^{+0.84}_{-0.40}$ & 0.84$^{+0.19}_{-0.26}$  & 4.74$^{+1.67}_{-0.72}$ & 2.25$^{+0.99}_{-0.40}$ & 0.79$^{+0.13}_{-0.13}$  & 4.12$^{+0.67}_{-0.51}$ & 2.12$^{+1.43}_{-0.27}$ & 0.86$^{+0.36}_{-0.26}$ \\
SMC\_2 & 5.46$^{+7.18}_{-0.72}$ & 3.24$^{+0.31}_{-0.14}$ & 1.19$^{+0.15}_{-0.25}$  & 8.17$^{+4.48}_{-2.79}$ & 3.00$^{+0.89}_{-0.02}$ & 1.04$^{+0.21}_{-0.08}$  & 5.36$^{+5.59}_{-0.33}$ & 3.24$^{+0.56}_{-0.39}$ & 1.18$^{+0.19}_{-0.30}$ \\
SMC\_7 & 4.44$^{+1.87}_{-0.11}$ & 3.54$^{+0.65}_{-0.95}$ & 2.03$^{+0.62}_{-0.72}$  & 5.44$^{+1.32}_{-1.35}$ & 3.61$^{+1.56}_{-1.22}$ & 1.86$^{+0.56}_{-0.27}$  & 5.33$^{+1.45}_{-0.71}$ & 3.32$^{+1.40}_{-0.86}$ & 1.92$^{+0.36}_{-0.21}$ \\
SMC\_8 & 3.86$^{+1.07}_{-0.62}$ & 2.06$^{+0.51}_{-0.55}$ & 0.27$^{+0.20}_{-0.09}$  & 5.04$^{+1.37}_{-0.85}$ & 2.28$^{+0.82}_{-0.55}$ & 0.32$^{+0.05}_{-0.11}$  & 4.02$^{+1.39}_{-0.36}$ & 2.15$^{+1.24}_{-0.38}$ & 0.44$^{+0.14}_{-0.14}$ \\
SMC\_9 & 3.93$^{+1.31}_{-0.63}$ & 2.21$^{+0.46}_{-0.48}$ & 0.99$^{+0.27}_{-0.17}$  & 4.76$^{+1.11}_{-1.03}$ & 2.18$^{+1.42}_{-0.42}$ & 0.96$^{+0.07}_{-0.06}$  & 4.22$^{+1.59}_{-0.50}$ & 2.29$^{+1.38}_{-0.43}$ & 1.09$^{+0.41}_{-0.23}$ \\
SMC\_10 & 4.24$^{+0.82}_{-0.67}$ & 2.16$^{+1.23}_{-0.62}$ & 0.17$^{+0.36}_{-0.00}$  & 4.79$^{+1.40}_{-0.32}$ & 1.92$^{+1.83}_{-0.20}$ & 0.29$^{+0.26}_{-0.07}$  & 4.43$^{+0.88}_{-0.57}$ & 1.98$^{+1.66}_{-0.38}$ & 0.51$^{+0.49}_{-0.24}$ \\
SMC\_11 & 4.76$^{+1.39}_{-0.61}$ & 2.34$^{+1.99}_{-0.77}$ & 0.80$^{+0.32}_{-0.29}$  & 5.03$^{+1.42}_{-0.88}$ & 2.81$^{+1.47}_{-0.88}$ & 0.71$^{+0.26}_{-0.26}$  & 4.83$^{+1.58}_{-0.62}$ & 2.35$^{+2.24}_{-0.35}$ & 0.69$^{+0.48}_{-0.18}$ \\
SMC\_13 & 3.96$^{+0.57}_{-0.83}$ & 2.25$^{+1.03}_{-0.22}$ & 0.82$^{+0.49}_{-0.29}$  & 3.94$^{+1.68}_{-0.05}$ & 2.48$^{+1.39}_{-0.19}$ & 0.82$^{+0.12}_{-0.08}$  & 3.83$^{+1.26}_{-0.30}$ & 2.26$^{+1.03}_{-0.39}$ & 0.86$^{+0.30}_{-0.24}$ \\
SMC\_14 & 4.14$^{+1.18}_{-0.67}$ & 2.53$^{+0.74}_{-0.25}$ & 1.03$^{+0.33}_{-0.17}$  & 4.89$^{+2.01}_{-1.06}$ & 2.25$^{+1.83}_{-0.22}$ & 0.98$^{+0.10}_{-0.11}$  & 4.40$^{+1.22}_{-0.62}$ & 2.61$^{+1.13}_{-0.40}$ & 1.16$^{+0.93}_{-0.39}$ \\
SMC\_15 & 4.22$^{+0.64}_{-0.65}$ & 2.52$^{+1.25}_{-0.42}$ & 1.26$^{+0.31}_{-0.14}$  & 4.75$^{+0.32}_{-1.04}$ & 2.48$^{+1.82}_{-0.39}$ & 1.11$^{+0.96}_{-0.06}$  & 4.53$^{+1.46}_{-0.43}$ & 2.47$^{+1.74}_{-0.18}$ & 1.40$^{+0.23}_{-0.08}$ \\
SMC\_16 & 3.43$^{+1.14}_{-0.50}$ & 2.11$^{+0.70}_{-0.49}$ & 0.70$^{+0.14}_{-0.08}$  & 4.26$^{+1.56}_{-0.67}$ & 1.94$^{+1.50}_{-0.09}$ & 0.75$^{+0.16}_{-0.05}$  & 3.89$^{+1.03}_{-0.69}$ & 2.18$^{+0.53}_{-0.34}$ & 0.81$^{+0.09}_{-0.20}$ \\
SMC\_17 & 4.16$^{+0.85}_{-0.55}$ & 2.12$^{+0.51}_{-0.52}$ & 0.89$^{+0.37}_{-0.08}$  & 4.22$^{+0.68}_{-0.41}$ & 1.90$^{+0.99}_{-0.27}$ & 0.91$^{+0.04}_{-0.20}$  & 3.84$^{+1.12}_{-0.72}$ & 2.09$^{+0.45}_{-0.45}$ & 0.88$^{+0.38}_{-0.23}$ \\
SMC\_18 & 3.71$^{+1.06}_{-0.37}$ & 2.10$^{+1.16}_{-0.22}$ & 0.65$^{+0.58}_{-0.17}$  & 3.97$^{+1.33}_{-1.13}$ & 2.30$^{+0.44}_{-0.61}$ & 0.61$^{+0.28}_{-0.20}$  & 3.92$^{+0.95}_{-1.00}$ & 2.22$^{+0.63}_{-0.45}$ & 0.67$^{+0.28}_{-0.18}$ \\
SMC\_19 & 4.24$^{+0.34}_{-1.16}$ & 2.27$^{+1.14}_{-0.26}$ & 0.94$^{+0.33}_{-0.38}$  & 4.61$^{+1.82}_{-0.96}$ & 2.59$^{+0.93}_{-0.60}$ & 0.89$^{+0.07}_{-0.24}$  & 4.50$^{+0.73}_{-0.99}$ & 2.14$^{+1.46}_{-0.27}$ & 1.02$^{+0.24}_{-0.38}$ \\
SMC\_20 & 4.19$^{+1.19}_{-1.14}$ & 2.04$^{+0.71}_{-0.31}$ & 0.92$^{+0.30}_{-0.35}$  & 5.06$^{+1.61}_{-1.68}$ & 1.92$^{+1.16}_{-0.04}$ & 0.90$^{+0.05}_{-0.24}$  & 4.16$^{+2.53}_{-0.85}$ & 2.01$^{+0.99}_{-0.16}$ & 0.81$^{+0.45}_{-0.27}$ \\
SMC\_21 & 4.23$^{+0.64}_{-1.03}$ & 2.09$^{+0.80}_{-0.59}$ & 0.71$^{+0.33}_{-0.14}$  & 3.85$^{+1.27}_{-0.90}$ & 1.82$^{+1.03}_{-0.12}$ & 0.74$^{+0.25}_{-0.24}$  & 4.44$^{+1.46}_{-0.91}$ & 2.12$^{+0.96}_{-0.41}$ & 0.69$^{+0.57}_{-0.15}$ \\
SMC\_22 & 8.20$^{+1.22}_{-2.10}$ & 3.95$^{+1.69}_{-0.53}$ & 1.14$^{+0.32}_{-0.17}$  & 8.29$^{+0.42}_{-2.56}$ & 3.69$^{+1.28}_{-1.08}$ & 1.03$^{+0.44}_{-0.19}$  & 6.33$^{+2.56}_{-1.60}$ & 3.53$^{+0.85}_{-1.18}$ & 1.29$^{+0.25}_{-0.53}$ \\
SMC\_23 & 4.38$^{+0.26}_{-0.94}$ & 2.17$^{+0.17}_{-0.57}$ & 0.57$^{+0.11}_{-0.18}$  & 5.14$^{+0.33}_{-1.52}$ & 2.06$^{+0.75}_{-0.53}$ & 0.50$^{+0.45}_{-0.16}$  & 4.35$^{+0.90}_{-0.60}$ & 1.91$^{+0.26}_{-0.24}$ & 0.54$^{+0.45}_{-0.05}$ \\
SMC\_25 & 3.00$^{+0.92}_{-0.16}$ & 1.62$^{+0.35}_{-0.20}$ & 0.61$^{+0.25}_{-0.01}$  & 4.47$^{+0.62}_{-1.60}$ & 1.82$^{+0.37}_{-0.16}$ & 0.67$^{+0.14}_{-0.05}$  & 3.47$^{+0.89}_{-0.20}$ & 1.82$^{+0.35}_{-0.23}$ & 0.66$^{+0.22}_{-0.06}$ \\
SMC\_27 & 3.35$^{+2.19}_{-0.33}$ & 2.05$^{+0.79}_{-0.22}$ & 0.64$^{+0.19}_{-0.03}$  & 3.50$^{+2.50}_{-0.39}$ & 1.97$^{+0.86}_{-0.06}$ & 0.67$^{+0.23}_{-0.03}$  & 3.16$^{+2.40}_{-0.17}$ & 2.03$^{+0.40}_{-0.19}$ & 0.60$^{+0.16}_{-0.02}$ \\
SMC\_28 & 8.06$^{+3.21}_{-2.40}$ & 4.12$^{+1.12}_{-0.66}$ & 1.66$^{+0.30}_{-0.64}$  & 7.86$^{+4.15}_{-2.04}$ & 4.58$^{+0.83}_{-1.00}$ & 1.41$^{+0.59}_{-0.30}$  & 6.99$^{+4.74}_{-1.62}$ & 3.36$^{+1.63}_{-0.29}$ & 1.30$^{+0.56}_{-0.23}$ \\
SMC\_29 & 3.35$^{+1.82}_{-0.33}$ & 2.06$^{+0.28}_{-0.13}$ & 0.68$^{+0.31}_{-0.07}$  & 4.37$^{+2.54}_{-0.66}$ & 1.83$^{+0.79}_{-0.08}$ & 0.82$^{+0.23}_{-0.24}$  & 3.87$^{+2.93}_{-0.51}$ & 1.89$^{+0.55}_{-0.05}$ & 0.59$^{+0.71}_{-0.04}$ \\
SMC\_30 & 3.37$^{+0.86}_{-0.17}$ & 2.04$^{+0.46}_{-0.32}$ & 0.98$^{+0.24}_{-0.26}$  & 4.13$^{+1.50}_{-0.35}$ & 1.82$^{+1.55}_{-0.10}$ & 0.99$^{+0.17}_{-0.06}$  & 3.81$^{+0.71}_{-0.70}$ & 2.04$^{+0.82}_{-0.16}$ & 1.00$^{+0.44}_{-0.27}$ \\
SMC\_33 & 3.60$^{+0.75}_{-0.25}$ & 2.38$^{+0.41}_{-0.16}$ & 0.90$^{+0.27}_{-0.12}$  & 3.64$^{+0.69}_{-1.04}$ & 2.04$^{+0.48}_{-0.21}$ & 0.86$^{+0.21}_{-0.23}$  & 3.51$^{+0.60}_{-0.34}$ & 2.16$^{+0.24}_{-0.16}$ & 0.82$^{+0.28}_{-0.14}$ \\
SMC\_34 & 3.34$^{+1.26}_{-0.41}$ & 2.13$^{+0.47}_{-0.50}$ & 0.64$^{+0.16}_{-0.03}$  & 3.15$^{+1.42}_{-0.55}$ & 1.97$^{+0.29}_{-0.17}$ & 0.59$^{+0.08}_{-0.04}$  & 5.33$^{+0.92}_{-2.50}$ & 2.12$^{+0.19}_{-0.27}$ & 0.58$^{+0.15}_{-0.03}$ \\
SMC\_35 & 4.02$^{+0.83}_{-0.91}$ & 2.62$^{+0.61}_{-0.44}$ & 1.44$^{+0.72}_{-0.52}$  & 4.72$^{+1.71}_{-0.88}$ & 2.27$^{+1.44}_{-0.09}$ & 1.59$^{+0.36}_{-0.61}$  & 4.65$^{+1.11}_{-1.11}$ & 2.49$^{+1.06}_{-0.28}$ & 1.77$^{+0.30}_{-0.83}$ \\
SMC\_36 & 3.44$^{+1.95}_{-0.43}$ & 2.02$^{+0.38}_{-0.28}$ & 0.66$^{+0.25}_{-0.09}$  & 4.55$^{+1.01}_{-1.50}$ & 2.00$^{+0.34}_{-0.27}$ & 0.74$^{+0.13}_{-0.22}$  & 3.96$^{+1.86}_{-0.66}$ & 2.11$^{+0.50}_{-0.27}$ & 0.77$^{+0.20}_{-0.22}$ \\
SMC\_37 & 3.70$^{+2.18}_{-0.28}$ & 2.14$^{+1.16}_{-0.37}$ & 0.61$^{+0.36}_{-0.01}$  & 3.96$^{+0.37}_{-1.11}$ & 2.01$^{+0.43}_{-0.29}$ & 0.64$^{+0.10}_{-0.09}$  & 3.75$^{+0.58}_{-0.75}$ & 2.27$^{+0.41}_{-0.50}$ & 0.63$^{+0.06}_{-0.13}$ \\
SMC\_38 & 6.77$^{+0.88}_{-1.89}$ & 3.23$^{+0.53}_{-0.80}$ & 0.77$^{+0.56}_{-0.15}$  & 7.37$^{+2.22}_{-2.21}$ & 3.26$^{+1.20}_{-1.04}$ & 0.70$^{+0.70}_{-0.05}$  & 6.28$^{+1.29}_{-2.57}$ & 3.18$^{+1.35}_{-0.85}$ & 0.82$^{+0.31}_{-0.26}$ \\
SMC\_40 & 2.91$^{+0.87}_{-0.13}$ & 1.80$^{+0.32}_{-0.18}$ & 0.60$^{+0.16}_{-0.00}$  & 3.06$^{+1.57}_{-0.32}$ & 1.64$^{+0.51}_{-0.11}$ & 0.66$^{+0.10}_{-0.06}$  & 3.16$^{+1.55}_{-0.39}$ & 1.76$^{+0.54}_{-0.07}$ & 0.60$^{+0.40}_{-0.03}$ \\
SMC\_41 & 3.76$^{+0.90}_{-0.32}$ & 1.38$^{+0.32}_{-0.32}$ & 0.63$^{+0.19}_{-0.03}$  & 4.01$^{+1.85}_{-0.23}$ & 1.54$^{+0.37}_{-0.39}$ & 0.68$^{+0.13}_{-0.04}$  & 3.79$^{+0.43}_{-0.83}$ & 1.59$^{+0.22}_{-0.67}$ & 0.59$^{+0.09}_{-0.02}$ \\
SMC\_42 & 3.06$^{+1.01}_{-0.05}$ & 1.91$^{+0.76}_{-0.20}$ & 0.82$^{+0.16}_{-0.12}$  & 3.06$^{+1.58}_{-0.10}$ & 1.83$^{+0.33}_{-0.25}$ & 0.76$^{+0.14}_{-0.09}$  & 3.38$^{+0.88}_{-0.11}$ & 1.94$^{+0.47}_{-0.14}$ & 0.77$^{+0.49}_{-0.09}$ \\
SMC\_43 & 4.97$^{+0.66}_{-0.78}$ & 2.48$^{+0.47}_{-0.37}$ & 0.79$^{+0.20}_{-0.17}$  & 5.23$^{+0.89}_{-0.71}$ & 2.40$^{+0.89}_{-0.42}$ & 0.77$^{+0.10}_{-0.17}$  & 5.24$^{+0.89}_{-1.63}$ & 2.37$^{+0.65}_{-0.24}$ & 0.89$^{+0.20}_{-0.28}$ \\
SMC\_44 & 4.66$^{+0.90}_{-1.12}$ & 2.60$^{+0.53}_{-0.42}$ & 0.70$^{+0.24}_{-0.04}$  & 4.71$^{+1.22}_{-1.33}$ & 2.31$^{+0.86}_{-0.42}$ & 0.64$^{+0.31}_{-0.05}$  & 5.17$^{+1.58}_{-1.81}$ & 2.40$^{+0.63}_{-0.34}$ & 0.64$^{+0.24}_{-0.04}$ \\
SMC\_45 & 3.10$^{+1.79}_{-0.13}$ & 1.78$^{+0.17}_{-0.55}$ & 0.67$^{+0.11}_{-0.01}$  & 5.11$^{+1.33}_{-2.06}$ & 1.81$^{+0.35}_{-0.01}$ & 0.69$^{+0.07}_{-0.07}$  & 3.26$^{+0.67}_{-0.58}$ & 1.54$^{+0.44}_{-0.18}$ & 0.52$^{+0.11}_{-0.01}$ \\
SMC\_46 & 3.02$^{+1.69}_{-0.09}$ & 1.86$^{+0.66}_{-0.32}$ & 0.56$^{+0.08}_{-0.13}$  & 4.54$^{+2.63}_{-0.70}$ & 2.10$^{+0.64}_{-0.28}$ & 0.53$^{+0.11}_{-0.10}$  & 2.62$^{+1.22}_{-0.16}$ & 1.56$^{+0.52}_{-0.08}$ & 0.44$^{+0.08}_{-0.06}$ \\
SMC\_47 & 5.27$^{+1.44}_{-1.20}$ & 1.76$^{+1.54}_{-0.10}$ & 0.46$^{+0.59}_{-0.06}$  & 5.68$^{+1.47}_{-1.75}$ & 2.05$^{+1.65}_{-0.47}$ & 0.39$^{+0.66}_{-0.03}$  & 3.01$^{+1.38}_{-0.20}$ & 2.02$^{+0.24}_{-0.84}$ & 0.52$^{+0.48}_{-0.23}$ \\
SMC\_48 & 5.91$^{+1.82}_{-1.47}$ & 1.92$^{+1.93}_{-0.70}$ & 0.27$^{+0.20}_{-0.05}$  & 6.15$^{+4.36}_{-1.48}$ & 2.24$^{+1.21}_{-0.49}$ & 0.27$^{+0.17}_{-0.06}$  & 5.53$^{+2.77}_{-2.62}$ & 1.60$^{+0.81}_{-0.16}$ & 0.12$^{+0.25}_{-0.00}$ \\
SMC\_49 & 3.34$^{+1.51}_{-0.16}$ & 2.04$^{+1.16}_{-0.17}$ & 0.95$^{+0.40}_{-0.28}$  & 4.18$^{+1.94}_{-0.85}$ & 2.01$^{+0.67}_{-0.01}$ & 1.05$^{+0.13}_{-0.18}$  & 2.99$^{+0.96}_{-0.44}$ & 1.74$^{+0.68}_{-0.06}$ & 0.55$^{+0.79}_{-0.03}$ \\
SMC\_52 & 5.11$^{+1.27}_{-1.80}$ & 1.41$^{+1.80}_{-0.29}$ & 0.26$^{+0.42}_{-0.03}$  & 4.27$^{+2.30}_{-1.17}$ & 2.02$^{+1.21}_{-0.90}$ & 0.24$^{+0.42}_{-0.02}$  & 4.59$^{+0.93}_{-1.46}$ & 1.55$^{+1.23}_{-0.75}$ & 0.21$^{+0.48}_{-0.01}$ \\
SMC\_53 & 4.21$^{+3.01}_{-1.07}$ & 2.15$^{+1.40}_{-0.06}$ & 0.66$^{+0.34}_{-0.02}$  & 4.31$^{+1.84}_{-0.41}$ & 2.16$^{+1.50}_{-0.07}$ & 0.67$^{+0.18}_{-0.06}$  & 3.50$^{+4.14}_{-0.29}$ & 1.91$^{+0.71}_{-0.06}$ & 0.49$^{+0.40}_{-0.00}$ \\
SMC\_54 & 2.85$^{+0.57}_{-0.24}$ & 1.85$^{+0.33}_{-0.57}$ & 0.70$^{+0.09}_{-0.07}$  & 3.43$^{+0.93}_{-0.64}$ & 1.73$^{+0.52}_{-0.10}$ & 0.71$^{+0.21}_{-0.08}$  & 2.76$^{+0.68}_{-0.27}$ & 1.77$^{+0.42}_{-0.29}$ & 0.54$^{+0.19}_{-0.08}$ \\
SMC\_55 & 2.84$^{+1.91}_{-0.03}$ & 2.03$^{+0.82}_{-0.25}$ & 0.84$^{+0.30}_{-0.11}$  & 4.36$^{+2.17}_{-1.00}$ & 2.17$^{+0.61}_{-0.23}$ & 0.93$^{+0.77}_{-0.11}$  & 2.64$^{+2.37}_{-0.17}$ & 1.86$^{+0.64}_{-0.08}$ & 0.65$^{+0.31}_{-0.15}$ \\
SMC\_W1 & 4.18$^{+0.75}_{-0.76}$ & 2.04$^{+0.51}_{-0.64}$ & 0.83$^{+0.45}_{-0.15}$  & 5.09$^{+0.35}_{-1.33}$ & 1.86$^{+1.30}_{-0.25}$ & 0.90$^{+0.14}_{-0.17}$  & 3.66$^{+0.89}_{-0.64}$ & 1.93$^{+1.08}_{-0.22}$ & 0.70$^{+0.48}_{-0.09}$ \\
SMC\_W2 & 4.99$^{+1.71}_{-0.09}$ & 4.52$^{+1.31}_{-0.93}$ & 3.23$^{+2.09}_{-0.94}$  & 8.33$^{+1.51}_{-2.08}$ & 6.49$^{+2.03}_{-1.52}$ & 3.44$^{+3.94}_{-0.60}$  & 5.59$^{+2.10}_{-1.14}$ & 4.49$^{+1.84}_{-1.84}$ & 3.65$^{+1.41}_{-1.40}$ \\
SMC\_W3 & 6.47$^{+0.55}_{-1.49}$ & 4.76$^{+1.49}_{-0.80}$ & 3.56$^{+1.67}_{-1.43}$  & 8.49$^{+0.25}_{-2.24}$ & 5.70$^{+1.95}_{-1.05}$ & 4.46$^{+1.61}_{-2.32}$  & 7.33$^{+0.63}_{-1.56}$ & 5.24$^{+1.16}_{-1.29}$ & 3.03$^{+2.86}_{-0.56}$ \\
SMC\_W4 & 8.46$^{+1.14}_{-1.51}$ & 7.96$^{+0.15}_{-3.29}$ & 3.82$^{+1.18}_{-0.69}$  & 8.49$^{+0.17}_{-2.07}$ & 5.45$^{+1.57}_{-1.22}$ & 3.34$^{+2.61}_{-0.72}$  & 7.76$^{+1.76}_{-1.34}$ & 4.66$^{+3.39}_{-0.86}$ & 3.35$^{+3.00}_{-0.96}$ \\
SMC\_W5 & 4.89$^{+1.71}_{-0.20}$ & 3.14$^{+2.30}_{-0.22}$ & 2.81$^{+0.86}_{-0.51}$  & 7.92$^{+0.29}_{-2.58}$ & 3.10$^{+3.53}_{-0.32}$ & 2.62$^{+2.92}_{-0.30}$  & 7.84$^{+0.88}_{-1.51}$ & 4.09$^{+2.72}_{-0.44}$ & 3.30$^{+2.72}_{-0.49}$ \\
SMC\_W6 & 4.18$^{+0.60}_{-0.78}$ & 2.32$^{+0.36}_{-0.40}$ & 0.83$^{+0.13}_{-0.12}$  & 5.04$^{+0.66}_{-1.09}$ & 2.17$^{+1.40}_{-0.31}$ & 0.81$^{+0.16}_{-0.05}$  & 3.95$^{+2.10}_{-0.21}$ & 2.08$^{+1.43}_{-0.26}$ & 0.80$^{+0.13}_{-0.07}$ \\
SMC\_W7 & 3.64$^{+0.25}_{-0.57}$ & 2.16$^{+0.27}_{-0.27}$ & 0.62$^{+0.28}_{-0.03}$  & 4.35$^{+0.44}_{-0.64}$ & 2.85$^{+0.57}_{-1.07}$ & 0.69$^{+0.21}_{-0.13}$  & 3.65$^{+1.05}_{-0.26}$ & 1.91$^{+0.80}_{-0.18}$ & 0.65$^{+0.23}_{-0.11}$ \\
SMC\_A1 & 4.52$^{+0.37}_{-0.76}$ & 2.79$^{+0.80}_{-0.16}$ & 1.48$^{+0.27}_{-0.34}$  & 5.13$^{+0.86}_{-0.90}$ & 3.29$^{+0.95}_{-0.60}$ & 1.23$^{+0.24}_{-0.18}$  & 4.44$^{+0.98}_{-0.37}$ & 2.98$^{+1.20}_{-0.50}$ & 1.48$^{+0.33}_{-0.14}$ \\
SMC\_A2 & 4.12$^{+0.65}_{-0.81}$ & 2.25$^{+0.16}_{-0.45}$ & 0.82$^{+0.28}_{-0.26}$  & 4.39$^{+0.79}_{-0.47}$ & 2.48$^{+0.25}_{-0.65}$ & 0.81$^{+0.07}_{-0.27}$  & 4.20$^{+1.00}_{-0.71}$ & 2.03$^{+0.67}_{-0.25}$ & 0.81$^{+0.25}_{-0.31}$ \\
SMC\_A3 & 3.86$^{+0.50}_{-0.42}$ & 2.55$^{+0.82}_{-0.37}$ & 0.89$^{+0.31}_{-0.03}$  & 4.64$^{+0.81}_{-0.69}$ & 3.02$^{+0.89}_{-0.78}$ & 0.92$^{+0.23}_{-0.10}$  & 3.96$^{+0.64}_{-0.88}$ & 2.22$^{+1.16}_{-0.37}$ & 0.97$^{+0.13}_{-0.30}$ \\
SMC\_A4 & 2.66$^{+0.32}_{-0.11}$ & 1.15$^{+0.10}_{-0.29}$ & 0.26$^{+0.19}_{-0.01}$  & 3.82$^{+0.39}_{-0.76}$ & 1.00$^{+0.27}_{-0.23}$ & 0.26$^{+0.15}_{-0.01}$  & 3.22$^{+0.49}_{-1.06}$ & 1.09$^{+0.24}_{-0.32}$ & 0.27$^{+0.14}_{-0.02}$ \\
SMC\_A5 & 4.07$^{+0.86}_{-0.53}$ & 2.45$^{+1.24}_{-0.50}$ & 1.08$^{+0.32}_{-0.30}$  & 4.51$^{+0.73}_{-1.07}$ & 2.22$^{+1.17}_{-0.47}$ & 0.87$^{+0.11}_{-0.15}$  & 4.59$^{+0.36}_{-1.50}$ & 2.28$^{+1.36}_{-0.45}$ & 1.04$^{+0.38}_{-0.40}$ \\
SMC\_A6 & 4.02$^{+0.54}_{-0.87}$ & 2.31$^{+0.27}_{-0.35}$ & 1.13$^{+0.35}_{-0.21}$  & 4.54$^{+0.42}_{-0.93}$ & 2.61$^{+1.21}_{-0.75}$ & 0.94$^{+0.20}_{-0.09}$  & 3.77$^{+1.12}_{-0.41}$ & 2.13$^{+1.50}_{-0.20}$ & 1.01$^{+0.39}_{-0.19}$ \\
SMC\_A7 & 3.79$^{+0.56}_{-0.98}$ & 2.22$^{+0.39}_{-0.55}$ & 0.82$^{+0.39}_{-0.14}$  & 4.58$^{+0.28}_{-1.31}$ & 2.01$^{+1.38}_{-0.39}$ & 0.84$^{+0.10}_{-0.22}$  & 3.80$^{+0.85}_{-0.68}$ & 2.02$^{+1.42}_{-0.65}$ & 0.72$^{+0.57}_{-0.15}$ \\
SMC\_A8 & 6.06$^{+0.93}_{-1.14}$ & 4.16$^{+1.73}_{-0.48}$ & 3.24$^{+1.81}_{-0.27}$  & 6.85$^{+1.22}_{-0.66}$ & 4.74$^{+2.53}_{-0.38}$ & 3.46$^{+2.69}_{-0.22}$  & 6.98$^{+1.77}_{-1.19}$ & 4.43$^{+2.25}_{-0.94}$ & 3.31$^{+2.13}_{-0.68}$ \\
SMC\_A9 & 5.19$^{+0.99}_{-0.90}$ & 4.00$^{+0.78}_{-0.92}$ & 1.90$^{+1.02}_{-0.85}$  & 6.74$^{+1.58}_{-1.41}$ & 4.37$^{+1.36}_{-0.89}$ & 1.61$^{+1.41}_{-0.31}$  & 5.99$^{+1.84}_{-1.67}$ & 3.61$^{+0.81}_{-0.71}$ & 1.75$^{+1.15}_{-0.78}$ \\
SMC\_A10 & 4.28$^{+0.30}_{-1.23}$ & 2.15$^{+0.10}_{-0.30}$ & 0.66$^{+0.16}_{-0.23}$  & 4.91$^{+0.42}_{-0.92}$ & 2.60$^{+0.70}_{-0.92}$ & 0.78$^{+0.12}_{-0.18}$  & 4.60$^{+0.54}_{-1.06}$ & 1.91$^{+0.24}_{-0.15}$ & 0.73$^{+0.13}_{-0.13}$ \\
SMC\_A11 & 4.60$^{+0.21}_{-0.89}$ & 2.28$^{+0.09}_{-0.42}$ & 0.56$^{+0.21}_{-0.08}$  & 5.14$^{+0.21}_{-0.95}$ & 2.41$^{+0.14}_{-0.79}$ & 0.54$^{+0.45}_{-0.06}$  & 4.53$^{+0.22}_{-0.93}$ & 1.91$^{+0.25}_{-0.19}$ & 0.59$^{+0.58}_{-0.09}$ \\
SMC\_A12 & 4.42$^{+1.56}_{-0.51}$ & 2.57$^{+1.04}_{-0.47}$ & 0.59$^{+0.35}_{-0.00}$  & 4.18$^{+2.16}_{-0.26}$ & 2.83$^{+0.77}_{-0.79}$ & 0.79$^{+0.05}_{-0.22}$  & 4.96$^{+1.26}_{-1.06}$ & 2.32$^{+1.24}_{-0.23}$ & 0.64$^{+0.24}_{-0.08}$ \\
SMC\_A13 & 4.52$^{+0.30}_{-1.17}$ & 2.64$^{+0.70}_{-0.52}$ & 1.14$^{+0.49}_{-0.24}$  & 5.03$^{+0.46}_{-1.40}$ & 2.96$^{+0.84}_{-0.72}$ & 1.01$^{+0.19}_{-0.08}$  & 4.64$^{+0.37}_{-1.40}$ & 2.72$^{+0.62}_{-0.51}$ & 1.28$^{+0.26}_{-0.35}$ \\
SMC\_A14 & 4.54$^{+0.12}_{-1.09}$ & 2.33$^{+0.38}_{-0.34}$ & 1.03$^{+0.26}_{-0.31}$  & 5.04$^{+0.20}_{-1.28}$ & 2.46$^{+0.61}_{-0.65}$ & 0.88$^{+0.13}_{-0.09}$  & 5.02$^{+0.13}_{-1.40}$ & 2.17$^{+1.42}_{-0.12}$ & 1.09$^{+0.35}_{-0.28}$ \\
SMC\_A15 & 4.12$^{+0.41}_{-0.95}$ & 2.52$^{+0.42}_{-0.55}$ & 0.90$^{+0.33}_{-0.28}$  & 5.07$^{+0.27}_{-1.38}$ & 2.86$^{+0.88}_{-0.78}$ & 0.99$^{+0.08}_{-0.24}$  & 4.59$^{+0.17}_{-1.30}$ & 2.34$^{+0.89}_{-0.30}$ & 1.04$^{+0.37}_{-0.49}$ \\
SMC\_A16 & 4.12$^{+0.39}_{-0.55}$ & 2.27$^{+1.27}_{-0.21}$ & 0.60$^{+0.54}_{-0.08}$  & 4.93$^{+0.62}_{-0.83}$ & 2.14$^{+1.42}_{-0.28}$ & 0.63$^{+0.22}_{-0.10}$  & 5.15$^{+0.34}_{-1.44}$ & 1.98$^{+1.61}_{-0.13}$ & 0.60$^{+0.37}_{-0.06}$ \\
SMC\_A17 & 4.16$^{+0.58}_{-0.82}$ & 2.89$^{+0.41}_{-0.44}$ & 1.51$^{+0.24}_{-0.17}$  & 5.26$^{+0.51}_{-1.41}$ & 3.10$^{+1.03}_{-0.83}$ & 1.24$^{+0.88}_{-0.03}$  & 5.50$^{+0.01}_{-1.96}$ & 2.86$^{+0.67}_{-0.31}$ & 1.54$^{+0.19}_{-0.10}$ \\
SMC\_A18 & 4.80$^{+0.87}_{-0.44}$ & 2.47$^{+0.51}_{-0.30}$ & 0.66$^{+0.23}_{-0.30}$  & 5.41$^{+1.88}_{-0.57}$ & 2.51$^{+1.09}_{-0.56}$ & 0.55$^{+0.35}_{-0.20}$  & 4.82$^{+1.42}_{-0.80}$ & 2.15$^{+0.73}_{-0.09}$ & 0.60$^{+0.15}_{-0.16}$ \\
SMC\_A19 & 4.75$^{+1.42}_{-0.52}$ & 3.34$^{+1.78}_{-0.28}$ & 1.38$^{+0.72}_{-0.28}$  & 5.90$^{+0.54}_{-1.72}$ & 4.17$^{+0.52}_{-1.34}$ & 1.48$^{+0.59}_{-0.33}$  & 4.98$^{+0.59}_{-1.32}$ & 3.59$^{+0.53}_{-1.09}$ & 1.52$^{+0.34}_{-0.65}$ \\
SMC\_A20 & 4.11$^{+0.49}_{-0.68}$ & 2.58$^{+0.83}_{-0.20}$ & 1.43$^{+0.39}_{-0.26}$  & 4.68$^{+0.57}_{-0.67}$ & 2.22$^{+1.48}_{-0.07}$ & 1.11$^{+0.57}_{-0.01}$  & 4.97$^{+1.01}_{-0.84}$ & 2.57$^{+1.23}_{-0.03}$ & 1.39$^{+0.41}_{-0.02}$ \\
SMC\_A21 & 4.88$^{+0.25}_{-1.54}$ & 2.37$^{+0.96}_{-0.15}$ & 0.90$^{+0.40}_{-0.39}$  & 5.66$^{+0.01}_{-1.84}$ & 2.10$^{+1.75}_{-0.16}$ & 1.02$^{+0.05}_{-0.49}$  & 4.57$^{+1.70}_{-0.92}$ & 2.20$^{+1.56}_{-0.15}$ & 1.26$^{+0.21}_{-0.72}$ \\
SMC\_A22 & 5.14$^{+0.48}_{-1.00}$ & 3.26$^{+0.73}_{-0.38}$ & 1.57$^{+0.51}_{-0.10}$  & 5.34$^{+0.86}_{-0.58}$ & 3.26$^{+1.91}_{-0.31}$ & 1.38$^{+1.12}_{-0.10}$  & 4.62$^{+0.99}_{-0.93}$ & 3.12$^{+0.70}_{-0.39}$ & 1.56$^{+0.29}_{-0.14}$ \\
SMC\_A23 & 5.84$^{+0.90}_{-0.80}$ & 3.78$^{+1.43}_{-0.31}$ & 1.91$^{+1.63}_{-0.31}$  & 5.68$^{+1.59}_{-0.64}$ & 3.67$^{+2.10}_{-0.56}$ & 1.69$^{+1.32}_{-0.27}$  & 6.26$^{+2.01}_{-0.54}$ & 3.44$^{+2.89}_{-0.22}$ & 1.70$^{+1.97}_{-0.11}$ \\
SMC\_A24 & 5.15$^{+0.20}_{-1.16}$ & 2.78$^{+0.68}_{-0.29}$ & 1.18$^{+0.51}_{-0.35}$  & 6.58$^{+1.52}_{-1.50}$ & 3.13$^{+1.60}_{-0.21}$ & 1.06$^{+0.77}_{-0.20}$  & 6.68$^{+1.46}_{-2.26}$ & 3.16$^{+1.27}_{-0.47}$ & 1.41$^{+0.45}_{-0.43}$ \\
SMC\_A25 & 4.83$^{+1.15}_{-0.64}$ & 3.82$^{+0.57}_{-0.71}$ & 2.37$^{+0.88}_{-0.44}$  & 5.35$^{+1.61}_{-1.06}$ & 3.05$^{+2.14}_{-0.48}$ & 2.25$^{+1.26}_{-0.52}$  & 5.90$^{+0.63}_{-2.12}$ & 3.06$^{+0.81}_{-0.30}$ & 2.02$^{+0.63}_{-0.22}$ \\
SMC\_A26 & 3.44$^{+0.76}_{-0.02}$ & 1.90$^{+0.35}_{-0.12}$ & 0.73$^{+0.21}_{-0.09}$  & 4.18$^{+0.70}_{-0.61}$ & 1.99$^{+0.64}_{-0.38}$ & 0.67$^{+0.17}_{-0.01}$  & 3.77$^{+0.75}_{-0.06}$ & 1.83$^{+0.18}_{-0.19}$ & 0.73$^{+0.21}_{-0.04}$ \\
SMC\_A27 & 4.64$^{+0.90}_{-1.49}$ & 2.61$^{+0.53}_{-0.35}$ & 1.79$^{+0.54}_{-0.39}$  & 5.08$^{+1.90}_{-1.19}$ & 2.48$^{+0.67}_{-0.50}$ & 1.78$^{+0.12}_{-0.31}$  & 6.21$^{+2.48}_{-2.39}$ & 2.72$^{+1.16}_{-0.43}$ & 1.87$^{+0.31}_{-0.25}$ \\
SMC\_A28 & 5.80$^{+0.21}_{-1.38}$ & 3.29$^{+0.50}_{-0.84}$ & 1.34$^{+0.12}_{-0.44}$  & 7.14$^{+0.68}_{-2.35}$ & 3.20$^{+1.30}_{-0.99}$ & 1.09$^{+0.38}_{-0.16}$  & 6.35$^{+2.00}_{-1.92}$ & 3.18$^{+1.15}_{-0.63}$ & 1.32$^{+0.24}_{-0.36}$ \\
SMC\_A29 & 5.35$^{+1.42}_{-0.94}$ & 4.47$^{+0.03}_{-1.51}$ & 1.33$^{+0.36}_{-0.50}$  & 6.43$^{+0.76}_{-1.73}$ & 3.02$^{+2.06}_{-0.36}$ & 1.09$^{+0.74}_{-0.13}$  & 7.98$^{+0.81}_{-3.04}$ & 3.00$^{+2.52}_{-0.08}$ & 1.21$^{+0.38}_{-0.17}$ \\
SMC\_A30 & 7.26$^{+0.21}_{-1.62}$ & 4.83$^{+1.23}_{-0.42}$ & 3.66$^{+2.03}_{-0.45}$  & 8.72$^{+0.07}_{-2.44}$ & 4.81$^{+2.42}_{-0.31}$ & 3.83$^{+2.64}_{-0.63}$  & 8.11$^{+1.56}_{-1.37}$ & 4.78$^{+3.02}_{-0.31}$ & 3.74$^{+2.17}_{-0.49}$ \\
SMC\_A31 & 8.58$^{+0.32}_{-2.23}$ & 5.80$^{+1.69}_{-0.83}$ & 3.89$^{+2.06}_{-0.64}$  & 9.62$^{+1.24}_{-1.53}$ & 6.24$^{+2.21}_{-0.75}$ & 3.66$^{+2.75}_{-0.39}$  & 8.55$^{+1.27}_{-1.23}$ & 5.62$^{+2.14}_{-0.41}$ & 3.68$^{+2.09}_{-0.61}$ \\
SMC\_A32 & 11.4$^{+0.74}_{-5.24}$ & 4.60$^{+1.62}_{-1.04}$ & 2.96$^{+1.03}_{-0.52}$  & 12.2$^{+0.64}_{-6.17}$ & 4.56$^{+2.23}_{-1.13}$ & 3.32$^{+0.91}_{-0.92}$  & 12.7$^{+0.06}_{-6.57}$ & 4.70$^{+2.09}_{-1.05}$ & 3.01$^{+0.90}_{-0.50}$ \\
SMC\_A33 & 12.7$^{+0.02}_{-4.24}$ & 4.60$^{+2.16}_{-0.64}$ & 3.05$^{+1.39}_{-0.28}$  & 12.6$^{+0.04}_{-4.41}$ & 4.31$^{+2.73}_{-0.14}$ & 2.80$^{+2.19}_{-0.22}$  & 12.6$^{+0.18}_{-3.69}$ & 4.91$^{+1.24}_{-1.03}$ & 2.76$^{+1.44}_{-0.36}$ \\
\enddata
\end{deluxetable}

\end{document}